\documentclass[prd,aps,amsfonts,eqsecnum,superscriptaddress,nofootinbib,longbibliography,notitlepage,twocolumn]{revtex4-2}

\usepackage{framed}
\usepackage{graphicx}
\usepackage[dvipsnames]{xcolor}
\usepackage{rotating}
\usepackage{amsmath,amssymb,graphics,amsthm} 
\usepackage{bbm}
\usepackage{bm}
\usepackage[hidelinks]{hyperref}
\usepackage{cleveref}
\usepackage{mdframed}
\usepackage{mathrsfs}
\DeclareUnicodeCharacter{210F}{\hbar}

\crefformat{equation}{(#2#1#3)}
\crefformat{align}{(#2#1#3)}
\crefformat{section}{Sec.~#2#1#3}
\crefformat{appendix}{App.~#2#1#3}
\crefformat{subsection}{Sec.~#2#1#3}
\crefmultiformat{equation}{(#2#1#3,}{ #2#1#3)}{ #2#1#3,}{ #2#1#3)}
\crefmultiformat{align}{(#2#1#3,}{ #2#1#3)}{ #2#1#3,}{ #2#1#3)}
\crefmultiformat{section}{Secs.~(#2#1#3,}{ #2#1#3)}{ #2#1#3,}{ #2#1#3)}

\usepackage{slashed}
\usepackage{comment}

\usepackage{tikz}
\usetikzlibrary{shapes.geometric, arrows}
\usetikzlibrary{shapes.multipart, arrows}
\usetikzlibrary{decorations.pathreplacing}

\allowdisplaybreaks

\numberwithin{thm}{section}

\newcommand{\nn}{\nonumber\\}

\makeatletter
\renewcommand{\p@subsection}{}
\renewcommand{\p@subsubsection}{}
\makeatother

\usepackage[dvipsnames]{xcolor}
\usepackage{mathtools}
\usepackage{subcaption}

\newcommand{\secref}[1]{Section\;\ref{#1}}

\newcommand{\ud}{\mathrm{d}}

\newcommand\cO{\mathcal{O}}
\newcommand\cP{\mathcal{P}}

\newcommand\cF{\mathcal{F}}

\newcommand{\ii}{\mathrm{i}}

\newcommand{\dv}{\partial}
\newcommand{\wmn}{\omega^{\mu\nu}}
\newcommand{\wab}{\omega^{\alpha\beta}}
\newcommand{\wsl}{\omega^{\sigma\lambda}}
\newcommand{\sacomm}[2]{\left\{#1\,\overset{\star},\,#2\right\}}
\newcommand{\sacommu}[2]{\{#1\,\overset{\star},\,#2\}}
\newcommand{\scomm}[2]{\left[#1\,\overset{\star},\,#2\right]}
\newcommand{\scommu}[2]{[#1\,\overset{\star},\,#2]}
\newcommand{\comm}[2]{\left[#1\,,#2\right]}
\newcommand{\commu}[2]{[#1\,,#2]}
\newcommand{\acomm}[2]{\left\{#1\,,#2\right\}}
\newcommand{\acommu}[2]{\{#1\,,#2\}}
\newcommand{\sff}{{\sf F}}
\newcommand{\sfh}{{\sf H}}
\newcommand{\dd}{\textrm{d}}
\DeclareMathOperator{\diag}{diag}

\DeclareMathOperator{\sexp}{E}

\let\Re\relax
\let\Im\relax
\DeclareMathOperator{\Re}{Re}
\DeclareMathOperator{\Im}{Im}

\makeatletter
\DeclareRobustCommand{\cev}[1]{%
  {\mathpalette\do@cev{#1}}%
}
\newcommand{\do@cev}[2]{%
  \vbox{\offinterlineskip
    \sbox\z@{$\m@th#1 x$}%
    \ialign{##\cr
      \hidewidth\reflectbox{$\m@th#1\vec{}\mkern4mu$}\hidewidth\cr
      \noalign{\kern-\ht\z@}
      $\m@th#1#2$\cr
    }%
  }%
}
\makeatother

\newcommand{\im}{\mathrm{Im}~}
\newcommand{\re}{\mathrm{Re}~}
\newcommand{\tr}{\mathrm{tr}}

\newcommand{\p}{\partial}

\newcommand{\abs}[1]{\lvert{#1}\rvert}

\newcommand{\F}{\mathrm{F}}
\newcommand{\FS}{\mathrm{FS}}

\newcommand{\ul}{\underline}

\newcommand{\red}[1]{#1}

\usepackage{dsfont}

\begin{document}

\title{Quantum geometry from the Moyal product: quantum kinetic equation and non-linear response}

\author{Takamori Park}
\affiliation{Department of Physics, University of California, Santa Barbara, CA 93106, USA}

\author{Xiaoyang Huang}
\affiliation{Department of Physics and Center for Theory of Quantum Matter, University of Colorado, Boulder, CO 80309, USA}
\affiliation{Kavli Institute for Theoretical Physics, University of California, Santa Barbara, CA 93106, USA}

\author{Lucile Savary}
\affiliation{French American Center for Theoretical Science, CNRS,
  KITP, Santa Barbara, CA 93106-4030, USA}
\affiliation{Kavli Institute for Theoretical Physics, University of
  California, Santa Barbara, CA 93106, USA}
\affiliation{\'Ecole Normale Supérieure de Lyon, CNRS, Laboratoire de
  physique, 46, all\'ee d’Italie, 69007 Lyon, France}

\author{Leon Balents}
\affiliation{Kavli Institute for Theoretical Physics, University of
  California, Santa Barbara, CA 93106, USA}
\affiliation{French American Center for Theoretical Science, CNRS,
  KITP, Santa Barbara, CA 93106-4030, USA}
\affiliation{Canadian Institute for Advanced Research, Toronto, Ontario, M5G 1M1, Canada}

\date{\today}

\begin{abstract}

We systematically derive the dissipationless quantum kinetic equation for a multi-band free fermionic system with U(1) symmetry.  Using the Moyal product formalism, we fully band-diagonalize the dynamics.  Expanding to the second order in gradients, which is beyond the semiclassical limit, we give a complete analysis of the band-resolved thermodynamics and transport properties, especially those arising from the quantum geometric tensor. We apply our framework to a Bloch band theory under electric fields near equilibrium and find the linear and nonlinear transport coefficients.   We also obtain the dynamical density-density response functions in the metallic case, including quantum metric corrections.  Our results and approach can be applied very generally to multi-band problems even in situations with spatially varying Hamiltonians and distributions.

\end{abstract}

\maketitle

\section{Introduction}

Kinetic equations describe the time
evolution of the phase space density of an ensemble of
particles. Such equations have quantum and classical versions, and can provide a means to investigate transport and responses of many body systems.  Kinetic equations provide an alternative to the method of Kubo formulae, which are formally exact perturbative calculations of time-dependent response to applied forces.  Compared to the latter, kinetic equations have some advantages: they are more intuitive, can more easily treat inhomogeneous systems, and directly yield the non-equilibrium distribution function of the system under study. 

While the most famous kinetic equation is the semiclassical Boltzmann equation
for phase space densities, capturing the quantum coherence of particles requires
using and ``diagonalizing'' quantum kinetic equations which involve full density
{\em matrices}. Although these equations can be formulated exactly in phase phase
using the Moyal formalism \cite{Moyal_1949},
diagonalizing them exactly is usually not
possible. However, this formalism is well suited to the case of small
gradients, for which one can {\em systematically} expand the equations order by order
in gradients, subsequently diagonalize them, and express observables using
``single-band quantities.''
Under this procedure, the quantum coherence effects,
in particular the multiple-band effects, manifest in each individual band
through ``geometric terms'', which involve derivatives of the diagonalization
matrix (and in turn eigenstates of all bands).  At first order in the gradients,
one recovers the usual Boltzmann's equation with the proper velocity and force
using this formalism \cite{culcer_geometrical,gosselin2007,gosselinRecursiveDiagonalizationQuantum2008,wickles2013,Sekine_quantum,mangeolle2024quantum}. At
that order, the real, momentum and mixed space components of the Berry curvature
enter. Identifying some coefficients of phase space derivatives of the distribution function with phase space velocities, one can also recover the single particle
equations of motion of the center of a wavepacket in the ``wavepacket
formalism'' \cite{Niu_review}, which can be further formulated as symplectic mechanics \cite{Son_chiral,Son:2012wh,Chen:2016lra,Huang_effective}.

In addition to the Berry curvature, the \emph{quantum metric} describes another fundamental aspect of quantum geometry.  It arises as the real part of the quantum geometric tensor (while the Berry curvature constitutes its imaginary part) \cite{provost1980,resta2011,savary2025analytic,zhang2025identifying}. The quantum metric quantifies the distance between neighboring quantum states and plays a crucial role in non-adiabatic responses. Although identifying physical observables that depend solely on the quantum metric remains challenging, significant efforts have been made to characterize quantum-metric-induced linear and nonlinear responses, employing both perturbation theory and the wavepacket formalisms \cite{Gao_field,Lapa_nonuniform,gaoNonreciprocalDirectionalDichroism2019a,Yan_prl_2024}. 

Here, we push the Moyal product formalism to second order in the gradients for 
particle-conserving systems of non-interacting fermions
without assuming translation symmetry. \red{Our approach and results are closely related to the pioneering works of Blount\cite{Blount_bloch,Blount_extension,blount_1962_formalisms}, but extend and systematize the procedure to greater generality and to higher accuracy.}  We show that the Berry curvature
gains corrections at this order, and that the quantum metric tensor and a symmetric
interband coherence tensor modify the kinetic equation and the observables.  Our results are general and exact to second order in the gradients, and provide a unified framework that can be applied to many different problems including transport and dynamical response.  Notably, using this
systematic approach, we find in particular that the wavepacket
formalism does {\em not} capture all contributions to response functions,
and working up to second order also allows us to highlight, even at
first order, an ambiguity in the definition of band diagonal quantities.

\subsection{Main results in the case of a separable Hamiltonian}
\label{sec:main_results}

The general setting of this paper is a $d+1$-dimensional non-interacting fermion system with the
particle-conserving Hamiltonian%
\footnote{We define $\int_x\equiv\int \dd^d x$ and
$\int_p\equiv\int\frac{\dd^d p}{(2\pi\hbar)^d}$ for integrals.}
\begin{equation}
  \hat{\rm H}(t)=\sum_{n,m=1}^N
  \int_{x,x'}\hat\psi_n^\dagger(x){\sf H}_{nm}(x,x',t)\hat\psi_m(x'),
  \label{eq:hamiltonian}
\end{equation}
where $x=(x_1,\ldots,x_d)$ and $\hat\psi^\dagger_n,\hat\psi_n$ $(n=1,\ldots,N)$ are fermionic fields that
satisfy the anti-commutation relations
$\acomm{\hat\psi_n(x)}{\hat\psi_m^\dagger(x')} =\delta_{nm}\delta^d(x-x')$.

Within our formalism, this problem defined with quantum mechanical operators is
mapped to a problem defined on phase space. We define a
diagonalized Hamiltonian $h(x,p,t)$ and fermion distribution function $f(x,p,t)$
which are both $N\times N$ diagonal matrices defined on phase space in such a
way that they satisfy the equations
\begin{equation}
  \label{eq:8}
  {\rm N}(t)=\int_{x,p}\tr[f(x,p,t)],
\end{equation}
\begin{equation}
  {\rm E}(t)=\langle  {\rm \hat H}(t)\rangle
  =\int_{x,p}\tr[h(x,p,t)f(x,p,t)],
  \label{eq:toten_intro}
\end{equation}
where ${\rm N}$  and ${\rm E}$ are the total number of fermions and the
expectation value of the Hamiltonian. Band structures, in the conventional sense,
can only be defined for systems with translation symmetry, but by defining
$f(x,p,t)$ and $h(x,p,t)$ in this way, we can think of the diagonal elements of each
as the phase-space distribution functions and energies of a ``band''
even for systems without translation symmetry.

While this formalism that we will fully develop in \cref{sec:formalism} applies to any
particle-conserving non-interacting Hamiltonian of fermions, we are particularly
interested in the case of a separable Hamiltonian of the form
\begin{equation}
  \label{eq:5}
  \hat{\rm H}(t)=\int_x \hat\psi^\dagger(x) \left( {\sf H}_0(-i\hbar \nabla_x) -e
  V(x,t)I_N\right)\hat\psi(x)
\end{equation}
that describes a translationally invariant system with Hamiltonian
${\sf H}_0$ probed by an external electric potential $V$.
We assume ${\sf H}_0$ is a $N\times N$ matrix Hamiltonian with
matrix eigenvalues $\{\varepsilon_n\}$ and eigenstates $\{|u_n\rangle\}$
for $n=1,\ldots N$,
and we show that the diagonal distribution function $f(x,p=\hbar k,t)$ satisfies
the kinetic equation
\begin{align}
    \dv_t f=&-v\cdot\nabla_x f
    -\frac{e}{\hbar}\nabla_x V\cdot\nabla_p f
-\frac{e}{\hbar}\dv_{x_i}f {\sf \Omega}_{ij}\dv_{x_j} V\notag\\
&+\frac{1}{24\hbar}\dv^3_{x_i x_j x_l}f
    \dv^3_{k_i k_j k_l} \varepsilon+\frac{e}{24\hbar}\dv^3_{k_i k_j k_l}f
    \dv^3_{x_i x_j x_l}V\notag\\
&-\dv_{x_i}\bigg[f\bigg(
    e\,{\sf t}_{ij}(\dv_{t}+v\cdot\nabla_x)\dv_{x_j}V\notag\\
    &\hspace{40pt}+\frac{e^2}{2\hbar}\left(2\dv_{k_j}{\sf t}_{i l}
    -\dv_{k_i} {\sf t}_{jl}\right)\dv_{x_j}V \dv_{x_l}V
    \bigg)\bigg]\notag\\
    & +\frac{e}{2\hbar}\dv_{x_i}(f\dv^2_{x_jx_l}V)\dv_{k_i}{\sf g}_{jl}
    +\frac{e}{2\hbar}\dv^2_{x_i x_l}(f
    \dv_{x_j} V)\dv_{k_j}{\sf g}_{il}\notag\\
    &-\frac{e}{\hbar}\dv^2_{k_i x_j}(f
    {\sf g}_{jl}\dv^2_{x_i x_l}V)
    +\mathcal{O}(\dv_x^4),
    \label{eq:intro_ke}
\end{align}
where $i,j,l=1,..,d$. Let us describe \eqref{eq:intro_ke}. All quantities involved are either diagonal matrices or scalars. Hence, the
equation above actually describes $N$ separate kinetic equations for each band.
We will routinely use this approach to express equations for each band as a
single equation using diagonal matrices. In addition, we assumed the Einstein
summation convention which we will continue to adopt for the rest of the paper. 
It should also be noted that the kinetic equation is expressed using wave numbers
$k=p/\hbar$ rather than momenta to adhere to the standard convention when working
with Bloch states.
In \cref{eq:intro_ke}, $v_i=\hbar^{-1}\dv_{k_i}\varepsilon$ is the group velocity;
${\sf\Omega}_{ij}={\sf\Omega}_{k_ik_j}$ is the Berry curvature; ${\sf g}_{ij}={\sf g}_{k_ik_j}$ is the quantum metric;
and ${\sf t}_{ij}$ is an interband coherence
term with $n$-th diagonal element
\begin{equation}
  {\sf t}_{n,ij}\equiv-\sum_{m\neq n}\frac{\langle\dv_{k_i}u_n|u_m\rangle
  \langle u_m|\dv_{k_j}u_n\rangle}
  {\varepsilon_n-\varepsilon_m}+(i\leftrightarrow j).
  \label{eq:bnqm_intro}
\end{equation}
Existing literature refers to this quantity as 
the Berry connection polarizability \cite{Gao_field,lai_third-order_2021} or
the band-normalized quantum metric \cite{wang_quantum-metric-induced_2023,Yan_prl_2024}.
The first line in the kinetic equation is the usual Boltzmann equation
with the Berry curvature contribution from the anomalous velocity \cite{Niu_review}.
The second line is a band-geometry independent correction to the Boltzmann
equation coming from expanding to second order in gradients. The
third and fourth lines describe interband effects and the last two lines describe
the effects of the quantum metric.

Using a relaxation time approximation to approximate the collision integral
$\mathcal{I}[f]$, we solve the kinetic equation perturbatively in $e$ to
calculate the electric current; see \eqref{eq:J2} and \eqref{eq:J3} for the whole expression. At the order of linear response, we 
find the latter contains the usual Drude and Hall term, and additionally, a contribution
coming from the quantum-metric dipole (QMD) tensor, $\dv_{k_l}{\sf g}_{ij}$, defined as:
\begin{equation}
    \widehat{J}_{i}^{\textrm{QMD}}(q,\omega)=e^2\left(\sum_{n=1}^N\int_k  n_{F,n}
    \dv_{k_l}{\sf g}_{ij}\right)q_jq_l\widehat{V}(q,\omega)
    \label{eq:QMDcurrent}
\end{equation}
where $n_{F,n}\equiv n_F(\varepsilon_n)$ is the Fermi-Dirac distribution function for
the $n$-th band and $\widehat{V}$ is the Fourier transformed potential $V$ \cite{gaoNonreciprocalDirectionalDichroism2019a}.
In addition, going to second-order response, we find the
nonlinear Hall effect as expected \cite{sodemann2015} as well as contributions
from the ``band-normalized quantum metric'' \cite{Yan_prl_2024}.

A similar calculation by Lapa and Hughes in Ref.~\cite{Lapa_nonuniform}
identified a contribution to the linear electrical response from the QMD tensor
as well. Lapa and Hughes generalized the semiclassical wavepacket formalism to
include effects of the quantum metric into the equations of motion of the wave
packet, which they used as input to the Boltzmann equation in solving for the electric
current. However, we observe that the overall prefactor as well as the
permutation of indices of the QMD tensor in their result are different from
\cref{eq:QMDcurrent}. At the same time, our results of second-order response are distinct at various terms from Ref.~\cite{Yan_prl_2024}, which is also a generalization of the wavepacket formalism by quantum geometry. It is important to note that the above differences are not subject to the ambiguity in defining the current operator by divergence-free terms. 
We believe that this discrepancy can be explained by the
fact that, in Ref.~\cite{Lapa_nonuniform,Yan_prl_2024}, the Boltzmann equation was not properly generalized along with the wave
packet equations of motions. In general, the
semiclassical wavepacket approach is difficult to systematically generalize in
contrast to the formalism we develop in this paper.

Moreover, by expanding the distribution function perturbatively in external fields, we obtain the dynamical $n$-point density correlation functions in \eqref{eq:C2} and \eqref{eq:C3} with both intraband and interband contributions. By integrating over the frequency, we obtain the static structure factor with (i) a term $\propto q$ as in usual single-band Fermi liquid and (ii) a term $\propto q^3$ due to the quantum metric in \eqref{eq:Sq intra}.

\subsection{Outline of the manuscript}
The main body of this paper is divided into three parts and is organized as
follows. In \cref{sec:formalism}, we begin by first introducing the Wigner
transformation and the Moyal product. The results in that section are obtained
by expanding the Moyal product to second-order in $\hbar$. We then define the matrix-valued
phase-space distribution function $\sff$ (whose Moyal diagonalization will lead to $f$ introduced above) and its kinetic equation which together
form the starting point of our formalism. We introduce the
``Moyal diagonalization'' of the Hamiltonian which gives us a definition for
bands in phase-space (see however the discussion in \cref{sec:gauge_invt_hf}) and also leads to an emergent gauge structure. We show that
for a time-independent Hamiltonian or for a slowly-varying time-dependent
Hamiltonian, ``Moyal diagonalization'' transforms the kinetic equation for the
matrix-valued $\sff$ into a set of $N$ decoupled scalar kinetic equations where
$N$ is the number of bands in phase-space. We also develop all of the necessary
machinery to perform practical calculations using our formalism and derive an
expression for the electric current. We make several assumptions to develop our
formalism further, namely assume that the Hamiltonian is 1) non-interacting; 2) slowly
varying in space which is necessary for the validity of the $\hbar$ expansion;
3) slowly varying in time which allows us to ignore interband transitions in time; 4)
$U(1)$ charge symmetric; and 5) non-degenerate. As we will discuss in
\cref{sec:conclusion}, some of these assumptions can be dropped by generalizing
our formalism.

In \cref{sec:application}, we apply the developed formalism to the case of a
translationally-invariant system that is perturbed by an external electric
potential (which we introduced earlier in \cref{sec:main_results}). Assuming a
time-independent electric potential, we calculate the equilibrium electric
current. We also solve the kinetic equation for the case of a time-dependent
electric potential using the relaxation time approximation and we calculate the
non-equilibrium electric current from which we can identify the
contribution from the QMD. We also consider the collisionless limit in order to calculate the corrections to the two-point and three-point density correlation functions 
of a Fermi liquid coming from band geometry.

We conclude in \cref{sec:conclusion} with an extended discussion of different aspects of our formalism, a summary of results, and potential future applications and extensions.

\section{Formalism}\label{sec:formalism}

\subsection{Setup}\label{sec:setup}
We begin by considering a non-interacting fermion system with the Hamiltonian
from \cref{eq:hamiltonian}.  We define the density matrix
\begin{equation}\label{eq: def density matrix}
  \begin{split}
    {\sf F}_{nm}(x,x',t)
    &\equiv\langle\hat\psi_m^\dagger(x',t)\hat\psi_n^{\vphantom{\dagger}}(x,t)\rangle,
  \end{split}
\end{equation}
where $\hat\psi_{n}(x,t),\hat\psi_m^{\dagger}(x',t)$ are defined in the
Heisenberg picture.  The time-dependence of the density matrix is dictated by
the Liouville-von Neumann equation
\begin{align}
  \label{eq:lvn}
  &i\hbar\dv_t\sff (x,x',t)\\
  &\qquad=\int_y \left(\sfh(x,y,t)\sff(y,x',t)-\sff(x,y,t)\sfh(y,x',t)\right),\notag
\end{align}
where matrix multiplication is implicit.  In principle, we can solve for the
density matrix using \cref{eq:lvn} and calculate a variety of observables such
as total energy, number (energy) density, number (energy) current, etc. with
the solution, but this is an intractable task for a general matrix Hamiltonian.

In order to make the problem tractable, following
\cite{mangeolle2024quantum}, we use the Wigner transformation to express the
density matrix and other bilocal fields such as $\sfh$ as functions defined over
phase space. The Moyal or ``star'' product which appears in place of operator products is
denoted by $\star$ and is defined as
\begin{equation}
  \begin{split}
  \star =&
  \exp\left(\frac{i\hbar}{2}\left(\cev{\nabla}_x\cdot\vec{\nabla}_p
  -\cev{\nabla}_x\cdot\vec{\nabla}_p\right)\right)\\
  =&\exp\left(\frac{i\hbar}{2}\wab\cev{\dv}_\alpha\vec{\dv}_\beta\right),
  \end{split}
\end{equation}
where $\omega$ is an antisymmetric tensor with indices that run over both
position and momentum coordinates and is defined as $\omega^{x_i
p_j}=-\omega^{p_i x_j}=\delta_{ij}$ and $\omega^{x_i x_j}=\omega^{p_i p_j}=0$ [note that here we do not carry out a Wigner transformation in time but rather consider a single time].
In general, Greek letters, e.g.\ $\alpha,\mu=1,..,2d$, will be used to denote indices that run over
phase-space coordinates and Roman letters, e.g.\ $i=1,..,d$, will be used to denote the indices of
either position or momentum coordinates. Hereafter, we assume all functions are
defined over phase space unless otherwise stated.

Typically, Moyal products are not evaluated exactly. Rather, they are evaluated
by treating $\hbar$ as a small parameter and expanding the exponential as a
series. Since each factor of $\hbar$ comes with a single position and momentum
derivative, the power of $\hbar$ equals the number of additional position and
momentum derivatives introduced in an expression after expanding the Moyal
product. If the system of interest varies sufficiently slowly in space, this
expansion in $\hbar$ becomes controlled. In this work, we will carry out this
expansion to second order in $\hbar$. In \cref{sec:application}, we will
choose to express our formulas using wave numbers $k=p/\hbar$ rather than momenta
which leads to the loss of the $\hbar$ in the Moyal product. In that case,
the expansion can be controlled by directly counting the number of position
derivatives.

Under the Wigner transformation, \cref{eq:lvn} becomes
\begin{equation}
  \label{eq:eom f star}
  \begin{split}
    i\hbar\dv_t \sff=&\scomm{\sfh}{\sff}
  \end{split}
\end{equation}
where $\sff$ and $\sfh$ are now matrix-valued functions defined on phase space
and $\scomm{\cdot}{\cdot}$ is defined as the commutator with respect to the
matrix and Moyal products. The density matrix is now defined on phase space, and
we refer to $\sff$ as the \emph{distribution function} and the Liouville-von
Neumann equation as its kinetic equation. In this work, we operate under the assumption that $\sff\rightarrow0$ as $\abs{x},\abs{p}\rightarrow\infty$ so that the boundary terms of phase-space integrals vanish when integrating by parts. 

Since $\sff$ is matrix valued, by expanding the Moyal product in
\cref{eq:eom f star}, this kinetic equation becomes a set of coupled
differential equations. We will now discuss the
diagonalization of the Hamiltonian such that this kinetic equation can be
transformed into a set of $N$ decoupled differential equations for $N$
scalar-valued distribution function.

\subsection{Moyal diagonalization}\label{sec:moyal-diag}
Following Refs.~\cite{gosselin2007,wickles2013,mangeolle2024quantum}, we introduce
a unitary matrix $U$, which in the phase-space formalism satisfies the equation
\begin{equation}
  U\star U^\dagger=U^\dagger\star U=I_N,
  \label{eq:unitaritycondition}
\end{equation}
where $I_N$ is the $N\times N$ identity matrix.  We then define the 
``Moyal-diagonalization''
of $\sfh$ as
\begin{align}\label{eq:diagonalization}
     U^\dagger \star  {\sf H}\star  U
     \equiv \tilde h
     =\begin{pmatrix}
       \tilde h_1 & &\\
       &\ddots&\\
       &&\tilde h_N
     \end{pmatrix}
\end{align}
where $\tilde h=\tilde h(x,p)$ is a diagonal matrix that is a priori not a
constant but a function on phase space.\footnote{In general, lowercase letters will be used to denote diagonal
quantities except for the Berry connection $A$
which will be introduced shortly.} In this work, $\{\tilde h_1,\cdots,\tilde h_N\}$ are assumed to be non-degenerate.
If the Hamiltonian is
translation-invariant, $\tilde h=\tilde h(p)$ is the dispersion relations of the
band structure in the system. For an inhomogeneous system, $\tilde h=\tilde h(x,p)$ is a
function on phase space, and it gives us an operational definition for
a band structure even in the absence of translation symmetry. These bands
defined on phase space can be thought of as semi-classically corrected bands.
Further discussion on this can be found in \cref{sec:conclusion_band_geometric}.

Just like $\tilde h$, in general, quantities that are associated with a single
band such as the Berry curvature $\Omega_{n,\alpha\beta}$ or the quantum metric
$g_{n,\alpha\beta}$, which we will introduce shortly, will usually be expressed
using diagonal matrices. Hence, a trace over these diagonal quantities is a summation
over all the bands.

The diagonalization equation \cref{eq:diagonalcurrent} can be solved by
expanding the Moyal product order-by-order in $\hbar$. At zeroth order, it is
\begin{equation}
  U^{(0)\dagger}\sfh U^{(0)}=\tilde h^{(0)},
\end{equation}
where $U^{(0)},U^{(0)\dagger}, \tilde h^{(0)}$ denote the solutions for the
zeroth-order equation. Therefore, at lowest order, $\tilde h_n^{(0)}$ are simply
the matrix eigenvalues of $\sfh$. By going to finite orders in $\hbar$, we can
calculate corrections to these eigenvalues that come from the spatial variation
of $\sfh$.  Closed form expressions for these corrections will be derived in \cref{sec:diagonalizing_h}.

We can take the star-unitary operators that are used to diagonalize $\sfh$ to
define $\tilde F\equiv U^\dagger\star \sff\star U$. Although $\sfh$ is
diagonalized by $U$, $\tilde F$ is generally not a diagonal matrix. However only the diagonal
components of $\tilde F$ contribute to
observables such as the total particle number and energy. For convenience, we separate the diagonal
and off-diagonal components of $\tilde F$ labeling them $\tilde f$ and $\tilde{\mathcal{F}}$
respectively such that $\tilde F=\tilde f+\tilde{\mathcal{F}}$.
 
\subsection{Gauge symmetry}\label{sec:gauge_transform}
In diagonalizing the Hamiltonian in \eqref{eq:diagonalization}, we introduced a
U(1)$^{\otimes N}$ \emph{gauge redundancy} meaning there are multiple $\star$-unitary
transformations that can diagonalize the Hamiltonian.  For simplicity, if we
assume the diagonalized Hamiltonian does not have any degeneracies, i.e.
$\tilde h_n\neq \tilde h_m$ if $n\neq m$, we define the gauge transformation as
\begin{equation}\label{eq:gauge symmetry}
  \begin{split}
  U&\to U\star \sexp^{i\theta},\\
  U^\dagger&\to \sexp^{-i\theta}\star U^\dagger,
  \end{split}
\end{equation}
where $\theta=\theta(x,p)$ is a diagonal matrix and $\sexp^{(\cdot)}$ is the
``star-exponential'' defined as
\begin{equation}\label{eq:star E}
  \begin{split}
    \mathrm{E}^{\lambda}\equiv&1+\lambda + \frac{1}{2}\lambda\star \lambda+\cdots 
      + \frac{1}{n!}\lambda^{\star n}+\cdots\\
      =&\mathrm{e}^\lambda+\mathcal{O}(\hbar^2).
  \end{split}
\end{equation}
To first order in $\hbar$, the ``star-exponential'' is equivalent to the
exponential function defined using regular multiplication.  Under this gauge
transformation, the diagonalized Hamiltonian is still diagonal,
$\tilde{h}=U^\dagger\star \sfh\star U\rightarrow \sexp^{-i\theta}
\star\tilde{h}\star\sexp^{i\theta}$, but $\tilde h$ is {\em not}
gauge invariant. By expanding in powers of $\hbar$, we can show that under this gauge transformation, the diagonalized Hamiltonian
$\tilde{h}$ as well as the diagonal component of the distribution function $\tilde{f}$
transform as
\begin{align}
  \tilde{d}\rightarrow&\sexp^{-i\theta}\star \tilde{d}\star \sexp^{i\theta}\notag\\
  &=\tilde{d}-\hbar\wab \dv_\alpha
  \tilde{d}\left(\dv_\beta\theta+\frac{\hbar}{2}\wsl
  \dv^2_{\beta\lambda}\theta\dv_\sigma\theta\right)\notag\\
  &\quad+\frac{\hbar^2}{2}\wab\wsl
  \dv^2_{\alpha\sigma}\tilde{d}\dv_\beta\theta\dv_\lambda\theta
  +\mathcal{O}(\hbar^3),
  \label{eq:diagonaltransform}
\end{align}
where $\tilde{d}=\tilde{h}$ or $\tilde{f}$. Note that at zeroth order in
$\hbar$, $\tilde d$ is invariant under this transformation. In \eqref{eq:diagonaltransform}, we only considered the diagonal components of the distribution function
but the off-diagonal components are also gauge
dependent. In this work, we use a tilde to clearly denote quantities that are
gauge-dependent. 

Since we have a gauge symmetry, the gauge field associated with this gauge symmetry is
\begin{equation}
  \Lambda_\alpha\equiv -i U^\dagger\star \dv_\alpha U,
\end{equation}
and we define the diagonal components of this gauge field as the Berry connection
\begin{equation}
  \label{eq:connection}
  A_\alpha=\diag[\Lambda_\alpha].
\end{equation}
Under the gauge transformation \cref{eq:gauge symmetry}, the gauge field
transforms as
\begin{align}
  \Lambda_\alpha&\rightarrow \sexp^{-i\theta}\star
  \left(\Lambda_\alpha+i\sexp^{i\theta}\star\dv_\alpha\sexp^{-i\theta}\right)
  \star\sexp^{i\theta}\notag\\
  &=e^{-i\theta}\left(\Lambda_\alpha
  +\frac{\hbar}{2}\wsl\left(i\dv_\sigma\theta\Lambda_\alpha\dv_\lambda\theta
  -\acomm{\dv_\sigma\Lambda_\alpha}{\dv_\lambda\theta}\right)\right)e^{i\theta}\notag\\
  &\qquad+\dv_\alpha\theta
  +\frac{\hbar}{2}\wsl\dv_\sigma\theta\dv^2_{\alpha\lambda}\theta
  +\mathcal{O}(\hbar^2),
  \label{eq:Lambda transform}
\end{align}
and the Berry connection transforms as
\begin{align}
  A_\alpha&\rightarrow \sexp^{-i\theta}\star
  \left(A_\alpha+i\sexp^{i\theta}\star\dv_\alpha\sexp^{-i\theta}\right)
  \star\sexp^{i\theta}\notag\\
  &=A_\alpha+\dv_\alpha\theta-\hbar\wsl\dv_\sigma A_\alpha \dv_\lambda\theta\notag\\
  &\qquad
  +\frac{\hbar}{2}\wsl\dv_\sigma\theta\dv^2_{\alpha\lambda}\theta
  +\mathcal{O}(\hbar^2).
  \label{eq:Atransform}
\end{align}

Now that we have an understanding of the gauge structure, let us construct some
basic gauge-invariant quantities that will come in handy later.  Firstly, the
simplest gauge-invariant object we can consider is the band projection
operator.  We define the semiclassical-band projection operator $P_n$ of the
$n$-th semiclassical band as
\begin{equation}
  \label{eq:projection}
  P_n\equiv U\star p_n\star U^\dagger,
\end{equation}
where $p_n$ is a diagonal matrix with matrix elements
$(p_n)_{ij}=\delta_{ni}\delta_{nj}$ \cite{PhysRevB.96.214514}. One can check explicitly that $P_n$ is
invariant under the gauge transformation \cref{eq:gauge symmetry}. It also
follows from the unitarity of $U$ that the projection operator is star-idempotent and
complete: $P_n\star P_m=\delta_{nm} P_n$, $\sum_{n=1}^N P_n=I_N$.

In analogy with the quantum geometric tensor (QGT) defined in band theory \cite{provost1980,resta2011,savary2025analytic,zhang2025identifying}, we
define the phase-space QGT as
\begin{equation}
  \begin{split}
  \mathcal{T}_{n,\alpha\beta}\equiv&\tr\left[\dv_\alpha P_n \star(I-P_n)\star
  \dv_\beta P_n\right]\\
  =&g_{n,\alpha\beta}+\frac{i}{2}\Omega_{n,\alpha\beta}.
  \end{split}
  \label{eq:QGT}
\end{equation}
The symmetric part which is real is the phase-space quantum metric
$g_{n,\alpha\beta}$, and the antisymmetric part which is imaginary is the
phase-space Berry curvature $\Omega_{n,\alpha\beta}$. For brevity, we simply
refer to them as the quantum metric and Berry curvature.

From the definition above, the Berry curvature and quantum metric must be 
\begin{equation}
  \begin{split}
  \Omega_{\alpha\beta}=&\dv_\alpha A_\beta-\dv_\beta A_\alpha
  -\hbar\wsl\dv_\sigma A_\alpha \dv_\lambda A_\beta\\
  &+\hbar \wsl\dv_\sigma((\dv_\alpha A_\beta-\dv_\beta A_\alpha)A_\lambda)
  +\mathcal{O}(\hbar^2),
  \end{split}
  \label{eq:berrycurvature}
\end{equation}
and
\begin{equation}
  g_{\alpha\beta}=\frac{1}{2}\diag(\Lambda_\alpha\Lambda_\beta
  +\Lambda_\beta\Lambda_\alpha)
  -A_\alpha A_\beta+\mathcal{O}(\hbar).
  \label{eq:quantummetric}
\end{equation}
We only derived the quantum metric to zeroth order in $\hbar$ since it only
appears at that order in observables and the kinetic equation at second order in $\hbar$ as we will see later.  In addition, notice that to calculate the Berry curvature $\Omega$ at first in $\hbar$, in \cref{eq:berrycurvature} we must use the the Berry connection $A$ up to first-order. In \cref{sec:diagonalizing_h}, we provide a formula, \cref{eq:A1}, to calculate this term.

\subsection{Gauge-invariant $h,f$}\label{sec:gauge_invt_hf}
As we discussed in the previous section, the diagonal quantities $\tilde h$ and
$\tilde f$ are gauge dependent. However, we can define a gauge-invariant form of
these quantities which we denote $h,f$ respectively. We make the following choice for the gauge-invariant
distribution function $f$ 
\begin{equation}
  \begin{split}
  f\equiv&\tilde f+\hbar\wab\dv_\alpha(\tilde f A_\beta)\\
  &+\frac{\hbar^2}{4}\wab\wsl
  \dv^2_{\alpha\sigma}(\tilde f \diag(\acomm{\Lambda_\beta}{\Lambda_\lambda}))
  +\mathcal{O}(\hbar^3),
  \end{split}
  \label{eq:invtf}
\end{equation}
and we choose the gauge-invariant diagonalized Hamiltonian $h$ to be
\begin{align}
    h\equiv&\tilde h+\hbar\wab\dv_\alpha\tilde h(A_\beta+\hbar\wsl A_\lambda
    \dv_\sigma A_\beta)\notag\\
    &-\frac{\hbar^2}{4}\wab\wsl\dv^2_{\alpha\sigma}\tilde h
    \left(\diag(\acomm{\Lambda_\beta}{\Lambda_\lambda})-4A_\beta
    A_\lambda\right)+\mathcal{O}(\hbar^3).
  \label{eq:invth}
\end{align}
Although it is tedious, \cref{eq:diagonaltransform,eq:Lambda
transform,eq:Atransform} can be used to show that $f,h$ are gauge-invariant.

As mentioned in the introduction \cref{eq:8,eq:toten_intro} the gauge-invariant distribution function $f$ is defined such that the total number of fermions is
\begin{equation}
  {\rm N}=\int_{x,p}\tr[\sff]=\int_{x,p}\tr[f],
  \label{eq:f_property}
\end{equation}
and $h$ is defined such that the total energy is given by
\begin{equation}\label{eq:E total}
  {\rm E}=\langle\hat H\rangle=
  \int_{x,p}\frac{1}{2}\tr[\{\sff\overset{\star}{,} \sfh\}]=\int_{x,p}\tr[fh].
\end{equation}
Note, the off-diagonal components of $\tilde{F}$, $\tilde{\mathcal{F}}$, do not contribute to either
of these quantities.
The details of how \cref{eq:invtf,eq:invth} are obtained are outlined in \cref{app:defining_fh}.

Let us finally emphasize the following point. While $f$ and $h$ defined in \cref{eq:invtf,eq:invth} are gauge invariant and satisfy \cref{eq:f_property,eq:E total}, they do not constitute a unique such {\em choice} and may therefore not  unequivocally represent {\em the} band filling and band energy, respectively. This is a manifestation of the fact that there exist no true ``bands'' when the system is not translationally invariant because momentum is in that case not a good quantum number. This notion of choice may seem surprising given the familiar semiclassical textbook equations. Indeed, up to first order in $\hbar$, some gauge-invariant choices for $f$ and $h$ are such that their relation to $\tilde{f}$ and $\tilde{h}$ can be recast in the form of a change of variables, $f(x,p)=\tilde{f}(x_i+\hbar A_{p_i},p_i-\hbar A_{x_i})$ (see e.g.\ Ref.~\cite{mangeolle2024quantum}). In that case, ${\rm N}\sim\int{\rm tr}[\mathfrak{J}f]$, with $\mathfrak{J}=1+\hbar\Omega_{x_ip_i}$ the Jacobian of this change of variables. One may also define ``single particle'' quantities such as $\left[\frac{dx}{dt}\right]$ and $\left[\frac{dk}{dt}\right]$ which directly enter the kinetic equation, which takes the simple form $\partial_tf+\hbar\left[\frac{dx_i}{dt}\right]\partial_{x_i}f+\hbar\left[\frac{dk_i}{dt}\right]\partial_{k_i}f=0$, thereby justifying a center-of-mass and mean-momentum wavepacket analysis and the picture of a single particle evolving adiabatically in the ``original'' (translationally invariant) bands with some corrections. This makes this choice of $f,h$ {\em natural}, but we emphasize that even in that case it remains a choice. Note that the choices we make in this manuscript, \cref{eq:invtf,eq:invth} do {\em not} reduce to the form of a change of variables when truncated to first order, because there exists no such ``change of variable'' simplification at second order, and we instead impose the  constraints \cref{eq:f_property,eq:E total}, which allow, at least at the level of these equations, the interpretation of $f$ as a ``band'' density which integrates to the total number of particles and consequently $h$ a local energy density such that the integral of $fh$ is the total energy.

\subsection{Kinetic equation}\label{sec:kinetic_equation}
In this section, we discuss the kinetic equation of the distribution function.
From \cref{eq:lvn}, $\tilde F=U^\dagger\star\sff\star U$ satisfies the kinetic equation
\begin{equation}
  i\hbar\dv_t \tilde{F}=\scommu{\tilde{h}+\hbar\Lambda_t}{\tilde{F}},
  \label{eq:bandlvn}
\end{equation}
where $\Lambda_t\equiv -i U^\dagger\star\dv_t U$. For time-independent
Hamiltonians and for a specific class of time-dependent Hamiltonians
that can be expressed as $\sfh(t)=\sum_{n=1}^N P_n\star h_n(t)\star P_n$ where $\{P_n\}$
are time-independent projection operators and $\{h_n(t)\}$ are time-dependent scalar functions
(c.f. \cref{app:projection_operator_representation}),
$\Lambda_t=0$.
In this case, if $\tilde F$ is diagonal at some past time, it will remain diagonal at later times.
However, here, we assume $\Lambda_t\neq0$ and proceed assuming $\tilde F$ is generally non-diagonal.
\cref{eq:bandlvn} can be separated into two coupled kinetic equations
for the diagonal ($\tilde f$) and off-diagonal ($\tilde \cF$) components of $\tilde F$:
\begin{align}
    i\hbar \dv_t\tilde f=&\scommu{\tilde h+\hbar A_t}{\tilde f}
    +\hbar\diag(\scommu{\Lambda_t}{\tilde{\mathcal{F}}}),
    \label{eq:tildef_kinetic_eq}\\[1em]
    i\hbar\dv_t\tilde{\mathcal{F}}=&
    \scommu{\tilde h+\hbar \Lambda_t}{\tilde{\mathcal{F}}}
    -\hbar\diag(\scommu{\Lambda_t}{\tilde{\mathcal{F}}})
    \label{eq:offdiagonal_eq}\\
    &+\hbar\scommu{\Lambda_t-A_t}{\tilde f}.\notag
\end{align}
From \cref{eq:tildef_kinetic_eq}, the kinetic equation for the gauge-invariant
diagonal distribution function $f$ is
\begin{widetext}
\begin{align}
  \label{eq:boltzmanneq}
  \dv_t f=&\frac{1}{i\hbar}\scommu{h}{f}-\dv_\alpha\left[\wab f\left(
  \hbar(\wsl\dv_\sigma h\Omega_{\lambda\beta}-\Omega_{t\beta})
   \left(1-\frac{\hbar}{2}\wmn\Omega_{\mu\nu}\right)
  +\frac{\hbar^2}{2}\wsl\wmn\dv^2_{\sigma\mu}h\dv_\beta g_{\nu\lambda}\right)\right]\\
  &-\dv^2_{\alpha\sigma}\left[\hbar^2\wab\wsl
  f\left(\frac{1}{2}\wmn\dv_\mu h\dv_\nu g_{\beta\lambda}+\wmn\dv^2_{\mu\beta}hg_{\nu\lambda}
  -\frac{1}{2}\dv_t g_{\beta\lambda}
  \right)
  \right]
  \notag\\
  &-i\diag\left(\scommu{\Lambda_t}{\tilde{\mathcal{F}}}
  +\hbar\wab\dv_\alpha(\scommu{\Lambda_t}{\tilde{\mathcal{F}}}A_\beta)
  +\frac{\hbar^2}{4}\dv^2_{\alpha\sigma}
  (\scommu{\Lambda_t}{\tilde{\mathcal{F}}}
  \diag(\sacomm{\Lambda_\beta}{\Lambda_\lambda}))
  \right)+\mathcal{O}(\hbar^3),\notag
\end{align}
\end{widetext}
where $\Omega_{t\beta}$ is defined by simply extending our definition of the
Berry curvature to also include time as an index. The last line of \eqref{eq:boltzmanneq} comes from the
off-diagonal components of the distribution function.

\cref{eq:offdiagonal_eq} and \cref{eq:boltzmanneq} form a complete
set of kinetic equations, and in general, both must be solved simultaneously.
However, in some cases, we can take a limit that that allows us to ignore the
off-diagonal components of the distribution function. For example, in
\cref{sec:application}, we will consider a translationally-invariant system that
is perturbed by an external electric potential. For frequencies much smaller
than the band gap nearest the Fermi energy, $ \omega\ll \omega_\textrm{gap}$,
the off-diagonal components of the distribution function are negligible, and 
only \cref{eq:boltzmanneq} which becomes a set of $N$
decoupled differential equations needs to be solved. On the other hand, for phenomena that involve
optical interband transitions such as the shift current or injection current,
the off-diagonal components are crucial and cannot be ignored. In this work, we
will primarily assume the off-diagonal components are negligible and 
leave the study of a general non-diagonal distribution function to future works.

Under this assumption, the last line in \cref{eq:boltzmanneq} can be ignored. Even in that case we note that terms proportional to $f$ as well as higher-order
derivatives of $f$ emerge which do not fit the form of a linear partial differential equation, i.e. $c_a \partial_a f = 0$ where $a$ runs over whole spacetime. The
particular form of the kinetic equation in \cref{eq:boltzmanneq} is due to the
choice of the gauge-invariant diagonal Hamiltonian and distribution functions
defined in \cref{eq:invth,eq:invtf}. Nevertheless, in general, the kinetic equation
cannot be written in the form of a collisionless Boltzmann equation because the {\em phase-space}
flow of a quantum system is compressible and does not obey Liouville's theorem
\cite{curtright_concise_2014}.

\subsection{Equilibrium distribution function $f_\textrm{eq}$}\label{sec:eq_dist}

In the previous section we derived the kinetic equation for the gauge-invariant
distribution function $f$ which can be used to calculate the distribution
function out of equilibrium. However, the exact, Moyal-diagonalized distribution function at equilibrium is already non-trivial at second order in $\hbar$, and is necessary for evaluating equilibrium quantities, as well as to provide a starting point for an expansion near-equilibrium.  In this section, we will derive the equilibrium
distribution function and show that it is also diagonal in the basis in which
the Hamiltonian is diagonal.

The equilibrium distribution function in the original (``orbital'') basis at temperature
$T=1/(k_B\beta)$ can be expressed using the imaginary time Green's function as
\begin{equation}
  \sff_{\textrm{eq}}(x,p)=\frac{1}{\beta}\sum_{i\omega_n} e^{i\omega_n
  0^+}{\sf G}(x,p;i\omega_n),
  \label{eq:Feq}
\end{equation}
where $\sf G$ is the imaginary-time Green's function in the Wigner
representation and the sum is over fermionic Matsubara frequencies,
$\omega_n=2\pi(n+1/2)/\beta$. ${\sf G}$ itself can be
formally defined by the equations
\begin{equation}
  \begin{split}
    \left(i\omega_n-\sfh+\mu\right)\star{\sf G}(x,p;i\omega_n)=&I_N,\\
    {\sf{G}}(x,p;i\omega_n)\star\left(i\omega_n-\sfh+\mu\right)=&I_N.
  \end{split}
\end{equation}
If we now apply the star-unitary operator $U$ which is used to diagonalize the
Hamiltonian and define $\tilde{G}\equiv U^\dagger\star {\sf G}\star U$, these
equations become
\begin{equation}
  \begin{split}
    (i\omega_n-\tilde{h}+\mu)\star \tilde G(x,p;i\omega_n)=&I_N\\
    \tilde G(x,p;i\omega_n)\star(i\omega_n-\tilde{h}+\mu)=&I_N.
  \end{split}
  \label{eq:imagGFdef}
\end{equation}
Since $i\omega_n-\tilde h+\mu$ is diagonal, $\tilde{G}$ and the equilibrium
distribution function must be diagonal in this basis.

It is straightforward to solve for $\tilde G$ by perturbatively
expanding in powers of $\hbar$. The distribution function can be obtained by summing 
the Green's functions over the Matsubara frequencies
$\tilde f_\textrm{eq}
=\beta^{-1}\sum_{i\omega_n}e^{i\omega_n0^+}\tilde{G}(x,p;i\omega_n)$. This gives us
\begin{align}\label{eq:tilde f eq}
  \tilde{f}_\textrm{eq}=& n_F(\tilde h)-\frac{\hbar^2}{16}
  n_F''(\tilde h)\wab\wsl\dv^2_{\alpha\sigma}\tilde{h}\dv^2_{\beta\lambda}\tilde{h}\\
  &-\frac{\hbar^2}{24}
  n_F'''(\tilde h)\wab\wsl\dv^2_{\alpha\sigma}\tilde{h}
  \dv_\beta\tilde{h}\dv_\lambda\tilde{h}
  +\mathcal{O}(\hbar^3),\notag
\end{align}
where $n_F(x)=1/(e^{\beta (x-\mu)}+1)$
is the Fermi-Dirac distribution function. Using \cref{eq:invtf} and \cref{eq:invth}, the
gauge-invariant form of the equilibrium distribution function is
\begin{widetext}
\begin{align}
  f_{\textrm{eq}}=&\left(1+\frac{\hbar}{2}\wab
  \left(\Omega_{\alpha\beta}+\hbar
  \wsl\left(\frac{1}{2}\Omega_{\alpha\sigma}\Omega_{\beta\lambda}
  -\frac{1}{4}\Omega_{\alpha\beta}\Omega_{\sigma\lambda}
  +\dv^2_{\alpha\sigma}g_{\beta\lambda}\right)\right)\right)n_F
  +\hbar^2\wab\wsl n_F'
  \dv_\sigma(\dv_{\alpha}hg_{\beta\lambda})\notag\\
  &+\frac{\hbar^2}{2}n_F''\wab\wsl\left(\dv_\alpha h\dv_\sigma h g_{\beta\lambda}
  -\frac{1}{8}\dv^2_{\alpha\sigma}h\dv^2_{\beta\lambda}h\right)
  -\frac{\hbar^2}{24}n_F'''\wab\wsl\dv^2_{\alpha\sigma}h\dv_\beta h
  \dv_\lambda h+\mathcal{O}(\hbar^3).
  \label{eq:invteqdist}
\end{align}

\end{widetext}
where $n_F=n_F(h)$.
We provide the details of the calculation leading to \eqref{eq:invteqdist} in \cref{app:eq_dist}. $f_{\rm eq}$ in \eqref{eq:invteqdist} solves \eqref{eq:boltzmanneq} with $\partial_tf_{\rm eq}=0$ (equilibrium) and $\Lambda_t=0$.

\red{The equilibrium distribution of an insulator can be written as the divergence of the polarization. In App.~\ref{app:polarization}, we extract the polarization from \eqref{eq:invteqdist} and find partial agreement with the results of \cite{Di_polarization}, which uses the semiclassical method.}  

\subsection{Gauge-invariant diagonalized Hamiltonian $h$}\label{sec:diagonalizing_h}
In \cref{sec:moyal-diag}, we introduced the Moyal diagonalization of the
Hamiltonian defined by \cref{eq:diagonalization}. As we discussed, this equation
can be solved by expanding the Moyal product in powers of $\hbar$. To zeroth
order in $\hbar$, the diagonalization equation becomes a matrix diagonalization
problem, and the diagonalized Hamiltonian is a diagonal matrix formed by the
matrix eigenvalues of the Hamiltonian, $\tilde{h}^{(0)}_n$. In addition, as we mentioned in
\cref{sec:gauge_transform}, this zeroth order contribution is gauge-invariant, $\tilde{h}^{(0)}_n=h^{(0)}_n$.
By going to finite orders of $\hbar$, we can obtain corrections to $\tilde{h}^{(0)}$ that
come from the variation of the Hamiltonian in phase  space. 
To
keep track of the order of these corrections, we introduce the notation 
\begin{equation}
  \begin{split}
    \tilde h &= \tilde h^{(0)} + \tilde h^{(1)}+\cdots,\\
    \Lambda_\alpha &= \Lambda_\alpha^{(0)} + \Lambda_\alpha^{(1)}+\cdots,
  \end{split}
\end{equation}
where $\tilde h^{(n)},\Lambda_\alpha^{(n)}\sim \hbar^n$. As we mentioned
earlier, the power of $\hbar$ counts the number of position and momentum
derivatives, so $\tilde h^{(n)},\Lambda ^{(n)}_\alpha$ are correction terms with
$n$ additional position and momentum derivatives. This notation will also be
applied to other quantities. In this section, we compute explicit
expressions of the corrections to $h^{(0)} = \tilde h^{(0)}$.

One convenient tool to expand the diagonalization equation
\cref{eq:diagonalization} and calculate the corrections is what we call
``\emph{flow equations}'' which we introduce in \cref{app:diagonalization}, but
here, we simply present our results. For simplicity, we assume the matrix
eigenvalues $\tilde h^{(0)}$ are non-degenerate. For the $n$-th band, we define
the tensor
$m_{n,\alpha\beta}$
\begin{equation}
    m_{n,\alpha\beta}\equiv\Im\tr[\dv_\alpha P_n\star (\sfh-h_n)\star\dv_\beta P_n].
    \label{eq:m}
\end{equation}
which defines a ``full'' orbital magnetization, i.e. an extension of the usual orbital magnetization to the Moyal formalism.  Then, the first-order correction to the gauge-invariant diagonalized Hamiltonian
written as a diagonal matrix is
\begin{equation}
  h^{(1)}= \frac{\hbar}{2}\wab m^{(0)}_{\alpha\beta}=
  \frac{i\hbar}{4}\wab
  \diag(\acommu{\Lambda_{\alpha}^{(0)}}{\commu{\Lambda_\beta^{(0)}}{h^{(0)}}}).
  \label{eq:h1}
\end{equation}
The second order correction is
\begin{align}
  &h^{(2)}=\frac{\hbar}{4}\wab m^{(1)}_{\alpha\beta}
  +\frac{\hbar^2}{16}\wab\wsl h^{(0)}\Omega_{\alpha\beta}^{(0)}
  \Omega_{\sigma\lambda}^{(0)}\notag
    \\
  &+\frac{\hbar}{2}\wab\dv_\alpha h^{(0)}
    \big(A_{\beta}^{(1)}+\hbar\wsl\big(\dv_\sigma A_{\beta}^{(0)} A_{\lambda}^{(0)}
    +\frac{1}{2} A_{\sigma}^{(0)}\dv_\beta A_{\lambda}^{(0)}
    \big)\big)\notag\\
    &
  -\frac{\hbar^2}{4}\wab\wsl 
  (\dv_\alpha h^{(0)}\dv_\sigma g^{(0)}_{\beta\lambda}
  +2\dv^2_{\alpha\sigma} h^{(0)} g^{(0)}_{\beta\lambda}).
\label{eq:h2}
\end{align}

To evaluate $h^{(2)}$ and $m^{(1)}_{\alpha\beta}$ we also need to know the first order correction
to the Berry connection and the projection operator. These can also be derived using
their flow equations. They are given by
\begin{equation}
  \begin{split}
  A_\alpha^{(1)}=&\diag(\commu{\Lambda_\alpha^{(0)}}{Y^{(1)}})\\
  &+\frac{\hbar}{4}\wsl\diag(\acommu{\Lambda_\sigma^{(0)}}{\dv_\lambda
  \Lambda_\alpha^{(0)}}),
  \end{split}
  \label{eq:A1}
\end{equation}
\begin{align}
    P_n^{(1)}=&
    U^{(0)}\left(\commu{Y^{(1)}}{p_n}
    +\frac{\hbar}{4}\wab \acommu{\dv_\alpha\Lambda^{(0)}_\beta}{p_n}\right)U^{(0)\dagger}\notag\\
    &+\frac{i\hbar}{2}\wab U^{(0)}\Lambda_\alpha^{(0)}
    p_n\Lambda^{(0)}_\beta U^{(0)\dagger},
\label{eq:P1}
\end{align}
where $Y^{(1)}$ is a matrix with only off-diagonal elements given by
\begin{align}
  \label{eq:Y}
  &(Y^{(1)})_{nm}=\\
  &\hspace{30pt}\hbar\,\wab\frac{(\acommu{\Lambda_\beta^{(0)}}{\dv_\alpha h^{(0)}}
  -\frac{i}{2}\acommu{\Lambda^{(0)}_\alpha}{\commu{\Lambda^{(0)}_\beta}
  {h^{(0)}}})_{nm}
  }{2(h^{(0)}_n-h^{(0)}_m)}.\notag
\end{align}
$A_\alpha^{(0)}$ is simply $A_\alpha^{(0)}={\rm diag}\Lambda_\alpha^{(0)}$.

\subsection{Electric current density}\label{sec:number_current}
In \cref{sec:gauge_invt_hf}, we discussed how the total fermion number and energy can be obtained
from the gauge-invariant diagonalized Hamiltonian and distribution function $h$ and $f$. Another physical observable of interest is the electric
current density. In this section, we derive a general expression for the electric current density (in real space, ``momentum-integrated'') and find an explicit expression of the contributions to the current coming from only the diagonal components of the distribution function.

The electric current density can be obtained by introducing an external $U(1)$ gauge
field $\mathcal{A}_i(x)$ and taking the derivative of the Hamiltonian with
respect to the gauge field, i.e.\ $J_i(x)\equiv-\delta\langle{\rm \hat H}\rangle/\delta
\mathcal{A}_i(x)$ (see \cref{sec:current_derivation}). What we find to second order in $\hbar$ is
\begin{align}
\label{eq:current}
  J_i(x)=&-e\int_p\Re\tr\left[\frac{\dv \sfh(x,\pi)}{\dv
  p_i}\star\sff(x,p)\right]\\
  &-\frac{e\hbar^2}{12}\frac{\dv^2}{\dv x_k\dv x_l}\int_p\tr\left[\frac{\dv^3
  \sfh(x,\pi)}{\dv p_i\dv p_k\dv p_l}\sff\right]
  +\mathcal{O}(\hbar^3),\notag
\end{align}
where $\pi_i(x,p)\equiv p_i+e \mathcal{A}_i(x)$.
For simplicity, we turn off the external gauge field $\mathcal{A}=0$.

Assuming the off-diagonal components $\tilde{\mathcal{F}}$ are negligible, we
can express the current density using the diagonalized Hamiltonian $h$ and the diagonal
components of the distribution function $f$. The details of this calculation
are provided in \cref{app:deriving_diagonalcurrent}.
We find that the current density can be expressed as
\begin{widetext}
    \begin{align}
        J_i(x)=&-e\int_p\tr\left[f\left(\dv_{p_i}h
        -\hbar\wab\dv_\alpha h\,\Omega_{p_i\beta}
    \left(1-\hbar\Omega_{x_jp_j}\right)
    +\frac{\hbar^2}{2}\wab\wsl\dv^2_{\alpha\sigma}h\dv_{p_i} g_{\beta\lambda}
    \right)\right]\notag\\
    &-e\hbar\dv_{x_j}\int_p\tr[f(b_{i j}^s+b_{ij}^a)]
    -\frac{e\hbar^2}{12}\dv^2_{x_j x_k}
    \int_p\tr[f(b^s_{ijk}+2b^m_{ij;k})]
    +\mathcal{O}(\hbar^3),
    \label{eq:diagonalcurrent}
    \end{align}
\end{widetext}
where $b^s_{ij}$ and $b^a_{ij}$ are symmetric and anti-symmetric tensors defined as
\begin{equation}
\label{eq:bs2}
  b_{ij}^s\equiv\frac{\hbar}{2}\wsl(\dv_\sigma h \dv_\lambda
  g_{p_i p_j}+\dv^2_{\sigma p_i}h g_{p_j\lambda}+\dv^2_{\sigma p_j}h g_{p_i\lambda}),
\end{equation}
\begin{align}
\label{eq:ba}
  b_{ij}^a&\equiv
  -m_{p_i p_j}\left(1-\frac{\hbar}{2}\wsl\Omega_{\sigma\lambda}\right)\\
  &+\frac{\hbar}{2}\wsl
  (\dv_{p_i}(\dv_\sigma h g_{ p_j\lambda})
  -\dv_{p_j}(\dv_\sigma h g_{ p_i\lambda})),\notag
\end{align}
where $m_{p_i p_j}$  was defined in \cref{eq:m}.
$b^s_{ijk}$ is a fully symmetric tensor defined as
\begin{equation}
\label{eq:bs3}
  b_{ijk}^s\equiv -\frac{1}{2}\dv^3_{p_ip_jp_k}h,
\end{equation}
and $b^m_{ij;k}$ is a tensor with mixed symmetry defined as
\begin{equation}
\label{eq:bm}
  b_{ij;k}^m\equiv
  \dv_{p_i} c_{jk}-\dv_{p_j} c_{ik}
  +2(c_{jk;i}-c_{ik;j})
\end{equation}
and satisfies $b^m_{ij;k}=-b^m_{ji;k}$. Here, $c_{ik}, c_{ik;j}$ are diagonal matrices with $n$-th diagonal elements
\begin{align}
\label{eq:cnij}
    c_{n,ij}&\equiv \Re\tr\left[\dv_{p_i} P_n \star (\sfh-h_n)\star
   \dv_{p_j} P_n\right],\\
   &\notag\\
    c_{n,ik;j}&\equiv \Re\tr\left[\dv_{p_ip_k}^2 P_n\star (\sfh-h_n)\star
   \dv_{p_j} P_n\right].
   \label{eq:cnijk}
\end{align}
The contributions to current density from the terms $b^a_{ij}, b^m_{ij;k}$ are
divergenceless, so they correspond to bound currents. $c_{ij}$, which appears in $b_{ij;k}^m$, has been
referred to as the ``band Drude weight'' in existing
literature \cite{kang_measurements_2025} and its momentum integral over occupied
\cite{resta_geometrical_2017,resta_drude_2018}. 

By substituting the expression for the equilibrium distribution function
\cref{eq:invteqdist} into this expression, we can also obtain an expression for
the equilibrium current density, $J_i^{\textrm{eq}}(x)$. The equilibrium current is
divergenceless and can generally be expressed in the form
$J_i^{\textrm{eq}}(x)=
\dv_{x_j}\mathcal{M}_{ij}(x)$ where $\mathcal{M}_{ij}=-\mathcal{M}_{ji}$
is an antisymmetric tensor given by
\begin{widetext}
\begin{align}
\label{eq:eq_magnetization}
    \mathcal{M}_{ij}=&-e\hbar\int_p
    \tr\bigg[n_F\bigg(\dv_{p_i} h\left(A_{p_j}+\frac{\hbar}{2}
    \wmn\dv_\mu(A_{p_j} A_\nu)\right)
    +\frac{\hbar}{2}\wmn\dv_\nu h(A_{p_j} \dv_{p_i} A_\mu+A_\mu\dv_{p_j} A_{p_i} +A_{p_i}\dv_\mu A_{p_j})\\
    &\qquad-\frac{1}{2}m_{p_ip_j}+\frac{\hbar}{2}\wsl
    \dv_{p_i}(\dv_\sigma h g_{p_j\lambda})
    \bigg)\bigg]
    -\frac{e\hbar^2}{2}\wmn\int_p\tr[\dv_\mu(n_F g_{\nu p_j})\dv_{p_i} h]\notag\\
    &+\frac{e\hbar^2}{24}\wmn\int_p\tr[n_F' \dv^2_{p_j\mu}h\dv^2_{p_i\nu}h]
    -\frac{e\hbar^2}{12}\dv_{x_k}\int_p\tr[n_F b^m_{ij;k}]
    -(i\leftrightarrow j)
    +\mathcal{O}(\hbar^3),
    \notag
\end{align}
\end{widetext}
where $n_F=n_F(h)$. 
The expression derived above is true when there are no degeneracies
in the energy. As a consequence, the equilibrium current does no work in the bulk: the integral of the Joule heating under an electric field $E_i = - e\partial_{x_i}V$ vanishes, $\int_x J^{\mathrm{eq}}_i E_i = e\int_x \p_{x_i}J^{\mathrm{eq}}_i V = 0 $ where we used integration by parts.

\subsection{Uniform magnetic field}\label{sec:magnetic field}
A uniform magnetic field can be introduced in our formalism relatively simply by
changing variables from the phase-space coordinates $(x,p)$ to $(x,\pi)$ where
$\pi_i=p_i+e\mathcal{A}_i$ is kinematic momentum, where $\mathcal{A}_i$ is the external vector potential, as before. Under this change of
variables, the Moyal product becomes \cite{onodaTheoryNonEquilibirumStates2006,wickles2013,
yeCoadjointorbitBosonizationFermi2024, Ye:2024pty}
\begin{align}
    \star _B=&\exp\left(
    \frac{i\hbar}{2}(\cev{\nabla}_x\cdot\vec{\nabla}_\pi-\cev{\nabla}_\pi\cdot\vec{\nabla}_x)
    -\frac{ie\hbar}{2}\mathfrak{F}_{ij}\cev{\dv}_{\pi_i}\vec{\dv}_{\pi_j}\right)\notag\\
    \equiv&\exp\left(\frac{i\hbar}{2}\wab_B\cev{\dv}_\alpha\vec{\dv}_\beta\right),
\end{align}
where $\mathfrak{F}_{ij}=\dv_{x_i}\mathcal{A}_i-\dv_{x_j}\mathcal{A}_j$ is the \emph{constant} magnetic
field tensor. Here, we defined a new anti-symmetric tensor $\wab_B$ such that
$\omega_B^{\pi_i\pi_j}\equiv -e\mathfrak{F}_{ij}$ and $\omega_B^{x_i
\pi_j}=1$. Therefore, introducing a uniform magnetic field is as
simple as making the replacement $p_i\rightarrow \pi_i$ and $\wab\rightarrow\wab_B$.

In \cref{app:chiral anomaly}, we apply this magnetic Moyal product to the equilibrium current, and find that the divergence of the latter is non-vanishing at a Weyl node  
due to the chiral anomaly effect \cite{Son_chiral,Son_kinetic}. \red{In \cref{app:blount}, we derive the band-diagonal gauge-invariant Hamiltonian, showing that in a uniform magnetic field, it only depends on the kinetic momentum.} A more detailed study of the transport under the uniform magnetic field to the second order in $\hbar$ is left for the future.

\section{Application: Transport driven by electric fields}\label{sec:application}
In this section, we apply the formalism developed in \cref{sec:formalism} to
study a translationally invariant system perturbed by an externally applied
electric potential. We first find an expression for the equilibrium current
assuming a static potential that is bounded from below. Then, we discuss
the assumption of ignoring the off-diagonal components of the distribution
function, $\tilde{\mathcal{F}}$, and solve the kinetic equation to find the
current out of equilibrium. Lastly, we demonstrate the power of our
formalism by calculating the two- and three-point density correlation functions
of a collisionless Fermi liquid with non-trivial band geometry.

The Hamiltonian in the Wigner representation is given by
\begin{align}\label{eq:seperable H}
    {\sf H}(x,p,t) = {\sf H}_0(p) - e V(x,t)I_N,
\end{align}
where $\sfh_0(p)$ is a $N\times N$ Hamiltonian
that is generally non diagonal and $-e<0$ is the electric charge of the fermion.
As mentioned before, at
zeroth order in $\hbar$, the diagonalization equation reduces to a matrix
diagonalization problem, so the columns of $U^{(0)}$ are the eigenvectors of
$\sfh_0(p)$. If the eigenvector and eigenvalues of $\sfh_0(p)$ are
given by $\left\{|u_n(p)\rangle\right\}$ and $\left\{\varepsilon_n(p)\right\}$,
\begin{equation}
  U^{(0)}(p)=\bigg[|u_1(p)\rangle\,,~|u_2(p)\rangle\,,~\cdots\,,~|u_N(p)\rangle\bigg],
\end{equation}
\begin{equation}
  h^{(0)}(p)=
  \begin{pmatrix}
    \varepsilon_1(p)&&\\&\ddots&\\&&\varepsilon_N(p)
  \end{pmatrix},
\end{equation}
and the gauge field at zeroth order must be
\begin{align}
  \left(\Lambda^{(0)}_{p_i}\right)_{nm}&=-i\langle u_n|\dv_{p_i}u_m\rangle,\\
  \left(\Lambda^{(0)}_{x_i}\right)_{nm}&=0.
\end{align}

From \cref{eq:Y}, we find for $n\neq m$
\begin{equation}
    (Y^{(1)})_{nm}=-e\hbar\frac{(\Lambda_{p_i}^{(0)})_{nm}}{\varepsilon_n-\varepsilon_m}\dv_{x_i}V\,,
\end{equation}
and using \cref{eq:A1} we find that the first-order correction to $A^{(0)}$ is
\begin{equation}\label{eq: a shift}
    A^{(1)}_{n,p_i}=-e\hbar\,t_{n,ij}\dv_{x_j}V,
\end{equation}
where $t_{n,ij}$ is the symmetric tensor already defined in the introduction, \eqref{eq:bnqm_intro},
\begin{align}
\label{eq:bnqn_applications}
  t_{n,ij}
  \equiv&\,2\Re\tr\left[\dv_{p_i} {\sf P}_n \frac{1-{\sf P}_n}{\sfh_0-\varepsilon_n}
  \dv_{p_j} {\sf P}_n\right] \\
  =&\sum_{m\neq n}\frac{\langle\dv_{p_i}u_n|u_m\rangle
  \langle u_m|\dv_{p_j}u_n\rangle+(i\leftrightarrow j)}
  {\varepsilon_m-\varepsilon_n}.\notag
\end{align}
Here, ${\sf P}_n$ is the projection operator to zeroth order in $\hbar$ given
by ${\sf P}_n\equiv P_n^{(0)}=|u_n\rangle\langle u_n|$. \red{We find \eqref{eq: a shift} is in agreement with the ``positional shift'' in \cite{Gao_field}.}
As mentioned in the introduction, $t_{n,ij}$ in existing literature is referred to as either the ``Berry connection
polarizability'' \cite{Gao_field, lai_third-order_2021} or as the
``band-normalized quantum
metric'' \cite{wang_quantum-metric-induced_2023,Yan_prl_2024}. We want to
emphasize that in general this quantity is not a band-geometric quantity since
it is a multi-band quantity as its calculation relies on the knowledge of all bands. Only in the case of a two-band model can it truly be considered a
band-geometric quantity since the energy denominator in the summation is simply
replaced by the energy difference of the two bands.

Using \cref{eq:P1,} the first-order correction to this projection operator is
\begin{equation}
    P_n^{(1)}=-ie\hbar\dv_{x_l}V\comm{\dv_{p_l}{\sf P}_n}{\frac{1-{\sf P}_n}{\sfh_0-\varepsilon}}.
\end{equation}
Next, using \cref{eq:h1} and \cref{eq:h2} we can calculate the first- and second-order
corrections to $h^{(0)}$ to get
\begin{align}\label{eq: separable h}
  h_n(x,p) =& \varepsilon_n(p)-e V(x)\\
  &+\frac{e^2\hbar^2}{2} t_{n,ij}(p)\dv_{x_i} V(x)\dv_{x_j} V(x)
  +\mathcal{O}(\hbar^3)\notag.
\end{align}
The quantum metric is only nonzero for momentum indices and we find that
\begin{align}
  g_{n,p_i p_j}=&\frac{1}{2}\left(\langle\dv_{p_i}u_n|\dv_{p_j}u_n\rangle
  -\langle \dv_{p_i}u_n|u_n\rangle\langle u_n|\dv_{p_j}u_n\rangle\right)\notag\\
  &+\left(i\leftrightarrow j\right)+\mathcal{O}(\hbar^2).
\end{align}
Similarly, the only finite components of the Berry curvature are
\begin{align}
\label{eq:berry_curvature}
  \Omega_{p_i p_j}=&\dv_{p_i}A^{(0)}_{p_j}
  -\dv_{p_j}A^{(0)}_{p_i}\\
  &-e\hbar\left(\dv_{p_i}t_{jl}-\dv_{p_j}t_{il}\right)
  \dv_{x_l}V+\mathcal{O}(\hbar^2),\notag\\
  \Omega_{x_i p_j}=&-e\hbar\,t_{jl}\dv^2_{x_i x_l}V
  +\mathcal{O}(\hbar^2).
\end{align}
The correction to the Berry curvature in \cref{eq:berry_curvature}
matches the correction calculated in \cite{Yan_prl_2024}.

Assuming $V$ is time dependent, $\Lambda_t$ has nonzero off-diagonal components given by
\begin{equation}
    (\Lambda_t)_{nm}=ie\hbar\frac{(\Lambda_{p_i}^{(0)})_{nm}}{\varepsilon_n-\varepsilon_m}
    \dv^2_{tx_i }V+\mathcal{O}(\hbar^2)
\end{equation}
for $n\neq m$, and vanishing diagonal components.

In our work, we have taken derivatives with respect to momentum $p$ so that we
can use $\hbar$ as a parameter to expand the Moyal product, but quantities such
as the Berry curvature and quantum metric are typically defined using
derivatives of wave number $k=p/\hbar$. Therefore, we redefine the zeroth order Berry
curvature, quantum metric, and $t_{ij}$ using wave number derivatives:
\begin{align}
  {\sf \Lambda}_{i}(k)&\equiv \hbar \Lambda_{p_i}^{(0)}(\hbar k),\\
  {\sf A}_{n,i}(k)&\equiv \hbar A_{n,p_i}^{(0)}(\hbar k),\\
  {\sf \Omega}_{n,ij}(k)&\equiv \hbar^2 \Omega_{n,p_i p_j}^{(0)}
  (\hbar k),\\
  {\sf g}_{n,ij}(k)&\equiv \hbar^2 g_{n,p_i p_j}^{(0)}
  (\hbar k),\\
  {\sf t}_{n,ij}(k)&\equiv \hbar^2 t_{n,p_i p_j}(\hbar k).
\end{align}
Effectively, all we did was multiply by factors of $\hbar$ since wave number and
momentum are related by $p=\hbar k$. We emphasize, that once these substitutions
are made, the power of $\hbar$ no longer corresponds to the order of the Moyal
product expansion. Instead, we will directly count the number of spatial
derivatives.

\subsection{Equilibrium current}
Let us consider the case of a static external potential that is bounded from
below such that an equilibrium state exists and calculate the equilibrium
current. Firstly, we find that $m_{n,p_ip_j}$ defined by \cref{eq:m} is 
\begin{align}
\label{eq:m_appli}
    m_{n,p_ip_j}=&\frac{1}{e\hbar}{\sf m}_{n,ij}^\textrm{orb}\\
    &-\frac{e}{\hbar^2}\dv_{x_l}V
    (\dv_{k_i} {\sf g}_{n,jl}-\dv_{k_j}{\sf g}_{n,il}+{\sf s}_{n,ij;l})
    +\mathcal{O}(\dv_x^2),
    \notag
\end{align}
where ${\sf m}^\textrm{orb}_{n,ij}$ is the anti-symmetric orbital
magnetization tensor defined as
\cite{xiao_berry_2005, thonhauser_orbital_2005, shi_quantum_2007}
\begin{align}
\label{eq:morb}
    {\sf m}^\textrm{orb}_{n,ij}&\equiv\frac{e}{\hbar} \Im \tr[\dv_{k_i}
    {\sf P}_n (\sfh_0-\varepsilon_n)\dv_{k_j} {\sf P}_n]\\
    &=-\frac{ie}{2\hbar}\langle \dv_{k_i} u_n|
    \sfh_0-\varepsilon_n|\dv_{k_j} u_n\rangle-(i\leftrightarrow j),
    \notag
\end{align}
and ${\sf s}_{n,ij;l}$ is defined as
\begin{align}
\label{eq:snijl}
    {\sf s}_{n,ij;l}\equiv&
    \Re\tr\left[\left[\sfh_0-\varepsilon_n,\dv_{k_j}{\sf P}_n\right]
    \dv_{k_l} {\sf P}_n\dv_{k_i}\left(
    \frac{1-{\sf P}_n}{\sfh_0-\varepsilon_n}\right)
    \right]\notag\\[1em]
    &-(i\leftrightarrow j).
\end{align}
Then, the equilibrium current is given by $J^\textrm{eq}_i(x)=\dv_{x_j} \mathcal{M}_{ij}$
where $\mathcal{M}_{ij}$ is (cf.\ \eqref{eq:eq_magnetization})
\begin{widetext}
\begin{align}
    \mathcal{M}_{ij}=&\sum_{n=1}^N\int_k f_n\left(-e(v_{n,i} {\sf A}_{n,j}-v_{n,j} {\sf A}_{n,i})
    +{\sf m}_{n,ij}^\textrm{orb}\right)
    +\sum_{n=1}^N\frac{e^2}{\hbar}\int_kf_n
    \left(\hbar v_{n,i}{\sf t}_{n,jl}-\hbar v_{n,j}{\sf t}_{n,il}
    +\dv_{k_i}{\sf g}_{n,jl}-\dv_{k_j}{\sf g}_{n,il}+{\sf s}_{n,ij;l}\right)\dv_{x_l}V\notag\\
    &+\sum_{n=1}^N\frac{e^2}{6\hbar}\int_k f_n'(\dv_{k_i}{\sf c}_{n,jl}
    -\dv_{k_j}{\sf c}_{n,il}
    +2({\sf c}_{n,jl;i}-{\sf c}_{n,il;j}))\dv_{x_l}V
    +\mathcal{O}( \dv_x^3).
    \label{eq:magn_application}
\end{align}
\end{widetext}
Here, $f_n\equiv n_F(\varepsilon_n-eV)$, $v_{n,i}=\hbar^{-1}\dv_{k_i}\varepsilon_n$,
and (cf.\ \cref{eq:cnij,eq:cnijk}))
\begin{align}
\label{eq:cnij_appli}
  {\sf c}_{n,jl}\equiv\Re\tr[\dv_{k_j}{\sf P}_n(\sfh_0-\varepsilon_n)
  \dv_{k_l}{\sf P}_n],
\end{align}
\begin{align}
  {\sf c}_{n,jl;i}\equiv\Re\tr[\dv^2_{k_j k_l}{\sf P}_n(\sfh_0-\varepsilon_n)
  \dv_{k_i}{\sf P}_n].
  \label{eq:cnijk_appli}
\end{align}
The first two terms in the first line are expressed using the Berry connection which 
is gauge dependent. 
 However, as mentioned before, the current itself is gauge-invariant.
Indeed the derivative $\dv_{x_j}$ acts on $f_n$ to produce $-f_n'
e\dv_{x_i}V$. Then, by integration by parts, these two terms together can be rewritten as
$-(e^2/\hbar)\sum_{n=1}^N\int_k f_n{\sf \Omega}_{n,ij}\dv_{x_j}V$ which, of
course, is the anomalous Hall current. \red{Moreover, by including the orbital magnetization (the third term in the first line), the first-order terms of \eqref{eq:magn_application} agree with the semiclassical results in \cite{Niu_review,thonhauser_orbital_2005}.}
 
As mentioned earlier, this expression is valid when $\{\varepsilon_n\}$
are non-degenerate. For the case of a Weyl fermion, discussed in \cref{app:chiral anomaly}, the Berry connection ${\sf A}$ is
ill-defined at the Weyl point, so the magnetization $\mathcal{M}_{ij}$
cannot be defined.

\subsection{Non-equilibrium current}
We now turn to the case of a time-dependent external potential. 

\subsubsection{Kinetic equation}
\label{sec:kin-eq-noneq}
We begin by
discussing the off-diagonal components of the distribution function
$\tilde{\mathcal{F}}$ which obey the kinetic equation \cref{eq:offdiagonal_eq}.
We can solve for the off-diagonal components by expanding
perturbatively in $e$. To first order in $e$, $\tilde{\mathcal{F}}$ must satisfy the kinetic
equation
\begin{align}
\label{eq:ke_offdiag_outeq}
    &i\hbar\dv_t(\tilde{\mathcal{F}})_{nm}-
    \varepsilon_n\star(\tilde{\mathcal{F}})_{nm}
    +(\tilde{\mathcal{F}})_{nm}\star \varepsilon_m\\
    &\approx\hbar (\Lambda_t)_{nm}\star n_{F,m}
    -n_{F,n}\star\hbar (\Lambda_t)_{nm}+\mathcal{O}(e^2),\notag
\end{align}
where we assumed $\tilde{\mathcal{F}}$ vanishes in the absence of an
external potential. If we define the Fourier transform
$\tilde{\mathcal{F}}(x,k,t)=\int_{q,\omega}\widehat{\tilde{\mathcal{F}}}_{nm}(q,k,\omega)
e^{iq\cdot x-i\omega t}$, we find to first order in small $q$ (the expansion in $q$ is equivalent to the expansion in spatial gradients)
\begin{align}
    (\widehat{\tilde{\mathcal{F}}})_{nm}=&
    -ie\hbar\frac{({\sf\Lambda}_i)_{nm}}{\varepsilon_n-\varepsilon_m}
    \frac{n_{F,n}-n_{F,m}}{\hbar\omega-\varepsilon_n+\varepsilon_m} 
    \omega q_i\widehat{V}(q,\omega)\notag\\
    &+ \mathcal{O}(e^2,q^2),
\end{align}
where $n_{F,n}=n_F(\varepsilon_n)$. At this order, if $\hbar\omega, eV\ll
\varepsilon_n -\varepsilon_m$, we see that
$(\widehat{\tilde{\mathcal{F}}})_{nm}\sim \hbar\omega e\widehat
V/(\varepsilon_n-\varepsilon_m)^2\ll 1$. Therefore, in the low-frequency limit
$\omega\ll  \omega_\textrm{gap}$ where $\hbar\omega_\textrm{gap}$ is the
smallest of the band gaps neighboring the Fermi energy, the contribution from
$\tilde{\mathcal{F}}$ is negligible. Physically, this simply means the
off-diagonal components can be ignored if we consider frequencies that are
sufficiently small such that the external potential cannot induce interband
transitions. This is the justification for the assumption that we made earlier that $\tilde{\mathcal{F}}=0$.

Now, we only have to solve the kinetic equation for $f$ given by
\cref{eq:boltzmanneq} which becomes (this is \eqref{eq:intro_ke} with the addition of a collision integral $\mathcal{I}[f]$)
\begin{widetext}
\begin{align}
    \dv_t f=&-\hbar^{-1}\nabla_k\varepsilon\cdot\nabla_x f-e\hbar^{-1}\nabla_x V\cdot\nabla_k f
-\frac{e}{\hbar}\dv_{x_i}f {\sf \Omega}_{ij}\dv_{x_j} V
+\frac{1}{24\hbar}\dv^3_{x_i x_j x_l}f
    \dv^3_{k_i k_j k_l} \varepsilon+\frac{e}{24\hbar}\dv^3_{k_i k_j k_l}f
    \dv^3_{x_i x_j x_l}V\notag\\
&-\dv_{x_i}\left[f\left(
    e\,{\sf t}_{ij}(\dv_{t}+v_l\dv_{x_l})\dv_{x_j}V
    +\frac{e^2}{2\hbar}\left(2\dv_{k_j}{\sf t}_{i l}
    -\dv_{k_i} {\sf t}_{jl}\right)\dv_{x_j}V \dv_{x_l}V
    \right)\right]\notag\\
    &+\frac{e}{2\hbar}\dv_{x_i}(f\dv^2_{x_jx_l}V)\dv_{k_i}{\sf g}_{jl}
    +\frac{e}{2\hbar}\dv^2_{x_i x_l}(f
    \dv_{x_j} V)\dv_{k_j}{\sf g}_{il}
    -\frac{e}{\hbar}\dv^2_{k_i x_j}(f
    {\sf g}_{jl}\dv^2_{x_i x_l}V)+\mathcal{I}[f]
    +\mathcal{O}(\dv_x^4).
\label{eq:trans_inv_kinetic_eq}
\end{align}
\end{widetext}
We remind the reader that all the quantities mentioned above are diagonal
matrices. 

\subsubsection{Definition of the relaxation time approximation}
We added a collision integral to incorporate scattering effects which
we will approximate using the relaxation time approximation (RTA),
$\mathcal{I}[f]=-\tau^{-1}(f-f^r)$. The RTA can be understood as modeling
relaxation by coupling the system to a large external bath that maintains a
distribution function $f^r$ and allowing fermions to scatter between the system
and bath with a rate $\tau^{-1}$ (in particular, the total number of particles is not conserved). We choose the distribution function of the
bath such that the system would be in thermal equilibrium with the bath in the
absence of an external potential. The density matrix of the external bath in the
original basis before diagonalizing is $\sff^r=\sum_{n=1}^N{\sf P}_n
n_{F,n}$. If we diagonalize this and express it in its
gauge-invariant form we get the $n$-th diagonal component of $f^r$,
\begin{align}
\label{eq:f_r}
    &f^r_{n}=(1-e\dv^2_{x_ix_j} V {\sf t}_{n,ij})n_{F,n}\\
    &-e^2\sum_{m\neq n} \frac{n_{F,n}-n_{F,m}}
    {(\varepsilon_n-\varepsilon_m)^2}({\sf \Lambda}_i)_{nm}({\sf \Lambda}_j)_{mn}
    \dv_{x_i}V\dv_{x_j}V+\mathcal{O}(\dv_x^3).\notag
\end{align}
The details of the derivation are described
in \cref{app:f_r}.

\subsubsection{Solutions to the kinetic equation and current (expansion in small potential)}\label{sec:solutions-kin-eq}
\cref{eq:trans_inv_kinetic_eq} is then solved by letting $f=f^r+\delta f$ and solving
for $\delta f$ perturbatively in powers of $e$, $\delta f=\delta f_{e=1}+\delta
f_{e=2}+\cdots$ where $\delta f_{e=i}\sim e^i$ for $i=1,2,\ldots$. If we
define the Fourier transform, $\delta
f_{e=i}(x,k,t)=\int_{q,\omega}\widehat{\delta f}_{e=i}(q,k,\omega) e^{i q\cdot
x-i\omega t}$, we find
\begin{widetext}
\begin{align}
   \widehat{\delta f}_{e=1}(q,k,\omega)&= 
    \frac{(n_F(k+q/2)-n_F(k-q/2))\,(1- {\sf g}_{jl}q_jq_l)}
    {\hbar\omega-\varepsilon(k+q/2)+\varepsilon(k-q/2)+i\hbar\tau^{-1}}
    e\widehat{V}(q,\omega)+\mathcal{O}(q^3),
\end{align}
\begin{align}
    \widehat{\delta f}_{e=2}(q,k,\omega)&=
    e\int'_{\substack{q'q''\\\omega'\omega''}}
    \frac{i{\sf\Omega}_{ij}q_iq_j'+\hbar(\omega'-\vec{v}\cdot{\vec{q}\,}')
    {\sf t}_{ij}q_iq_j'
    +\frac{1}{2}\dv_{k_i}{\sf g}_{jl} (q''_iq'_jq'_l+q_i'q_j''q_l'')}
    {\hbar(\omega-\vec{v}\cdot\vec{q}+i\tau^{-1})}
    \widehat V(q',\omega')\widehat{\delta f}_{e=1}(q'',k,\omega'')\\[1em]
    &+e\int'_{\substack{q'q''\\\omega'\omega''}}
    \frac{1-{\sf g}_{ij}q_jq'_l}
    {\hbar(\omega-\vec{v}\cdot\vec{q}+i\tau^{-1})}
    \widehat V(q',\omega')
    \left(\widehat{\delta f}_{e=1}\left(q'',k+\frac{q'}{2},\omega''\right)-
    \widehat{\delta f}_{e=1}\left(q'',k-\frac{q'}{2},\omega''\right)\right)\notag\\[1em]
    &+e^2\int'_{\substack{q'q''\\\omega'\omega''}}
    \frac{n_F'v_i {\sf t}_{jl} q'_iq''_jq''_l}
    {\hbar(\omega-\vec{v}\cdot\vec{q}+i\tau^{-1})}
    \widehat{V}(q',\omega')\widehat V(q'',\omega'')-\frac{\omega-\vec{v}\cdot\vec{q}}
    {\omega-\vec{v}\cdot\vec{q}+i\tau^{-1}}\widehat{f}^r_{e=2}(q,k,\omega)+\mathcal{O}(q^3)\notag
\end{align}
\end{widetext}
where the primed integral indicates an implicit delta function
$\int'_{\substack{q'q''\\\omega'\omega''}}=\int_{\substack{q'q''\\\omega'\omega''}}
(2\pi)^{d+1}\delta^d(q-q'-q'')\delta(\omega-\omega'-\omega'')$, and $\widehat
V$, $\widehat{f}^r_{e=2}$ are the Fourier transforms of $V$ and the $e^2$ term
of $f^r$ respectively. For $\omega+i\hbar\tau^{-1}\neq0$, the expressions above are valid to order $q^4$, but in the limit $\omega+i\hbar\tau^{-1}\rightarrow0$, we find
they are valid to order $q^3$.
The solutions above can be used to calculate the electric
current in \eqref{eq:diagonalcurrent}. The $e$ contributions to the current vanish. The $e^2$ and $e^3$ contributions to the current are
\begin{widetext}
\begin{align}
    \widehat{J}_{i,e=2}(q,\omega)=&-e\int_k\tr\left[\widehat{\delta f}_{e=1}
    (q,k,\omega)
    \left(v_i-ie^{-1}{\sf m}_{ij}^\textrm{orb}q_j
    +\frac{1}{12\hbar}\left(\frac{1}{2}\dv_{k_ik_jk_l}^3
    \varepsilon-2(\dv_{k_i}{\sf c}_{jl}
    -\dv_{k_j}{\sf c}_{il}+2({\sf c}_{jl;i}
    -{\sf c}_{il;j}))\right)q_jq_l\right)\right]\notag\\
    &-\frac{e^2}{\hbar}\int_k\tr[n_F(i{\sf\Omega}_{ij}q_j
    +\hbar(v_i{\sf t}_{jl}-v_j{\sf t}_{il})q_jq_l
    +(\dv_{k_l}{\sf g}_{ij}-{\sf s}_{ij;l})q_jq_l)]\widehat{V}(q,\omega)
    +\mathcal{O}(q^3),
    \label{eq:Je2}
\end{align}
\begin{align}
    \widehat{J}_{i,e=3}(q,\omega)=&-e\int_k\tr\left[\widehat{\delta f}_{e=2}(q,k,\omega)
    \left(v_i-ie^{-1}{\sf m}_{ij}^\textrm{orb}q_j
    +\frac{1}{12\hbar}\left(\frac{1}{2}\dv_{k_ik_jk_l}^3
    \varepsilon-2(\dv_{k_i}{\sf c}_{jl}
    -\dv_{k_j}{\sf c}_{il}+2({\sf c}_{jl;i}
    -{\sf c}_{il;j}))\right)q_jq_l\right)\right]\notag\\
    &-\frac{e^2}{\hbar}\int'_{\substack{q'q''\\\omega'\omega''}}
    \int_k\widehat{V}(q',\omega')\tr\left[\widehat{\delta f}_{e=1}(q'',k,\omega'')
    \left(i{\sf\Omega}_{ij}q'_j-\hbar v_j{\sf t}_{il}q_j'q_l'-\frac{1}{2}
    \dv_{k_i}{\sf g}_{jl}q''_jq'_l
    +\frac{1}{2}(\dv_{k_l}{\sf g}_{ij}+\dv_{k_j}{\sf g}_{il}
    -2{\sf s}_{ij;l})q_jq_l'\right)\right]\notag\\
    &-\frac{e^3}{\hbar}\int'_{\substack{q'q''\\\omega'\omega''}}
    \int_k\widehat{V}(q',\omega')\widehat{V}(q'',\omega'')
    \tr\left[n_F\left(\frac{1}{2}\dv_{k_i}{\sf t}_{jl}
    -\dv_{k_j}{\sf t}_{il}\right)\right]q_j'q_l''
    -e\int_k\tr[\widehat{f}^r_{e=2}(q,k,\omega)v_i]
    +\mathcal{O}(q^3).
    \label{eq:Je3}
\end{align}
\end{widetext}
Here, $\widehat{J}_{i,e=2}, \widehat{J}_{i,e=3}$ describe linear and second-order electrical responses respectively, to second-order in $q$.

\subsubsection{$v_iq_i\ll|\omega+i\tau^{-1}|$ limit}
Let us now consider the limit in which $v_i
q_i,v_iq_i',v_iq_i''\ll\abs{\omega+i\tau^{-1}}$, so that we can expand the
denominators of $\widehat{\delta f}_{e=i}$ in powers of $q$.
In this limit $\widehat{J}_{i,e=2}(q,\omega)$ becomes
\begin{widetext}
\begin{align}\label{eq:J2}
    \widehat{J}_{i,e=2}(q,\omega)=&
    \frac{e^2}{\hbar}\left(
    \frac{\hbar\tau \langle v_i v_j\rangle_\textrm{FS}}
    {1-i\omega\tau}
    -\sum_{n=1}^N \int_k n_{F,n} {\sf \Omega}_{n,ij} \right)i q_j
    \widehat{V}(q,\omega)
    +e^2\left(
    \frac{\tau^2\langle v_i v_jv_l\rangle_\textrm{FS}}
    {(1-i\omega\tau)^2}
    -\langle v_l {\sf g}_{ij}\rangle_\textrm{FS}\, \right)
    q_jq_l\widehat{V}(q,\omega)\\
    &+\frac{e^2}{\hbar}\left(\frac{\hbar\tau/e}{1-i\omega\tau}
    \langle v_l {\sf m}^\textrm{orb}_{ij}\rangle_\textrm{FS}
    +\sum_{n=1}^N \int_k n_{F,n}
    (\hbar(v_{n,j}{\sf t}_{n,il}-v_{n,i}{\sf t}_{n,jl})
   +{\sf s}_{n,ij;l})\right)q_j q_l\widehat{V}(q,\omega)
    +\mathcal{O}(q^3),\notag
\end{align}
\end{widetext}
where we defined $\langle d\rangle_\textrm{FS}\equiv\sum_{n=1}^N\int_k n_{F,n}'
d_n$ as an integral over the Fermi surface. The first two terms are the familiar
Drude term ($\frac{e^2}{\hbar}
    \frac{\hbar\tau \langle v_i v_j\rangle_\textrm{FS}}
    {1-i\omega\tau}$) and Hall effect term ($-\frac{e^2}{\hbar}\sum_{n=1}^N \int_k n_{F,n} {\sf \Omega}_{n,ij}$) respectively. The other terms describe the
electric current linear response to a non-uniform electric field. Notice that all the terms
in the last line are divergenceless terms that describe edge currents and they
are also multi-band quantities (meaning their calculation requires knowledge of the
wavefunctions of multiple bands) in contrast to the terms in the
first line which are all single-band quantities (whose calculation only requires knowledge of
the wavefunction of a single band). Also note that a few terms in \eqref{eq:J2} remain nonzero even when the Fermi
energy is in a band gap at $T=0$. This does not however contradict what we expect for a band insulator since those terms which are nonzero
are divergenceless currents and conserve the local charge
distribution. 

Among the ``divergenceful'' terms is the contribution from the quantum
metric ($e\langle v_l{\sf g}_{ij}\rangle_{\rm FS}$). Physically, this contribution comes from the electric quadrupole moment
of the Bloch states  
which experiences a net force in a non-uniform electric
field \cite{Lapa_nonuniform}. Rewriting this term as an integral over the Fermi sea, it can also be expressed
as a contribution from the quantum-metric dipole (QMD), see \eqref{eq:QMDcurrent}.

A similar term was previously discussed in Ref.~\cite{Lapa_nonuniform} by Lapa
and Hughes. There, the authors generalized the semiclassical wave packet
formalism of Ref.~\cite{Niu_review} by introducing a correction to the wave packet energy that comes from
the electric quadrupole moment coupling to a non-uniform component of an electric field.
They then derived the corrected equations of motion of the wave packet and
solved the Boltzmann equation to find
the contribution to the current $\widehat{J}_{\textrm{geom},i}\sim
\frac{e^2}{2}\langle v_i {\sf g}_{jl}\rangle_ \textrm{FS} q_jq_l \widehat{V}$,
(rewritten to match our notation). 
Contrasting this with our result, we see the
numerical prefactor as well as the permutation of the indices are different.\footnote{Note that the choice of $f^r$ does not affect the quantum metric term which comes from an intrinsic (i.e.\ not collision-integral related) mechanism.}. Experimental detection of such a non-uniform response has been proposed using nonreciprocal directional dichroism in Ref.~\cite{gaoNonreciprocalDirectionalDichroism2019a}, although the prediction there also differs from ours due to the above discrepancy.

Similarly, $\widehat{J}_{i,e=3}(q,\omega)$ becomes
\begin{widetext}
\begin{align}
\label{eq:J3}
    \widehat{J}_{i,e=3}(q,\omega)=&
    -\frac{e^3}{\hbar}\int'_{\substack{q'q''\\\omega'\omega''}}
    \bigg(\frac{\tau^2\langle v_l\dv_{k_j} v_i\rangle_\textrm{FS}}
    {(1-i\omega\tau)(1-i\omega''\tau)}
    +\frac{\tau\langle v_l{\sf \Omega}_{ij}\rangle_\textrm{FS}}
    {1-i\omega''\tau}
    -\frac{\hbar}{2}\langle v_i{\sf t}_{jl}\rangle_\textrm{FS}
    +\hbar\langle v_j{\sf t}_{il}\rangle_\textrm{FS}\\
    &\hspace{30pt}-\frac{1}{1-i\omega\tau}\sum_{\substack{n,m=1\\n\neq m}}^N
    \int_k
    n_{F,n}\dv_{k_i}(\varepsilon_n-\varepsilon_m)^{-1}
    ({\sf \Lambda}_j)_{nm}({\sf \Lambda}_l)_{mn}
    \bigg)q'_jq''_l
    \widehat{V}(q',\omega')\widehat{V}(q'',\omega'')
    +\mathcal{O}(q^3).\notag
\end{align}
\end{widetext}
Note that the second term of \cref{eq:J3}, i.e.\ $-\frac{e^2}{\hbar}\int'_{\substack{q'q''\\\omega'\omega''}}\frac{\tau\langle v_l{\sf \Omega}_{ij}\rangle_\textrm{FS}}
    {1-i\omega''\tau}$, which is a transverse contribution to the
current, is exactly the nonlinear Hall effect first proposed by Sodemann and Fu
\cite{sodemann2015} that originates from the Berry curvature dipole. 

If the Fermi energy is in a band gap at $T=0$, all terms except the last vanish,
but unlike in the case of linear electrical response this finite term is not
divergenceless. This occurs because the RTA violates local charge conservation
in the system. The RTA couples the system to an external bath using a constant
scattering rate, so even if the Fermi energy is in the band gap, a
``divergenceful'' current can still be generated. A practical remedy for this issue is to
take the collisionless limit $\tau^{-1}\rightarrow0$ when the Fermi energy is in the band gap.

A general expression for second-order conductivity that ignored interband
transitions was also studied in Ref.~\cite{Yan_prl_2024} by Kaplan et al.\ in which they chose $f^r=n_F(\varepsilon)$. \red{If we use this choice of $f^r$ in our formalism for $\widehat{J}_{i,e=3}$, our result is modified to include only the first line of \cref{eq:J3}, which should be directly comparable to the results of
Ref.~\cite{Yan_prl_2024}. }
What we find is that only the first and last terms of the first line in \cref{eq:J3} match the results of Ref.~\cite{Yan_prl_2024} exactly. Contributions proportional to the second and third term are also
present in the results of Ref.~\cite{Yan_prl_2024}, but we find the numerical
prefactors are different. \red{Our results also extend those of Ref.~\cite{Gao_field}%
\footnote{
The nonlinear Hall effect in \cite{Gao_field} can be seen as the transverse part of the nonlinear current that is proportional to ${\sf t}_{ij}$, $\widehat{J}_{i,e=3}\sim\frac{1}{2} e^3\int \langle v_i {\sf t}_{jl} - v_j {\sf t}_{il} \rangle_{\FS} q'_j q''_l \widehat{V}(q')\widehat{V}(q'')  $.}}

\subsection{Fermi liquids with nontrivial band geometry}

The quantum kinetic equation is powerful enough to calculate dynamical responses for quantities other than the current. Here we examine the dynamical density response functions, which in a metal display typical non-analytic frequency and momentum dependence at low energy.  We recover the standard forms known for free electrons, and observe corrections arising from quantum geometry.  

The real space fermion density $\rho(x,t)\equiv\langle\psi^\dagger(x,t)\psi(x,t)\rangle=\int_k
\tr[\sff(x,k,t)]$ can be calculated from the solution to the kinetic equation.
Since the external potential $V(x,t)$ is coupled to
$\psi^\dagger(x,t)\psi(x,t)$, the coefficients of the expansion of $\rho(x,t)$ in powers of $V$
are be density correlation functions.

Here, we consider the collisionless limit $\mathcal{I}[f]=0$, and assume
$\omega\ll\omega_\textrm{gap}$ such that $\tilde{\mathcal{F}}$ is negligible.
Then, $\rho(x,t)=\int_k\tr[f(x,k,t)]$, and we solve for $f$
using the kinetic equation \cref{eq:trans_inv_kinetic_eq}. Expanding $f$ perturbatively in powers of $e$
such that $f=n_F(\varepsilon)+
f_{e=1}+f_{e=2}+\cdots$ where $f_{e=i}\sim e^i$ as in \cref{sec:solutions-kin-eq}, we find
\begin{align}
    \widehat f_{e=1}(q,k,\omega)
    =&\frac{\hbar n_F' \vec{v}\cdot\vec{q}\,
    (1-{\sf g}_{ij}q_i q_j)}
    {\hbar\omega-\varepsilon(k+q/2)+\varepsilon(k-q/2) }e\widehat V\notag\\
    &+n_F  {\sf t}_{ij}q_iq_j e\widehat V+\mathcal{O}(q^3),
\label{eq:FL1}
\end{align}
and
\begin{widetext}
\begin{align}
\label{eq:FL2}
    \widehat f_{e=2}(q,k,\omega)=&-\frac{e^2}{2}
    \int'_{\substack{q'q''\\\omega'\omega''}}\mathscr{F}(\omega q;\omega'q';\omega''q'';k)
    (1+{\sf g}_{ij}(q_i q'_j+q'_i q''_j+q''_i q_j))\widehat V(-q',-\omega')\widehat V(-q'',-\omega'')\\
   &-\frac{e^2}{2\hbar}
  \int'_{\substack{q'q''\\\omega'\omega''}}
  n_F'\Bigg[\left(\frac{\vec{v}\cdot{\vec{q}\,}'}
   {(\omega'-\vec{v}\cdot {\vec{q}\,}')(\omega-\vec{v}\cdot \vec{q})}
   -\frac{\vec{v}\cdot{\vec{q}\,}''}
   {(\omega''-\vec{v}\cdot{\vec{q}\,}'')(\omega-\vec{v}\cdot \vec{q})}\right)
   (\dv_{k_l} {\sf g}_{ij}q''_iq''_jq'_l+i{\sf \Omega}_{ij}q''_iq'_j)\notag\\
   &\hspace{50pt}+\left(
   \frac{q_iq'_jq''_l} {\omega''-\vec{v}\cdot{\vec{q}\,}''}
   -\frac{q'_iq''_jq'_l}{\omega-\vec{v}\cdot\vec{q}}
   \right) 2\hbar v_l{\sf t}_{ij}\Bigg]
   \widehat V(-q',-\omega')\widehat V(-q'',-\omega'')
   +\mathcal{O}(q^3),
   \notag
\end{align}
\end{widetext}
where $\widehat{f}_{e=i}$ is the Fourier transform of $f_{e=i}$ and
$n_F=n_F(\varepsilon)$. The primed integrals above indicate an implicit delta
function
$\int'_{\substack{q'q''\\\omega'\omega''}}=\int_{\substack{q'q''\\\omega'\omega''}}
(2\pi)^{d+1}\delta^d(q+q'+q'')\delta(\omega+\omega'+\omega'')$. Note that the
convention for the signs in the delta function differ from the convention used
in the previous subsection.  
In addition,
$\mathscr{F}(\omega,q;\omega'q';\omega''q'')$ is the integrand of the
three-point density correlation function in the case of trivial band geometry
which is defined as
\begin{widetext}
\begin{align}
    &\mathscr{F}(\omega,q;\omega',q';\omega'',q'';k)=
    -\sum_{r=L,R}\sum_{i=1}^3
    n_F(k+Q_{r,i})\prod_{j\neq i}
    \frac{1}{\hbar(\Omega_{r,j}-\Omega_{r,i})-\varepsilon(k+Q_{r,j})+\varepsilon(k+Q_{r,i})},
\end{align}
\end{widetext}
where $r=L,R$ labels the two contributing fermion loop diagrams and
$(\Omega_{R,1},\Omega_{R,2},\Omega_{R,3})=(\omega,-\omega'',0)$,
$(Q_{R,1},Q_{R,2},Q_{R,3})=(q/2,(q'-q'')/2,-q/2)$ and
$(\Omega_{L,1},\Omega_{L,2},\Omega_{L,3})=(\omega,-\omega',0)$,
$(Q_{L,1},Q_{L,2},Q_{L,3})=(q/2,-(q'-q'')/2,-q/2)$. 

We can then compute the $n$-point density correlation functions $C^{(n)}( x_1,t_1,\ldots,x_n,t_n)=\langle \left(\psi^\dagger \psi\right)(x_1,t_1)\cdots  \left(\psi^\dagger \psi\right)(x_n,t_n)\rangle$ for $n=2,3$. By differentiating \eqref{eq:FL1} and \eqref{eq:FL2} with respect to $\widehat V$, we obtain 
\begin{widetext}
\begin{subequations}
    \begin{align}
       \widehat{C}^{(2)}(\omega,q) &= -\ii e^{-1}\hbar \int_k \frac{\delta  \widehat f_{e=1}(q,k,\omega)}{\delta \widehat{V}(q,\omega) }
    \approx -\ii \int_k \left[\frac{\hbar^2 n_F' \vec{v}\cdot\vec{q}\,
    (1-{\sf g}_{ij}(k)q_i q_j)}
    {\hbar\omega-\varepsilon(k+q/2)+\varepsilon(k-q/2) } + \hbar n_F  {\sf t}_{ij}(k)q_iq_j\right], \label{eq:C2}  \\
    \widehat{C}^{(3)}(\omega',q',\omega'',q'') &= - e^{-2}\hbar^2 \int_k \frac{\delta^2  \widehat f_{e=2}(q,k,\omega)}{\delta \widehat{V}(q',\omega') \delta \widehat{V}(q'',\omega'') }\nn
    &\approx 
    \hbar^2\int_k \mathscr{F}(-\omega'-\omega'', -q'-q'';\omega'q';\omega''q'';k)
    (1-{\sf g}_{ij}(k)(q'_i q'_j+q'_i q''_j+q''_i q''_j)) \label{eq:C3}\\
   &\quad -\hbar
  \int_k
  n_F' \Bigg[\left(\frac{\vec{v}\cdot{\vec{q}\,}'}
   {(\omega'-\vec{v}\cdot {\vec{q}\,}')(\omega'+\omega''-\vec{v}\cdot (\vec{q}' + \vec{q}''))}
   -\frac{\vec{v}\cdot{\vec{q}\,}''}
   {(\omega''-\vec{v}\cdot{\vec{q}\,}'')(\omega'+\omega''-\vec{v}\cdot (\vec{q}' + \vec{q}''))}\right)\nn
   & \quad\quad  \times\left(\dv_{k_l} {\sf g}_{ij}(k)q''_iq''_jq'_l+i{\sf \Omega}_{ij}(k)q''_iq'_j \right) + \left(
   \frac{(q'_i + q''_i)q'_jq''_l} {\omega''-\vec{v}\cdot{\vec{q}\,}''}
   -\frac{q'_iq''_jq'_l}{\omega'+\omega''-\vec{v}\cdot(\vec{q}'+\vec{q}'')}
   \right) 2\hbar v_l{\sf t}_{ij}(k)  \Bigg],\notag 
    \end{align}
\end{subequations}
\end{widetext}
where $\widehat{C}^{(n)}$ is the Fourier transform of $C^{(n)}$. The static structure factor can be obtained from the two-point density correlation function through $S_q \equiv \re \int \frac{\ud \omega}{2\pi} \widehat{C}^{(2)}(\omega,q)$. Plugging in \eqref{eq:C2} into this expression, we find, for $d=2$,
\begin{equation}\label{eq:Sq intra}
\begin{split}
    S_q^{\mathrm{intra}} =&\im  \int \frac{\ud \omega}{2\pi} \int_k \frac{ \hbar n_F' \vec{v}\cdot\vec{q}\,
    (1-{\sf g}_{ij}(k)q_i q_j)}
    {\omega-\vec{v}\cdot\vec{q} +\ii\, \mathrm{sgn}(\omega)0^+ }\\
    =&\frac{\hbar^2 k_\F}{2\pi^2}\left(|\vec{q}\,|  -\frac{1}{4} \int_0^{2\pi}\ud\theta~ |\vec{n}\cdot\vec{q}\,|{\sf g}_{ij,\F}(\theta)q_i q_j
    \right),
\end{split}
\end{equation}
where $k_\F$ is the Fermi wavevector and ${\sf g}_{ij,\F}(\theta) = {\sf g}_{ij}(k = k_\F,\theta)$. Notice that, as we assumed $\omega\ll \omega_{\mathrm{gap}}$, \eqref{eq:Sq intra} only involves intraband contributions. In addition to the leading term, $\propto q$ ($\frac{\hbar^2k_{\rm F}}{2\pi^2}|\vec{q}|$), which is the single-band contribution to the Fermi gas, the quantum metric contributes at cubic order, $q^3$, to $S_q^{\mathrm{intra}}$ provided the $\theta$-integral in \eqref{eq:Sq intra} is non-vanishing. These results can be compared to a Green's function calculation of non-interacting multi-band fermions, which we carry out in \cref{app:density-density}.

\section{Conclusion}\label{sec:conclusion}

\subsection{Framework and comparison to other approaches}

The theory of electrons in solids rests on the concept of energy bands and Bloch states.  An essential element of the modern description of these bands is quantum geometry: not only the band energies but their wavefunctions enter into the dynamical response of electrons to various probes and forces, through quantities such as the Berry curvature and quantum metric.  The latter quantities can be understood as related to the motion of electrons through a constrained Hilbert space of a single band or group of bands, rather than through the full Hilbert space.  It is an axiom of the textbook semiclassical model \cite{AshcroftMermin} that the electronic response to weak field and forces, with slow spatial variations, can be described as independent motions within bands.  

(i) The most rigorous approach to derive the connection of quantum geometric quantities to physical responses is through linear and non-linear response theory, i.e. generalized Kubo formulas.  These formulas, based on time-dependent perturbation theory, often lead to simple expressions in terms of quantum geometric quantities, albeit through an obscure set of intermediate steps that make their origin opaque.  (ii) The simpler expressions can, in some cases, be reproduced through a semiclassical wavepacket approach, which treats the single particle Schr\"odinger equation in a quasi-classical approximation, and then importing the latter data into a Boltzmann equation phenomenologically.  This wavepacket approach \cite{Niu_review} is intuitive but involves two successive approximations (the semi-classical one for the wavepacket, and the reduction of the full density matrix to a diagonal Boltzmann equation) which may be non-trivial to improve consistently together at higher orders in semi-classics. There, the quantum geometric contributions arise through projection to the band manifold. (iii) A third approach is the quantum kinetic equation, essentially exploiting the equation of motion for the full quantum density matrix.  In a multi-band system the variables of this approach are matrix-valued, and include interband effects at the same level as intraband ones.  This approach is fully rigorous, but the simplicity of the semiclassical limit, i.e. the existence of a description in terms of intraband motion, is, like in the Kubo approach, hidden.  

In this paper, we have begun with the full matrix quantum kinetic equation and systematically derived a band-diagonal one to \emph{second} order in the semi-classical expansion, which can be regarded as a formal expansion in $\hbar$, but is practically controlled as an expansion in spatial gradients.  This is one order beyond the standard approach---first order leads to Berry curvature effects but not those of the quantum metric.\footnote{One may wonder why they appear at different orders even if they can be united in a QGT. This is because at the leading order, the Berry curvature is linear in the Berry connection but the quantum metric is not, which itself comes from cancellations due to the fact that the Moyal product is antisymmetric but the quantum metric is symmetric.}  The diagonal dynamics enjoys a reduction of degrees of freedom: for an $N$-band system, it is expressed in terms of $N$ phase space densities rather than $N^2$ matrix elements.  While this offers some computational gain, the main advantage is conceptual: one sees directly how all physics can be systematically expressed in terms of band-diagonal quantities, with quantum geometry emerging naturally through the diagonalization.  Our results can be seen as an extension of the pioneering approach of Blount\cite{Blount_bloch,Blount_extension,blount_1962_formalisms}, which also considered Moyal diagonalization, but to lower order and with less generality (and our presentation is significantly more self-contained than Blount's).  A more detailed comparison to some of Blount's work in found in Appendix~\ref{app:blount}.

A recent preprint by Mitscherling et al.\ takes an approach in which they define
a matrix-valued Wigner distribution function to study transport
\cite{mitscherling_orbital_2025}. Their formalism which is developed for lattice
systems, involves an expansion in spatial gradients which corresponds to
first-order in $\hbar$ of our results.

\subsection{Is semi-classical dynamics purely band geometric?}
\label{sec:conclusion_band_geometric}

A subtlety of band-projection is that a gauge freedom is introduced, corresponding to the freedom to redefine the phase of Bloch functions.  In the semi-classical approach, the associated phase can be both position and momentum dependent, so that the gauge freedom resides in the full semi-classical phase space.  Physical quantities must be gauge invariant.  They may or may not be band-geometric in the single band sense.  The Berry curvature and quantum metric are band geometric because they can be obtained from the wavefunctions of a single band only and do not require information about other bands or the Hamiltonian.  The orbital magnetization is a quantity which is gauge invariant but not band geometric: to obtain it one needs to know the Hamiltonian or the full set of wavefunctions of all the bands.  Many results in the literature involve quantities like the orbital magnetization which are not single-band geometric.  Another example is the ``Berry connection polarizability'' also known as the ``band normalized quantum metric''  derived in Refs.~\cite{Gao_field,lai_third-order_2021,wang_quantum-metric-induced_2023,Yan_prl_2024} as a contribution to non-linear conductivity.  This suggests that the axiom of independent band dynamics is violated by some semiclassical corrections.

A remarkable feature of our results is that, when the Hamiltonian lacks explicit time dependence, the diagonal quantum kinetic equation \emph{can} be expressed in a purely band-geometric form.  In this limit, \eqref{eq:boltzmanneq} becomes
\begin{widetext}
\begin{align}
  \label{eq:bltzdiag}
  \dv_t f=&\frac{1}{i\hbar}\scommu{h}{f}-\dv_\alpha\left[\wab f\left(
  \hbar \wsl\dv_\sigma h\Omega_{\lambda\beta}
   \left(1-\frac{\hbar}{2}\wmn\Omega_{\mu\nu}\right)
  +\frac{\hbar^2}{2}\wsl\wmn\dv^2_{\sigma\mu}h\dv_\beta g_{\nu\lambda}\right)\right]\\
  &-\dv^2_{\alpha\sigma}\left[\hbar^2\wab\wsl
  f\left(\frac{1}{2}\wmn\dv_\mu h\dv_\nu g_{\beta\lambda}+\wmn\dv^2_{\mu\beta}hg_{\nu\lambda}
  \right)
  \right] +\mathcal{O}(\hbar^3).\notag
\end{align}
\end{widetext}
One observes that this equation involves only explicitly single-band-geometric quantities: the band energy $h$, the Berry curvature $\Omega_{\mu\nu}$ and the quantum metric $g_{\mu\nu}$.  

This may be surprising given that it is well-known that multi-band quantities such as orbital magnetization and the quantum metric dipole do appear in physical responses.  The resolution is that the fully band-diagonal form of \cref{eq:bltzdiag} appears only when quantities are expressed in terms of semi-classically corrected bands, which are the ones defined through Moyal diagonalization.  When the kinetic equation is expressed in terms of quantities defined through the uncorrected bands (at zeroth order in $\hbar$), as worked out in \cref{sec:diagonalizing_h,sec:application} then the multi-band quantities appear, fully consistent with known results.  Thus, our formulation gives the physical interpretation that the electronic dynamics (at least to second order in semiclassics) can be considered fully band-diagonal and intrinsically band-geometric, provided that the suitably renormalized bands are defined, and that the ``magnetization-like'' quantities which are not single-band in nature are only symptoms of expressing the renormalized bands in terms of bare ones. 

\subsection{Ramifications of the gradient expansion}

The approach of this paper is based on the semi-classical $\hbar$ expansion, which amounts to an expansion in the smallness of spatial and temporal gradients.  Thus it is clearly limited to situations in which the Hamiltonian varies slowly in real space and in time.
The latter limits us to low frequencies, and generally ac response when external frequencies are comparable to or exceed interband transitions is not accessible.\footnote{The extension to ``fast'' {\em time} variations/arbitrary frequencies is feasible, and not a fundamental limitation of the method.  It would require solving for the full off-diagonal components of the distribution function $\tilde{\mathcal{F}}$, see \cref{eq:offdiagonal_eq,eq:boltzmanneq} and \cref{sec:kin-eq-noneq}.}  This means we cannot address in a general way quantities like the shift current.  There are, however, advantages to the present approach.  First, in the semiclassical regime, the reduction to diagonal form is completely general, and \cref{eq:bltzdiag} and its extensions can be applied to arbitrary multi-band problems.  Second, the approach allows treatment of general inhomogeneities in space.  This allows a calculation of momentum dependent response functions, or non-equilibrium solutions for arbitrary spatial geometries.  

By explicitly tracking the spatial dependence of physical quantities, the present approach directly confronts some of the subtle issues in transport theory.  One of the basic such issues relates to the definition of current.  A typical procedure to define a current is to extract it from the continuity equation.  This defines a ``transport current'': the current so-defined is guaranteed to describe the time dependence of the total charge in a closed region. That is, one takes a conserved quantity $Q$ and finds a local operator $\rho(x)$ such that $Q = \int_x \rho(x)$.  Then one calculates $\partial_t \rho = i[H,\rho]$ which is unambiguous for a given Hamiltonian.  One then finds a current $J_i(x)$ such that $\partial_t\rho = - \partial_{x_i} J_i$.  However, this procedure has two ambiguities: the choice for $\rho(x)$ is not unique, and neither is the choice of $J_i$.  The first ambiguity amounts to the freedom to shift $\rho \rightarrow \rho + \partial_{x_i} D_i$, where $D_i$ is a ``dipole density''.  If one makes this change to the definition of the density, then the definition of the current must change by $J_i \rightarrow J_i - \partial_t D_i$.  This can be regarded as a polarization current.    \emph{In a time-independent steady state} this change does not affect the average $\langle J_i\rangle$ (because $\langle \partial_t D_i\rangle = 0$).  The polarization current will in general affect the non-zero frequency current response.

The second ambiguity is more treacherous.  The (transport) current itself can be shifted by any curl, $J_i \rightarrow J_i + \partial_{x_j} M_{ij}$, where $M_{ij}$ is an anti-symmetric ``magnetization density'', even when the definition of the density $\rho$ is unchanged. Due to the spatial derivative, this is not expected to change the zero momentum current in a translationally invariant system (for which Fourier decomposition is sensible), but it does affect non-zero wavevector currents, and it does affect the local currents when the system is not translationally invariant.  The latter situation is important, for example, in the presence of even a constant temperature gradient, which leads to subtleties in calculations of the thermal Hall effect.  However, by construction, the magnetization currents do not contribute to net transport through any closed surface or a surface that terminates outside the sample.  This means that while they can and do affect calculations of the conductivity, the magnetization currents do not affect the \emph{conductance}.  In the formalism of this paper, it is possible to calculate a true conductance by treating a finite sample and obtaining the integrated current across a cross-section.  We emphasize that the ``Biot-Savart current'' density, which is defined as the one that appears in Maxwell's equations, is unambiguous.  This is the one we derived in \cref{sec:number_current}, and has been used in this work \footnote{\red{Although we do not attempt to distinguish the transport and magnetization current in this work, they may be separated by considering the Joule heating, c.f. Ref. \cite{Son_kinetic}. } }.  

\subsection{Applications}

In this paper, we applied our formalism to the case of a translationally-invariant Hamiltonian supplemented by a non-uniform electric field.  We derived a form of the equilibrium current and also calculated the non-equilibrium current for the case of a time-dependent
electric field.  Our results can be compared to several others in the literature.  We were inspired by Ref.~\cite{Lapa_nonuniform}, in which Lapa and Hughes generalized the wavepacket formalism
to incorporate the effects of a non-uniform electric field and identified a quantum metric contribution to finite-momentum conductivity. However, using our
formalism, we also find a quantum metric contribution but one that is distinct from theirs.
Our results also reproduce the non-linear Hall effect of
Ref.~\cite{sodemann2015}, but we find some discrepancies with our results and the
quantum geometric non-linear conductivity of Ref.~\cite{Yan_prl_2024}.
In addition to these transport quantities, we also obtain the dynamical density-density response functions.  The two-point correlator has the standard (non-interacting) Fermi-liquid form, corrected by a higher order in momentum factor from the quantum metric.  Our result evokes known quantum metric corrections for insulators discussed in Refs.~\cite{souzaPolarizationLocalizationInsulators2000,onishi2024universal}, but applies here at low frequencies.

\subsection{Extensions}

There are a number of interesting directions to extend this work.  One could \emph{apply} the formalism developed here to study thermal and thermo-electric currents and responses.  It is particularly well-suited for this because a spatially inhomogeneous temperature can be explicitly treated (as can a non-thermal non-equilibrium state) without recourse to the ``trick'' of inserting an artificial gravitational field \emph{\`a la} Luttinger, whose applicability beyond linear response is questionable.  It would be natural to extend the current formalism to superonducting states, i.e.\ Bogoliubov-de Gennes Hamiltonians.  An important direction is to include the effects of electron-electron or electron-phonon interactions.  This could be accomplished via the Keldysh method, in which the semi-classical expansion is well-established for {\em single band} systems and can be best carried out by extending the Wigner transformation from just the space/momentum domain to include as well the time/frequency variables.  The latter is essential to take into account frequency-dependent self-energy effects.  We believe the techniques developed here should be a foundation for such a study in multi-band Fermi liquids.
Another interesting direction is to extend our formalism to consider finite frequency responses beyond the low-frequency limit we considered in this work. This would involve solving for the off-diagonal components of the distribution function using \cref{eq:offdiagonal_eq}. 
This would allow us to study photovoltaic effects, shift and injection currents, and sum rules of optical conductivity which have been
shown to be related to the integrated quantum metric \cite{souzaPolarizationLocalizationInsulators2000,
jankowskiQuantizedIntegratedShift2024,
vermaInstantaneousResponseQuantum2024,vermaFrameworkMeasureQuantum2025}.
Lastly, our formalism was developed assuming non-degenerate energy states but it would be of interest to extend our formalism to be fully compatible with degeneracies which generically occur in crystalline systems.

We hope that in the future this method can be used profitably in novel comparisons to experiments.  Specifically, our formalism in phase space has a unique advantage of affording us both real
space and momentum space resolution. It would then be interesting to try to
apply this method to understand experimental measurements that probe
the momentum structure of materials \emph{locally} such as nano-angle-resolved photo-emission spectroscopy and interference patterns in scanning tunneling microscopy.

\section*{Acknowledgements}
X.H.\ thanks Jing-Yuan Chen for interesting discussions. L.S.\ and L.B.\ thank L\'eo Mangeolle for a previous collaboration which led to this work.
We especially thank Ren-Bo Wang and Kaixiang Su for many valuable discussions and collaboration at the early stage of this project.
  X.H.\ thanks the Kavli Institute for Theoretical Physics for hospitality; KITP is supported in part by the Heising-Simons Foundation, the Simons Foundation, and the National Science Foundation under Grant PHY-2309135.  T.P.\ was funded by the Air Force Office of Scientific Research (AFOSR) (award FA9550-22-1-0432). L.S.\ was funded by the European Research Council (ERC) under the European Union's Horizon 2020 research and innovation program (Grant agreement No.~853116, acronym TRANSPORT). L.B.\ was supported by the NSF CMMT program under Grants No. DMR-2419871, and the Simons Collaboration on Ultra-Quantum Matter, which is a grant from the Simons Foundation (Grant No. 651440).

\section*{Statement of author contributions}

T.P.\ and X.H.\ contributed equally to this work, and carried out the majority of calculations.  L.B.\ and L.S.\ conceptualized the project.  All authors contributed to and checked analytic results, and took part in writing of the manuscript.

\bibliography{moyal}

\onecolumngrid

  \appendix

\section{Notations}

\begin{itemize}
    \item non-diagonal quantities: sans-serif fonts, e.g. $\sf F,H$. We reserve the sans-serif fonts of other letters to distinguish for quantities defined in wave number. We also defined ${\sf P}_n=P_n^{(0)}$. 
    \item rotated gauge-dependent quantity : $\tilde F,\tilde H$
    \item rotated (non-diagonal) gauge-independent quantity : $F, H$
    \item gauge-dependent diagonal quantities: lowercase, e.g. $\tilde{f},\tilde{h}$
    \item off-diagonal quantities: $\tilde{\mathcal{F}}$
    \item gauge-invariant diagonal quantities: $f,h$
    \item Fourier transform with respect to the center of mass coordinate, $x\rightarrow q$: wide hat, e.g.\ $\widehat{V}$
    \item band indices: Roman letters $n,m,m',\cdots$ 
    \item phase space coordinate indices: Greek letters $\alpha,\beta,\lambda,\sigma,\cdots$ 
    \item real space coordinate indices: Roman letters $i,j,\cdots$
    \item Moyal algebra: $\omega^{\alpha\beta}$, and $\omega^{\alpha\beta}_B$ in the presence of a $B$ field. 
    \item Berry connection: $\Lambda_\alpha = -\ii U^\dagger\star \p_{\alpha} U$ and
    $A_\alpha=\diag(\Lambda_\alpha)$
    \item External vector potential: $\mathcal{A}_i$
    \item Physical current: $J_i(x)$
    \item $\hbar$ expansion: $h^{(n)}\sim \hbar^n$ such that $h=h^{(0)}+h^{(1)}+\cdots$
    \item external potential strength expansion: $f_{e=n}$ such that $f=f_{e=0}+f_{e=1}+f_{e=2}+f_{e=3}+\cdots$
\end{itemize}

Note that the notations in this manuscript differ significantly from those in a manuscript by two of us, Ref.~\cite{mangeolle2024quantum}. Moreover, beyond notations, it is noteworthy that (i) the convention for $\Lambda_\alpha$ is different in the two manuscripts (in Ref.~\cite{mangeolle2024quantum}, $\Lambda_\alpha=U^\dagger\star\partial_\alpha U$ while here $\Lambda_\alpha=-i U^\dagger\star\partial_\alpha U$) but the $A_\alpha$ agree ($A_\alpha={\rm Im}[{\rm diag}\Lambda_\alpha]$ in Ref.~\cite{mangeolle2024quantum} and here $A_\alpha={\rm diag}\Lambda_\alpha$), (ii) the matrices in Ref.~\cite{mangeolle2024quantum} are $2N\times2N$ matrices (compared to the $N\times N$ matrices used here) because of the lack of particle conservation in Ref.~\cite{mangeolle2024quantum}.

\section{Moyal product identities}
Here, we list two useful formulas that are repeatedly used in our formalism.
\vspace{10pt}

{\noindent\textbf{Identity 1}}\hspace{20pt}
Assuming $A,C$ are matrix-valued functions and $b$ a scalar-valued function,
\begin{align}
  A\star b\star C=&b\,(A\star C)-\frac{i\hbar}{2}\wab\dv_\alpha b\,(\dv_\beta
  A\star C-A\star \dv_\beta C)
  -\frac{\hbar^2}{8}\wab\wsl\dv^2_{\alpha\sigma}b\,(\dv^2_{\beta\lambda}(A\star
  C)-4\dv_\beta A\star \dv_ \lambda C)+\mathcal{O}(\hbar^3).
  \label{eq:id1}
\end{align}

\vspace{20pt}
{\noindent\textbf{Identity 2}}\hspace{20pt}
Assuming $A,B$ are matrix-valued functions,
\begin{align}
  \tr[A\star B]=\tr[B\star A]-i\hbar\wab\tr[\dv_\alpha B\star\dv_\beta A]
  -\frac{\hbar^2}{2}\wab\wsl\tr[\dv^2_{\alpha\sigma}B\star \dv^2_{\beta\lambda}A]
  +\mathcal{O}(\hbar^3).
  \label{eq:id2}
\end{align}

\section{Defining $f,h$}
\label{app:defining_fh}
Here, we define $f,h$ which are the gauge-invariant forms of $\tilde f, \tilde
h$ respectively. Firstly, we define $f$ such that it satisfies
\begin{equation}
    {\rm N}=\int_{x,p}\tr[\sff]=\int_{x,p}\tr[f].
    \label{eq:app_Ncondition}
\end{equation}
Note, $\int_{x,p}\tr[\sff]=\int_{x,p}\tr[U\star\tilde F\star U^\dagger]
=\int_{x,p}\tr[U\star\tilde f\star
U^\dagger]+\int_{x,p}\tr[U\star\tilde{\mathcal{F}}\star U^\dagger]
=\int_{x,p}\tr[U\star\tilde f\star U^\dagger]$.  The last equality was obtained
by using the fact that the terms in the trace can be cycled even in the presence
of star products upon taking the phase-space integral. The last equality tells
us that the off-diagonal components of the distribution function does not
contribute to the total particle number.
We can then rewrite \cref{eq:app_Ncondition} as
\begin{equation}
    \int_{x,p}\tr[f]=\sum_n\int_{x,p}\tr[U\star p_n\tilde f_n\star U^\dagger]
\end{equation}
where $p_n$ is a
diagonal matrix defined such that $(p_n)_{ij}=\delta_{in}\delta_{jn}$. The
simplest definition for $f$ that satisfies this condition is
\begin{equation}
    f_n\equiv \tr[U\star p_n\tilde f_n\star U^\dagger] = \tr[U\star p_n\tilde f\star U^\dagger] .
\end{equation}
Note that as a consequence of this definition, when $\tilde{\mathcal{F}}=0$, one has
\begin{equation}
    \tr[\mathsf{F}] = \tr[f], \qquad \tilde{\mathcal{F}}=0,
\end{equation}
so that the phase space density is given by $\tr[f]$ in this situation (note that here the identity holds without a phase space integral).

Using \cref{eq:id1,eq:id2}, the right-hand side can be expanded to second-order
in $\hbar$ to give us
\begin{equation}
    f_n=\tilde f_n+\hbar\wab\dv_\alpha(\tilde f_n A_{n\beta})
    +\frac{\hbar^2}{4}\wab\wsl\dv^2_{\alpha\sigma}
    (\tilde f_n (\acomm{\Lambda_\beta}{\Lambda_\lambda})_{nn})
    +\mathcal{O}(\hbar^3),
\end{equation}
and inverting this relation gives us
\begin{equation}
 \tilde f_n=f_n-\hbar\wab\dv_\alpha(f_n(A_{n\beta}+\hbar\wsl
  A_{n\lambda}\dv_\sigma A_{n\beta}))
  +\hbar^2\wab\wsl\dv^2_{\alpha\sigma}\left(f_n\left(A_{n\beta}A_{n\lambda}
  -\frac{1}{4}\acomm{\Lambda_\beta}{\Lambda_\lambda}_{nn}\right)\right)
  +\mathcal{O}(\hbar^3).
  \label{eq:tildef}
\end{equation}

Next, let us consider the definition for $h$. We require that $h$ satisfies
\begin{equation}
    {\rm E}=\frac{1}{2}\int_{x,p}\tr[\sfh\star\sff+\sff\star\sfh]
    =\int_{x,p}\tr[hf].
\end{equation}
The expectation value of the Hamiltonian can be rewritten in the following way.
\begin{align*}
    \langle\hat H\rangle=&
    \frac{1}{2}\int_{x,p}\tr[\sfh\star\sff+\sff\star\sfh]\\
    =&\sum_n\frac{1}{2}\int_{x,p}
    \tr[U\star p_n(\tilde h_n\star\tilde f_n+\tilde f_n\star\tilde h_n)
    \star U^\dagger]\\
    =&\sum_n\int_{x,p}
    \left(\tilde h_n\tilde f_n
    +\hbar\wab\dv_\alpha(\tilde h_n\tilde f_n A_{n\beta})
    -\frac{\hbar^2}{8}\wab\wsl\dv^2_{\alpha\sigma}\tilde h_n \dv^2_{\beta
    \lambda} \tilde f_n
    +\frac{\hbar^2}{4}\wab\wsl\dv^2_{\alpha\sigma}
    (\tilde h_n\tilde f_n(\acomm{\Lambda_\beta}{\Lambda_\lambda})_{nn})\right)
    +\mathcal{O}(\hbar^3)\\
    =&\sum_n\int_{x,p}
    f_n\left(\tilde h_n+\hbar\wab\dv_\alpha\tilde h_n(A_{n\beta}+\hbar\wsl
    A_{n\lambda}\dv_\sigma A_{n\beta})
    -\hbar^2\wab\wsl\dv^2_{\alpha\sigma}\tilde h_n
    \left(\frac{1}{4}\acomm{\Lambda_\beta}{\Lambda_\lambda}
    -A_{n\beta}A_{n\lambda}\right)\right)+\mathcal{O}(\hbar^3)
\end{align*}
Like before, the off-diagonal components of the distribution function contribute
nothing. The last line was obtained by expressing $\tilde f_n$ using $f_n$ and
integrating by parts. The last line gives us the definition of $h_n$.
\begin{equation}
    h_n\equiv \tilde h_n+\hbar\wab\dv_\alpha\tilde h_n(A_{n\beta}+\hbar\wsl
    A_{n\lambda}\dv_\sigma A_{n\beta})
    -\hbar^2\wab\wsl\dv^2_{\alpha\sigma}\tilde h_n
    \left(\frac{1}{4}\acomm{\Lambda_\beta}{\Lambda_\lambda}_{nn}
    -A_{n\beta}A_{n\lambda}\right)+\mathcal{O}(\hbar^3)
\end{equation}
Its inverse relation is
\begin{equation}
    \tilde h_n=h_n-\hbar\wab\dv_\alpha h_n A_{n\beta}
    +\frac{\hbar^2}{4}\wab\wsl\dv^2_{\alpha\sigma} h_n (\acomm{\Lambda_\beta}{\Lambda_\lambda})_{nn})
    +\mathcal{O}(\hbar^3).
\end{equation}

\section{Diagonalization by flow equations}\label{app:diagonalization}
The diagonalization equation \cref{eq:diagonalization} which we reproduce below
\begin{equation}
    \tilde h=U^\dagger\star {\sf H}\star U
    \label{eq:app_diag_eq}
\end{equation}
can be solved using what we call ``flow equations.'' The basic idea behind flow equations is to take derivatives with respect to $\hbar$.
First, we differentiate the unitarity condition of $U$
\cref{eq:unitaritycondition} to get
\begin{align}\label{eq:unitary flow}
    T^\dagger +T+\frac{\ii\hbar}{2}\omega^{\alpha\beta} \Lambda_\alpha \star
    \Lambda_\beta = 0
\end{align}
where we defined $T\equiv \hbar U^\dagger \star \p_{\hbar} U$. The second term
comes from the derivative acting on the $\hbar$ in the definition of the star
product.
\cref{eq:unitary flow} completely determines the hermitian part of $T$ so
we can express $T$ as
\begin{align}\label{eq:T total}
    T = Y - \frac{\ii\hbar}{4} \omega^{\alpha\beta}\Lambda_\alpha\star \Lambda_\beta
    = Y+\frac{\hbar}{4}\wab \dv_\alpha\Lambda_\beta
\end{align}
where $Y$ is the anti-Hermitian part of $T$, $Y^\dagger = - Y$.
Next, by differentiating the diagonalization equation \eqref{eq:app_diag_eq}
with respect to $\hbar$, we obtain, what we refer to as, the flow equation of 
$\tilde h$
\begin{equation}\label{eq:h_flow}
  \begin{split}
    \hbar\p_{\hbar} \tilde h =&
    \tilde{h}^{(1)}+2\tilde{h}^{(2)}+\cdots=
    \scommu{\tilde h}{Y} + \frac{\hbar}{2} \omega^{\alpha\beta}
    \sacommu{\Lambda_\alpha}{\p_\beta \tilde h}
    +\frac{i\hbar}{4} \omega^{\alpha\beta}
     \sacommu{\Lambda_\alpha}{ \scommu{\Lambda_\beta}{ \tilde h} }.
  \end{split}
\end{equation}
The first-order terms of the right-hand side gives us the first-order
correction to $\tilde h^{(0)}$.
\begin{equation}
  \begin{split}
  \tilde{h}^{(1)}&=\commu{\tilde{h}^{(0)}}{Y^{(1)}}
  +\frac{\hbar}{2}\wab\acommu{\Lambda_\alpha^{(0)}}{\dv_\beta\tilde{h}^{(0)}}
  \quad+\frac{i\hbar}{4}\wab\acommu{\Lambda^{(0)}_\alpha}
  {\commu{\Lambda^{(0)}_\beta}{\tilde{h}^{(0)}}}.
  \end{split}
\end{equation}
Since $\tilde h^{(1)}$ is supposed to be a diagonal matrix, this imposes a
constraint on $Y^{(0)}$ which determines its off-diagonal components.
\begin{align}
  Y^{(1)}_{nm}&\equiv-\frac{\hbar}{2(\tilde{h}^{(0)}_n-\tilde{h}^{(0)}_m)}\wab
  \bigg(\acommu{\Lambda_\alpha^{(0)}}{\dv_\beta\tilde{h}^{(0)}}
  +\frac{i}{2}\acommu{\Lambda^{(0)}_\alpha}{\commu{\Lambda^{(0)}_\beta}{\tilde{h}^{(0)}}}
  \bigg)_{nm}
  \,,\quad
  n\neq m.
\end{align}
Therefore $\tilde h^{(1)}$ is
\begin{equation}\label{eq:tilde h 1}
  \begin{split}
  \tilde{h}^{(1)}=&-\hbar\wab
  \dv_\alpha\tilde{h}^{(0)}A_{\beta}^{(0)}
  +\frac{i\hbar}{4}\wab\diag[\acommu{\Lambda^{(0)}_\alpha}
  {\commu{\Lambda^{(0)}_\beta}{\tilde{h}^{(0)}}}],
  \end{split}
\end{equation}
and from \cref{eq:invth}, the gauge-invariant form of this term is
\begin{equation}
  h^{(1)}=\frac{i\hbar}{4}\wab\diag[\acommu{\Lambda^{(0)}_\alpha}{\commu{\Lambda^{(0)}_\beta}{\tilde{h}^{(0)}}}].
\end{equation}
This term is a contribution from the phase-space orbital magnetization.
Note that the diagonal components of
$Y^{(1)}$ does not affect the diagonalization procedure, and we can let
$\diag(Y^{(1)})=0$ without consequence.

The second-order correction $\tilde{h}^{(2)}$ can be similarly calculated from
the flow equation \cref{eq:h_flow} by taking first-order terms of the right-hand
side. However, it is a tedious calculation. We first define $\tilde w$ as the
last term in the flow equation \cref{eq:h_flow}:
\begin{equation}
    \tilde w=\frac{i\hbar}{4}\wab \sacommu{\Lambda_\alpha}{\scommu{\Lambda_\beta}{\tilde h}}
\end{equation}
Consider the $n$-th diagonal of $\tilde w$. We find that it can be expressed as
\begin{align}
    \tilde w_n=&-\frac{i\hbar}{2}\wab\tr[\dv_\alpha P_n\star {\sf H}\star \dv_\beta P_n]
    +\frac{\hbar^2}{2}\wab\wsl\dv_\alpha(h_n\dv_\sigma g_{n,\beta\lambda})
    +\frac{\hbar^2}{8}h_n(\wab\Omega_{n,\alpha\beta})^2\\
    &+\hbar\wab\dv_\alpha \tilde{w}_n A_{n,\beta}+\frac{\hbar^2}{2}\wab\wsl\dv_\alpha h_n A_{n,\sigma}\dv_\beta A_{n,\lambda}+\mathcal{O}(\hbar^3)\notag.
\end{align}
Notice that the first-order component $\tilde w_n^{(1)}$ comes solely from the
first term on the right-hand side, and this term is defined using only
gauge-invariant objects. Therefore, $\tilde w_n^{(1)}$ is gauge-invariant and it
is exactly $h_n^{(1)}$. The two other terms in the first line of the
right-hand side are also gauge-invariant. Therefore, we define
\begin{equation}
    w_n\equiv-\frac{i\hbar}{2}\wab\tr[\dv_\alpha P_n\star {\sf H}\star \dv_\beta P_n]
\end{equation}
and write
\begin{equation}
    \tilde w_n=w_n
    +\frac{\hbar^2}{2}\wab\wsl\dv_\alpha(h_n\dv_\sigma g_{n,\beta\lambda})
    +\frac{\hbar^2}{8}h_n(\wab\Omega_{n,\alpha\beta})^2\\
    +\hbar\wab\dv_\alpha w_n A_{n,\beta}
    +\frac{\hbar}{2}\wab\wsl\dv_\alpha h_n A_{n,\sigma}\dv_\beta A_{n,\lambda}
    +\mathcal{O}(\hbar^3)\notag
\end{equation}
If we now take the second-order terms of the right-hand side of the flow equation \cref{eq:h_flow}, we see that
\begin{equation}\label{eq:tilde h 2}
    \tilde h_n^{(2)}=-\frac{\hbar}{2}\wab(\dv_\alpha\tilde h^{(1)} A_{n\beta}^{(0)}
    +\dv_\alpha\tilde h^{(0)} A_{n\beta}^{(1)})+\frac{1}{2}\tilde w_n^{(2)}
\end{equation}
Then using \cref{eq:invth}, we find
\begin{equation}
  \begin{split}
  h^{(2)}=&\frac{1}{2}w^{(2)}+\frac{\hbar}{2}\wab\dv_\alpha h^{(0)}
    \left(A_{\beta}^{(1)}+\hbar\wsl\left(\dv_\sigma A_{\beta}^{(0)} A_{\lambda}^{(0)}
    +\frac{1}{2} A_{\sigma}^{(0)}\dv_\beta A_{\lambda}^{(0)}
    +\frac{1}{2}\dv_\sigma g_{\beta\lambda}^{(0)}\right)\right)\\
    &+\frac{\hbar^2}{4}\wab\wsl h^{(0)}
    \left(\dv^2_{\alpha\sigma}g_{\beta\lambda}^{(0)}
    +\frac{1}{4}\Omega_{\alpha\beta}^{(0)}\Omega_{\sigma\lambda}^{(0)}\right)
    -\frac{\hbar^2}{2}\wab\wsl\dv^2_{\alpha\sigma}h^{(0)}g_{\beta\lambda}^{(0)}
  \end{split}
\end{equation}
Using the definition for $m_{n,\alpha\beta}$ given in \cref{eq:m} we see that
$w_n=\frac{\hbar}{2}\wab m_{n,\alpha\beta}-\frac{\hbar^2}{2}\wab\wsl
(h_n\dv^2_{\alpha\sigma} g_{n,\beta\lambda}+2
\dv_\alpha h_n \dv_\sigma g_{n,\beta\lambda})+\mathcal{O}(\hbar^3)$. If we now substitute
this into the equation above, we get \cref{eq:h2}.

In order to evaluate the second-order correction $h^{(2)}$ we also need to
calculate the first-order correction to the Berry connection and the projection
operator. We can do this by calculating their flow equations.
\begin{align}\label{eq:A flow}
    \hbar\p_{\hbar} \Lambda_\alpha 
     = \scomm{\Lambda_\alpha}{Y} -i\p_\alpha Y +
     \frac{\hbar}{4}\wsl \sacomm{\Lambda_\sigma}{ \p_\lambda
     \Lambda_\alpha} ,
\end{align}
and
\begin{equation}\label{eq:P flow}
  \begin{split}
  &\hbar\p_{\hbar} P_n=
  U\star\bigg(\scomm{Y}{p_n}
  +\frac{\hbar}{4}\wab \sacomm{\dv_\alpha\Lambda_\beta}{p_n}
  +\frac{i\hbar}{2}\wab \Lambda_\alpha\star
  p_n\Lambda_\beta\bigg)\star U^\dagger
  \end{split}
\end{equation}
which are obtained by taking the derivative of \cref{eq:connection} and
\cref{eq:projection} with respect to $\hbar$.
From these flow equations we get
\begin{equation}
  \begin{split}
  A_\alpha^{(1)}=&\diag\left(\commu{\Lambda_\alpha^{(0)}}{Y^{(1)}}\right)
  +\frac{\hbar}{4}\wsl\diag\left(\acommu{\Lambda_\sigma^{(0)}}{\dv_\lambda
  \Lambda_\alpha^{(0)}}\right),
  \end{split}
\end{equation}
and
\begin{equation}
  \begin{split}
    &P_n^{(1)}=
    U^{(0)}\bigg(\commu{Y^{(1)}}{p_n}
    +\frac{\hbar}{4}\wab \acommu{\dv_\alpha\Lambda^{(0)}_\beta}{p_n}
    +\frac{i\hbar}{2}\wab \Lambda_\alpha^{(0)}
    p_n\Lambda^{(0)}_\beta\bigg)U^{(0)\dagger}.
  \end{split}
\end{equation}

\section{Derivation of the equilibrium distribution}
\label{app:eq_dist}
Here, we derive the form for the equilibrium distribution function $\tilde f_\textrm{eq}$.
We begin by defining the imaginary-time Green's function ${\sf G}_{nm}(x_1,\tau_1;x_2,\tau_2)\equiv
-\langle T_\tau \psi_{n}(x_1,\tau)\psi^\dagger_m(x_2,\tau_2)\rangle_\beta$, where $\langle\cdots\rangle_\beta$
denotes an average with respect to a grand canonical ensemble at temperature $T=\frac{1}{\beta k_B}$.
The equilibrium density matrix in the position representation can then be expressed as
\begin{equation}
    {\sf F}_{\textrm{eq},nm}(x_1,x_2)\equiv
    \langle\psi^\dagger_m(x_1)\psi_n(x_2)\rangle_\beta
    ={\sf G}_{nm}(x_1,\tau;x_2,\tau+0^+)
    =\frac{1}{\beta}\sum_{i\omega_n}{\sf G}_{nm}(x_1,x_2;\omega_n)
    \label{eq:F_eq}
\end{equation}
where ${\sf G}_{nm}(x_1,x_2;\omega_n)$ is the Matsubara-frequency representation of the imaginary-time Green's function.
It is defined implicitly by the equations
\begin{align}
  \int_{y}\left[ (i\omega_n+\mu)\delta^d(x-y) -{\sf H}(x,y)\right]
  {\sf G}(y,x';\omega_n) =&\delta(x-x')I_N\\
  \int_{y}{\sf G}(x,y;\omega_n) 
  \left[ (i\omega_n+\mu)\delta^d(y-x') -{\sf H}(y,x')\right]
  =&\delta(x-x')I_N
\end{align}
where we have suppressed matrix indices for brevity.
Let's now transform this using the Wigner transformation.
\begin{align}
  \left[i\omega_n+\mu- {\sf H}(x,p)\right]\star {\sf G}(x,p;\omega_n)=&I_N\\
  {\sf G}(x,p;\omega_n)\star\left[i\omega_n+\mu- {\sf H}(x,p)\right]=&I_N
\end{align}
We now assume that all functions are defined on phase space and drop the $(x,p)$ dependence of functions.
The next step is to apply the transformation $U$ which diagonalizes ${\sf H}$.
\begin{align}
  \left[i\omega_n+\mu- \tilde h\right]\star \tilde G=&I_N
  \label{eq:geq1} \\
  \tilde G\star\left[i\omega_n+\mu- \tilde h\right]=&I_N
  \label{eq:geq2}
\end{align}
where $\tilde G\equiv U^\dagger\star {\sf G}\star U$.

We can now solve for $\tilde G$ in powers of $\hbar$. Let $\tilde G=\tilde G^{(0)}+\tilde G^{(1)}+\tilde G^{(2)}+\cdots$ where $\tilde G^{(n)}\sim \hbar^n$. To lowest order we find
\begin{equation}
    \tilde G^{(0)}=\frac{1}{i\omega_n- h^{(0)}+\mu}.
\end{equation}
A straightforward calculation also leads to the first- and second-order corrections:
\begin{equation}
  \tilde G^{(1)}
  =\frac{\tilde h^{(1)}}{\left(i\omega_n-h^{(0)}+\mu\right)^2}
\end{equation}
\begin{equation}
  \begin{split}
  \tilde G^{(2)}=&\frac{\tilde h^{(2)}}{\left( i\omega_n-h^{(0)}+\mu \right)^2}
  +\frac{(\tilde h^{(1)})^2}{\left( i\omega_n-h^{(0)}+\mu \right)^3}
  -\frac{1}{8}\wab\wsl
  \frac{\dv^2_{\alpha\sigma} h^{(0)}\dv^2_{\beta\lambda}h^{(0)}}
  {\left( i\omega_n-h^{(0)}+\mu \right)^3}
  -\frac{1}{4}\wab\wsl
  \frac{\dv^2_{\alpha\sigma} h^{(0)}\dv_\beta h^{(0)}\dv_\lambda h^{(0)}}
  {\left( i\omega_n-h^{(0)}+\mu \right)^4}
  \end{split}
\end{equation}
We have now solved for $\tilde G$ up to second-order in $\hbar$.

Going back to \cref{eq:F_eq}, if we apply $U$ to this we get
$\tilde f_\textrm{eq}=\beta^{-1}\sum_{i\omega_n}\tilde G(i\omega_n)$. Therefore, now all we need to do is sum over the Matsubara frequencies. 
We use the formula
\begin{equation}
    \frac{1}{\beta}\sum_{i\omega_n}\frac{e^{i\omega_n0^+}}{(i\omega_n-h^{(0)}+\mu)^k}
    =\frac{1}{(k-1)!} n_F^{(k-1)}(h^{(0)}-\mu)
\end{equation}
where $n_F^{(k-1)}$ is the $k-1$-th derivative of $n_F(x)=(e^{\beta(x-\mu)}+1)^{-1}$ which is the Fermi-Dirac distribution function. 
We then find
\begin{equation}\label{eq:f eq tilde app}
  \tilde{f}_\textrm{eq}= n_F(\tilde h)-\frac{\hbar^2}{16}
  n_F''(\tilde h)\wab\wsl\dv^2_{\alpha\sigma}\tilde{h}\dv^2_{\beta\lambda}\tilde{h}
  -\frac{\hbar^2}{24}
  n_F'''(\tilde h)\wab\wsl\dv^2_{\alpha\sigma}\tilde{h}
  \dv_\beta\tilde{h}\dv_\lambda\tilde{h}
  +\mathcal{O}(\hbar^3)
\end{equation}
Notably, this function can also be derived from the Moyal-generalization of the Fermi-Dirac distributions. Specifically, we define 
\begin{align}
    \tilde f_{\mathrm{eq}}  = n_{\mathrm{F}}(-\mu)+n_{\mathrm{F}}^{\prime}(-\mu) \tilde h+\frac{1}{2} n_{\mathrm{F}}^{\prime \prime}(-\mu) \tilde h \star \tilde h+\frac{1}{3!} n_{\mathrm{F}}^{\prime \prime \prime}(-\mu) \tilde h \star \tilde h \star \tilde h+\ldots,
\end{align}
which reduces to $n_\F(h-\mu)$ if all the star products are the normal products.
\red{By properly arranging the infinite series, we find it is precisely \eqref{eq:f eq tilde app} to the second order in $\hbar$; this agrees with a result of Blount\cite{Blount_bloch}.}

\section{Polarization in an insulator}\label{app:polarization}

Let us assume an insulator so that $n_\F = 1$ throughout the Brillouin zone. From \eqref{eq:invteqdist}, we find the bound charge density is given by
\begin{align}\label{eq: bound charge}
    \rho_b \equiv -e\int_p (f_{\mathrm{eq}} -n_\F)= -e\hbar \int_p \left(\Omega_{x^i p^i} +\frac{\hbar}{2}\left( \Omega_{x^i x^j}\Omega_{p^i p^j}- \Omega_{x^i p^j}\Omega_{p^i x^j} - \Omega_{x^i p^i}\Omega_{x^j p^j} + \p_{x^i x^j}^2 g_{p^i p^j}\right) \right) +\cO(\hbar^3).
\end{align}
The polarization vector $\cP_i$ is defined through $\rho_b = - \p_i \cP_i$. First, the single Berry curvature term in \eqref{eq: bound charge} gives
\begin{align}\label{eq:polarization 1}
    \cP_i^{(1)} = e\hbar \int_p \left( A_{p^i} +\hbar \left(A_{p^i} \p_{x^j}A_{p^j} + A_{x^j}\p_{p^j}A_{p^i} + A_{p^j}\p_{p^i}A_{x^j} \right) \right) +\cO(\hbar^3).
\end{align}
Second, the double Berry curvature terms give
\begin{align}
    \cP_i^{(2)} = \frac{e\hbar^2}{2} \int_p \left( A_{p^j} \p_{x^j}A_{p^i} - A_{p^i}\p_{x^j}A_{p^j} - 2 A_{p^j}\p_{p^i}A_{x^j} - 2 A_{x^j}\p_{p^j} A_{p^i} \right)  +\cO(\hbar^3).
\end{align}
Then, the quantum metric gives
\begin{align}
    \cP_i^{(3)} = \frac{e\hbar^2}{2} \int_p \p_{x^j}g_{p^i p^j}  +\cO(\hbar^3).
\end{align}
Gathering above, we find the total polarization vector as
\begin{align}
    \cP_i = e\hbar \int_p \left( A_{p^i} +\frac{\hbar}{2} \p_{x^j}\left(A_{p^i} A_{p^j}\right) + \frac{\hbar}{2} \p_{x^j}g_{p^i p^j}\right) +\cO(\hbar^3).
\end{align}

Comparing with the semiclassical result in \cite{Di_polarization}, we find that only the contribution in \eqref{eq:polarization 1} can be identified the latter, in particular with Eq.(1), Eq.(3) and Eq.(11) in \cite{Di_polarization}.  The remaining terms from $\cP_i^{(2)}$ and $\cP_i^{(3)}$ are distinct, but are not present in \cite{Di_polarization}.  Nevertheless, our final result passes all physical checks and the discrepancy is an interesting subject for further study.  In our opinion our calculation is more rigorous.

For a separable Hamiltonian (c.f. \eqref{eq:seperable H}), the gauge field $A_{p^i}$ and the quantum metric $g_{p^i p^j}$ are functions of only $p$ at the leading order, so the spatial derivatives vanish, and the resulting polarization $\cP_i = e\hbar \int_p A_{p^i}+\cO(\hbar^3)$ involves only the second order corrections to the gauge field $A_{p^i}$, recovering the semiclassical results \cite{Gao_field}. 

\section{Derivation of the electric current \cref{eq:current}}
\label{sec:current_derivation}
We assume our system is coupled to an external $U(1)$ gauge field $\mathcal{A}_\mu(x)$.
Then, the associated charge current is given by
\begin{equation}
    J_\mu(x)=-\frac{\delta\langle \hat H_\mathcal{A}\rangle}{\delta \mathcal{A}_\mu(x)}
\end{equation}
where $\hat H_\mathcal{A}$ is the Hamiltonian.
\begin{equation}
  \langle \hat H_\mathcal{A}\rangle=\int_{x,p}\tr[\sfh_\mathcal{A}(x,p)\sff(x,p)].
\end{equation}
Below we show that $\sfh_\mathcal{A}(x,p)$ can be expressed to second order in $\hbar$ as
\begin{equation}
  \sfh_\mathcal{A}(x,p)=\sfh(x,\pi)-\frac{e\hbar^2}{24}
  \frac{\dv^2\mathcal{A}_\mu(x)}{\dv x_\nu \dv x_\gamma}\frac{\dv^3\sfh(x,\pi)}{\dv\pi_\mu\dv\pi_\nu\dv
  \pi_\gamma}+\mathcal{O}(\hbar^3)
  \label{eq:ha_expansion}
\end{equation}
where $\sfh(x,p)$ is the Hamiltonian when $\mathcal{A}=0$ and
$\pi_\mu(x,p)=p_\mu+e\mathcal{A}_\mu(x)$. Then, the $U(1)$ current is
given by
\begin{equation}
  J_\mu(x)=-e\int_p \tr\left[\frac{\dv \sfh(x,\pi)}{\dv
  \pi_\mu}\sff(x,p)\right]
  +\frac{e\hbar^2}{24}\frac{\dv^2}{\dv x_\nu\dv x_\gamma}\int_p\tr\left[\frac{\dv^3
  \sfh(x,\pi)}{\dv\pi_\mu\dv\pi_\nu\dv\pi_\gamma}\sff\right]+\mathcal{O}(\hbar^3).
\end{equation}
We first prove \cref{eq:ha_expansion} by considering a separable Hamiltonian meaning $\hat H=H_0(\hat p)+ H_1(\hat x)$.
Then, we consider the more general case.

We assume spinless fermions and introduce the external field via minimal
coupling. Then,
  $\hat H_\mathcal{A}=H_0(\hat p+e\mathcal{A}(\hat x))+H_1(\hat x)$.
The second term $H_1(\hat x)$ does not contribute to the current, so we ignore it. In general
$H_0(p)$ can be expressed as
\begin{equation}
  H_0(\hat p)=\sum_{n=1}^\infty \frac{1}{n!}G_{\alpha_1\cdots\alpha _n}\,
  \hat p_{\alpha_1}\cdots \hat p_{\alpha_n}
\end{equation}
where $G_{\alpha_1\cdots \alpha_n}$ is a matrix-valued symmetric tensor. The
external gauge field can be introduced by replacing each $\hat p_\alpha$ with
$\hat \pi_\alpha=\hat p_\alpha+e\mathcal{A}(\hat x)$. Applying the Wigner
transformation we get
\begin{align}
  \hat \pi_\alpha\rightarrow&
  \,\pi_\alpha(x,p)=p_\alpha+e\mathcal{A}(x)\\
  H_0(\hat p)\rightarrow&
  \,\sfh_0(p)
  =\sum_{n=1}^\infty \frac{1}{n!}G_{\alpha_1\cdots\alpha _n}\,
  p_{\alpha_1}\cdots p_{\alpha_n}\\
  H_0(\hat \pi)\rightarrow&
  \,\sfh_{0\mathcal{A}}(x,p)
  =\sum_{n=1}^\infty \frac{1}{n!}G_{\alpha_1\cdots\alpha _n}\,
  \pi_{\alpha_1}\star\cdots\star\pi_{\alpha_n}
\end{align}
We don't have any Moyal products in the second line because $p_\alpha\star
p_\beta=p_\alpha p_\beta$ generally.

\begin{mdframed}
\textbf{Lemma}
\begin{align}
  G_{\alpha_1\cdots\alpha _n}\,
  \pi_{\alpha_1}\star\cdots\star\pi_{\alpha_n}
  =&G_{\alpha_1\cdots\alpha _n}\left(
  \pi_{\alpha_1}\cdots\pi_{\alpha_n}
  -\frac{e\hbar^2}{24}n(n-1)(n-2)
  \pi_{\alpha_4}\cdots\pi_{\alpha_n}
  \frac{\dv^2\mathcal{A}_{\alpha_1}(x)}{\dv x_{\alpha_2}\dv x_{\alpha_3}}
  \right)+\mathcal{O}(\hbar^3)\notag\\
  =&
  \left(1-\frac{e\hbar^2}{24}\frac{\dv^2\mathcal{A}_\mu(x)}{\dv x_\nu\dv x_\gamma}
  \frac{\dv^3}{\dv\pi_\mu\dv\pi_\nu\dv\pi_\gamma}\right)
  G_{\alpha_1\cdots\alpha _n}\,
  \pi_{\alpha_1}\cdots\pi_{\alpha_n}
  \label{eq:lemma}
\end{align}
{\underline{\emph{Proof:}}}
First, note $\sacomm{A}{B}/2=
AB-(\hbar^2/8)\wab\wsl\dv^2_{\alpha\sigma}A\dv^2_{\sigma\lambda}B
+\mathcal{O}(\hbar^3)$. Then, using the fact that $G_{\alpha_1\cdots\alpha_n}$
is symmetric under exchange of its indices we can get
\begin{align}
  G_{\alpha_1\cdots\alpha _n}\,
  \pi_{\alpha_1}\star\cdots\star\pi_{\alpha_n}=&
  \frac{1}{2}G_{\alpha_1\cdots\alpha _n}\,
  \sacomm{\pi_{\alpha_2}\star\cdots\star\pi_{\alpha_{n}}}{\pi_{\alpha_1}}\notag\\
  =&G_{\alpha_1\cdots\alpha _n}\,\left(
  (\pi_{\alpha_2}\star\cdots\star\pi_{\alpha_{n}})\pi_{\alpha_1}
  -\frac{\hbar^2}{8}\frac{\dv^2}{\dv p_\nu \dv p_\gamma}(\pi_{\alpha_1}\cdots\pi_{\alpha_{n-1}})
  \frac{\dv^2\pi_{\alpha_n}}{\dv x_\nu \dv
  x_\gamma}\right)+\mathcal{O}(\hbar^3)\notag\\
  =&G_{\alpha_1\cdots\alpha _n}\,\left(
  (\pi_{\alpha_2}\star\cdots\star\pi_{\alpha_{n}})\pi_{\alpha_1}
  -(n-1)(n-2)\frac{e\hbar^2}{8}(\pi_{\alpha_4}\cdots\pi_{\alpha_{n}})
  \frac{\dv^2\mathcal{A}_{\alpha_1}}{\dv x_{\alpha_2} \dv x_{\alpha_3}}\right)
  +\mathcal{O}(\hbar^3)\notag
\end{align}
Notice this sets up a recursive relation. The first term in the last line with
the $\star$ products is the left-hand side up to differences in the symmetric tensor. If we
continue this recursive relation we find
\begin{align}
  G_{\alpha_1\cdots\alpha _n}\,
  \pi_{\alpha_1}\star\cdots\star\pi_{\alpha_n}
  =&G_{\alpha_1\cdots\alpha _n}\,\left(
  \pi_{\alpha_1}\cdots\pi_{\alpha_{n}}
  -\sum_{m=1}^{n-2}\frac{(n-m)!}{(n-m-2)!}\frac{e\hbar^2}{8}(\pi_{\alpha_4}\cdots\pi_{\alpha_{n}})
  \frac{\dv^2\mathcal{A}_{\alpha_1}}{\dv x_{\alpha_2} \dv x_{\alpha_3}}\right)
  +\mathcal{O}(\hbar^3)\notag
\end{align}
Finally, we use the fact that
$\sum_{m=1}^{n-2}\frac{(n-m)!}{(n-m-2)!}=\frac{1}{3}n(n-1)(n-2)$ and we get
\eqref{eq:lemma}.
\end{mdframed}
Applying the lemma  \eqref{eq:lemma} we find
\begin{equation}
  \sfh_{0\mathcal{A}}(x,p)=
  \sfh_0(\pi)-\frac{e\hbar^2}{24}\frac{\dv^3
  \sfh_0(\pi)}{\dv\pi_\mu\dv\pi_\nu\dv\pi_\gamma}
  \frac{\dv^2 \mathcal{A}_\mu(x)}{\dv x_\nu\dv x_\gamma}+\mathcal{O}(\hbar^3).
  \label{eq:p_dependent_ham}
\end{equation}

We now consider when the Hamiltonian is not separable. We assume that the
Hamiltonian is ordered such that the $\hat x$
and $\hat p$ operators are separately grouped together. In other words, we
assume it takes the general form
\begin{equation}
  \hat H=\sum_n v_n(\hat x) k_n(\hat p)\Gamma_n+ \rm{h.c.}
\end{equation}
where $\{v_n\}, \{k_n\}$ are a set of polynomial functions and $\Gamma_n$ is a constant
matrix.
We now once again couple the system to an external gauge field by taking $\hat
p\rightarrow\hat \pi$, and calculate the Wigner transformation. We know what the
Wigner transformation of $k_n(\hat \pi)$ must be from \eqref{eq:p_dependent_ham}.
Therefore,
\begin{align}
  \sfh_\mathcal{A}(x,p)=&\sum_n v_n(x)\star \left(k_n(\pi)
  -\frac{e\hbar^2}{24}
  \frac{\dv^2\mathcal{A}_\mu(x)}{\dv x_\nu \dv x_\gamma}\frac{\dv^3k_n(\pi)}{\dv\pi_\mu \dv
  \pi_\nu\dv\pi_\gamma}
  \right)\Gamma_n+\rm{h.c.}+\mathcal{O}(\hbar^3)\notag\\
  =&\left(1-\frac{e\hbar^2}{24}
  \frac{\dv^2\mathcal{A}_\mu(x)}{\dv x_\nu \dv x_\gamma}
  \frac{\dv^3}{\dv\pi_\mu\dv\pi_\nu\dv\pi_\gamma}\right)\sum_n v_n(x)\star
  k_n(\pi)\Gamma_n+\rm{h.c.}+\mathcal{O}(\hbar^3)\notag\\
  =&\left(1-\frac{e\hbar^2}{24}
  \frac{\dv^2\mathcal{A}_\mu(x)}{\dv x_\nu \dv x_\gamma}
  \frac{\dv^3}{\dv\pi_\mu\dv\pi_\nu\dv\pi_\gamma}\right)\sfh(x,\pi)
  +\mathcal{O}(\hbar^3)
\end{align}
To get the second line, we simply move the derivatives outside of the sum since
$k_n(\pi)$ is the only part that's dependent on $\pi$. In addition, we recognize
that $\sum_n v_n(x)\star k_n(\pi)\Gamma_n+\textrm{h.c.}=\sum_n v_n(x)\star
k_n(p)\Gamma_n+\rm{h.c.}\vert_{p=\pi}=\sfh(x,\pi)$.
Therefore, \eqref{eq:p_dependent_ham} generalizes to arbitrary Hamiltonians
(that can be expressed as a polynomial of $\hat x$ and $\hat p$.)

\section{Projection operator identities}
\label{app:projection_identities}
The projection operator for the $n$-th band is defined as
\begin{equation}
    P_n\equiv U\star p_n\star U^\dagger
\end{equation}
where the matrix elements of $p_n$ are given by
$(p_n)_{ij}=\delta_{in}\delta_{jn}$. The projection operator is idempotent
$P_n\star P_n=P_n$ and sum to identity $\sum_{n=1}^NP_n=I_N$.

The trace of the projection operator is not $1$. Rather,
\begin{equation}
    \tr[P_n]=1
   +\frac{\hbar}{2}\wab\Omega_{n,\alpha\beta}
   +\frac{\hbar^2}{4}\wab\wsl\left(
   \Omega_{n,\alpha\sigma}\Omega_{n,\beta\lambda}
   -\frac{1}{2}\Omega_{n,\alpha\beta}\Omega_{n,\sigma\lambda}
   \right)
  +\frac{\hbar^2}{2}\wab\wsl\dv^2_{\alpha\sigma}g_{n,\beta\lambda}+\mathcal{O}(\hbar^3)
\end{equation}
which can be derived using \cref{eq:id2}.

Taking derivatives of the idempotence property $P_n\star P_n=P_n$ gives us
\begin{equation}
  \dv_\alpha P_n=\dv_\alpha P_n\star P_n+ P_n\star \dv_\alpha P_n,
\end{equation}
\begin{equation}
    \dv^2_{\alpha\sigma} P_n=\dv^2_{\alpha\sigma} P_n\star P_n +P_n\star \dv^2_{\alpha\sigma} P_n
    +\dv_\alpha P_n\star \dv_\sigma P_n+\dv_\sigma P_n\star \dv_\alpha P_n,
    \label{eq:d2P}
\end{equation}
\begin{align}
    \dv^3_{\alpha\sigma\mu} P_n=&\dv^3_{\alpha\sigma\mu} P_n\star P_n +P_n \star \dv^3_{\alpha\sigma\mu} P_n
    +\dv_\alpha P_n\star \dv^2_{\sigma\mu}P_n+\dv_\sigma P_n\star \dv^2_{\alpha\mu}P_n+\dv_\mu P_n\star \dv^2_{\alpha\sigma}P_n\\
    &+\dv^2_{\sigma\mu}P_n\star\dv_\alpha P_n+\dv^2_{\alpha\mu}P_n\star\dv_\sigma P_n+\dv^2_{\alpha\sigma}P_n\star\dv_\mu P_n.\notag\
\end{align}

Using these three relations we can derive many useful identities. Below we list some useful trace identities.
\begin{equation}
  \tr[P_n\star \dv_\alpha P_n\star P_n]=0
\end{equation}
\begin{equation}
  \tr[\dv_\alpha P_n\dv_\beta P_n\dv_\gamma P_n]=0+\mathcal{O}(\hbar)
\end{equation}
\begin{equation}
  \tr[\dv_\beta P_n\star P_n]=i\hbar\wsl \dv_\sigma \mathcal{T}_{n,\lambda\beta}+\mathcal{O}(\hbar^2)
\end{equation}
\begin{equation}
  \tr[\dv_\alpha P_n\dv_\beta P_n]=2g_{n,\alpha\beta}+\mathcal{O}(\hbar)
\end{equation}
\begin{equation}
  \tr[P_n\star\dv^2_{\beta\lambda}P_n\star P_n]
  =-2g_{n,\beta\lambda}
\end{equation}
\begin{equation}
  \tr[P_n\dv^3_{\beta\lambda\nu}P_n P_n]=-\dv_\nu g_{n,\beta\lambda}
  -\dv_\lambda g_{n,\nu\beta}-\dv_\beta g_{n,\nu\lambda}+\mathcal{O}(\hbar)
\end{equation}
\begin{equation}
    \tr[\dv_\beta P_n \dv^2_{\nu\lambda} P_n]=
    \dv_\nu g_{n,\beta\lambda}+\dv_\lambda g_{n,\nu\beta}-\dv_\beta g_{n,\nu\lambda}+\mathcal{O}(\hbar)
\end{equation}

\section{Projection operator representation of $\sfh$ and $\sff$}
\label{app:projection_operator_representation}
In this section, we show that $\sfh$ and $\sff$ can be expressed as
\begin{align}
    \sfh&=\sum_{n=1}^N P_n\star h_n\star P_n
    \label{eq:H_projection_rep}\\
    \sff&=\sum_{n=1}^N P_n\star \ul{f}_n\star P_n
    +U\star\tilde{\mathcal{F}}\star U^\dagger
\end{align}
where
\begin{align}
    \ul{f}=\bigg( 1-\frac{\hbar}{2}\wab\Omega_{\alpha\beta}
  -\frac{\hbar^2}{2}\wab\wsl\bigg(\frac{1}{2}\Omega_{\alpha\sigma}\Omega_{\beta\lambda}
  -\frac{3}{4}\Omega_{\alpha\beta}\Omega_{\sigma\lambda}
  +\dv^2_{\alpha\sigma}g_{\beta\lambda}\bigg)\bigg)f
  -\hbar^2\wab\wsl\dv_\sigma(\dv_\alpha f g_{\beta\lambda}).
  \label{eq:app_ulf}
\end{align}

The Hamiltonian can be expressed as $\sfh=U\star\tilde h\star U^\dagger$. From
this expression, it is straightforward to show that the Hamiltonian commutes
with the projection operator, $\scomm{\sfh}{P_n}=0$. Since $\sum_{n=1}^N
P_n=I_n$, we can write $\sfh=\sum_n P_n\star \sfh \star P_n=\sum_n U\star \tilde
h_n p_n \star U^\dagger$.  We now make an ansatz that there exists a $\ul{h}_n$
such that
\begin{equation}
    P_n\star\ul{h}_n\star P_n= U\star \tilde h_n p_n \star U^\dagger.
\end{equation}
We can take the trace of both sides and use \cref{eq:id1,eq:id2} to solve for $\ul{h}$.
\begin{equation}
    \tr[P_n\star\ul{h}_n\star P_n]= \ul{h}_n\tr[P_n]+\hbar^2\wab\wsl\dv_\sigma(\dv_\alpha \ul{h}_n g_{n,\beta\lambda})
    +\mathcal{O}(\hbar^3)
\end{equation}
and
\begin{equation}
    \tr[U\star\tilde h_n p_n\star U^\dagger]=\tilde h_n+\hbar\wab\dv_\alpha(\tilde h_n A_{n\beta})
    +\frac{\hbar^2}{4}\wab\wsl\dv^2_{\alpha\sigma}(\tilde h_n\acomm{\Lambda_\beta}{\Lambda_\lambda}_{nn})
    +\mathcal{O}(\hbar^3)
\end{equation}
If we then solve for $\ul{h}$ we find that $\ul{h}=h$ which gives us
\cref{eq:H_projection_rep}. 

Unlike the Hamiltonian, the distribution function is expressed as the sum of two
parts $\sff=U\star\tilde f\star U^\dagger+U\star\tilde{\mathcal{F}} \star
U^\dagger$. The first term can be expressed in the same way as the Hamiltonian
since $\tilde f$ is diagonal. Therefore, we define $\sff=\sum_{n=1}^N P_n\star
\ul{f}_n\star P_n+U\star\tilde{\mathcal{F}} \star U^\dagger$. We can recycle the
fact that $\ul{h}=h$ to determine that
\begin{equation}
    \ul{f}=
  \tilde f_n+\hbar\wab\dv_\alpha\tilde f_n(A_{n\beta}+\hbar\wsl
    A_{n\lambda}\dv_\sigma A_{n\beta})
    -\hbar^2\wab\wsl\dv^2_{\alpha\sigma}\tilde f_n
    \left(\frac{1}{4}\acomm{\Lambda_\beta}{\Lambda_\lambda}_{nn}
    -A_{n\beta}A_{n\lambda}\right)+\mathcal{O}(\hbar^3)
\end{equation}
by simply taking the definition for $h$ and replacing $\tilde h$ with $\tilde f$. Next, we use
\cref{eq:tildef} to express $\tilde f$ using $f$ and obtain \cref{eq:app_ulf}.

\section{Expressing electric current using $h,f$ (\cref{eq:diagonalcurrent})}
\label{app:deriving_diagonalcurrent}
Here, we outline the derivation of \cref{eq:diagonalcurrent}. First, we
generalize the current given by \cref{eq:current} to a phase-space current. This
is straightforward and we define
\begin{equation}
  \mathcal{J}_\mu(x,p)
  \equiv\wmn\Re\tr[\dv_\nu \sfh\star\sff]
  +\frac{\hbar^2}{12}\wmn\wab\wsl\dv^2_{\alpha\sigma}
  \tr[\dv^3_{\nu\beta\lambda}\sfh\sff]+\mathcal{O}(\hbar^3)
\end{equation}
such that $J_{i}(x)=-e\int_p \mathcal{J}_{x_i}(x,p)$. Assuming the off-diagonal
components of the distribution function, $\tilde{\mathcal{F}}$, are negligible,
we can write $\sff=\sum_n P_n\star \ul{f}_n\star P_n$ which is the projection
operator representation discussed in
\cref{app:projection_operator_representation}. Substituting this into the
expression for the phase-space current and using \cref{eq:id1}, we find
\begin{align}
  \mathcal{J}_\mu
  =&\wmn\sum_n \ul{f}_n V_{n,\nu}
  +\wmn\hbar\wab\dv_\alpha\sum_n\ul{f}_n V_{n,\nu\beta}
  +\frac{\hbar^2}{12}\wmn\wab\wsl\dv^2_{\alpha\sigma}
  \sum_n \ul{f}_n V_{n,\nu\beta\lambda}
  +\mathcal{O}(\hbar^3)
\end{align}
where 
\begin{equation}
  V_{n,\nu}\equiv \tr[P_n\star\dv_\nu\sfh\star P_n] 
\end{equation}
\begin{equation}
  V_{n,\nu\beta}\equiv
  -\Im\tr[\dv_\beta P_n\star\dv_\nu \sfh\star P_n]
\end{equation}
\begin{equation}
  V_{n,\nu\beta\lambda}\equiv
  -\frac{1}{2}\dv^3_{\nu\beta\lambda}h_n
  +\frac{1}{2}\dv_\lambda\tr[\dv_\beta P_n\dv_\nu
  \sfh]+\frac{1}{2}\dv_\beta\tr[\dv_\lambda P_n \dv_\nu\sfh]
  -2\tr[\dv^2_{\beta\lambda}P_n\dv_\nu \sfh]
  +6\Re\tr[\dv_\beta P_n\dv_\nu \sfh\dv_\lambda P_n]
\end{equation}
We can now rewrite each of these traces by substituting $\sfh=\sum_n P_n\star
h_n\star P_n$ and using the trace identities of the projection operator listed in
\cref{app:projection_identities}. We find
\begin{equation}
  \begin{split}
  V_\nu=&\dv_\nu h\left(1
   +\frac{\hbar}{2}\wab\Omega_{\alpha\beta}
   +\frac{\hbar^2}{4}\wab\wsl\left(
   \Omega_{\alpha\sigma}\Omega_{\beta\lambda}
   -\frac{1}{2}\Omega_{\alpha\beta}\Omega_{\sigma\lambda}
   \right)
  +\frac{\hbar^2}{2}\wab\wsl\dv^2_{\alpha\sigma}g_{\beta\lambda}\right)\\
  &-\hbar\wab\dv_\alpha
  h\Omega_{\nu\beta} +\hbar^2\wab\wsl\left(\frac{1}{2}\dv^2_{\alpha\sigma}h\dv_\nu
  g_{\beta\lambda}+\dv^3_{\nu\alpha\sigma}h g_{\beta\lambda}
  +\dv^2_{\nu\alpha}h \dv_\sigma g_{\beta\lambda} \right)
  +\mathcal{O}(\hbar^3)
\end{split}
\end{equation}
\begin{equation}
    V_{\nu\beta}=-m_{\nu\beta}-\hbar\wsl(\dv^2_{\nu\sigma} h g_{\lambda\beta}+\dv_\nu h \dv_\sigma g_{\lambda\beta})
    +\frac{\hbar}{2}\wsl\dv_\sigma h(\dv_\nu g_{\beta\lambda}-\dv_\beta g_{\lambda\nu}
    +\dv_\lambda g_{\nu\beta})+\mathcal{O}(\hbar^2)
\end{equation}
\begin{align}
    V_{\nu\beta\lambda}=-\frac{1}{2}\dv^3_{\nu\beta\lambda}h
    +12 \dv_\nu h g_{\beta\lambda}
    +2\dv_\nu c_{\beta\lambda}-\dv_\lambda c_{\beta\nu} -\dv_\beta c_{\lambda\nu}
    +4 c_{\beta\lambda;\nu}-2 c_{\beta\nu;\lambda}-2 c_{\lambda\nu;\beta}
    +\mathcal{O}(\hbar)
\end{align}
where we defined $m$ and $c$'s to be diagonal matrices whose $n$-th diagonal
elements are given by
\begin{equation}
    m_{n,\nu\beta}\equiv\Im\tr[\dv_\nu P_n\star (\sfh- h_n I_N)\star \dv_\beta P_n]
\end{equation}
\begin{equation}
    c_{n,\nu\beta}\equiv\Re\tr[\dv_\nu P_n\star (\sfh- h_n I_N)\star \dv_\beta P_n]
\end{equation}
\begin{equation}
    c_{n,\beta\lambda;\nu}\equiv\Re\tr[\dv^2_{\beta\lambda} P_n\star (\sfh- h_n I_N)\star \dv_\nu P_n]
\end{equation}
Next, we express $\ul{f}$ using $f$ by substituting
\begin{equation}
\ul{f}=\bigg( 1-\frac{\hbar}{2}\wab\Omega_{\alpha\beta}
  -\frac{\hbar^2}{2}\wab\wsl\bigg(\frac{1}{2}\Omega_{\alpha\sigma}\Omega_{\beta\lambda}\notag\\
  -\frac{3}{4}\Omega_{\alpha\beta}\Omega_{\sigma\lambda}
  +\dv^2_{\alpha\sigma}g_{\beta\lambda}\bigg)\bigg)f
  -\hbar^2\wab\wsl\dv_\sigma(\dv_\alpha f g_{\beta\lambda}).
\end{equation}
We then find
\begin{equation}
  \begin{split}
    \mathcal{J}_\mu=&\wmn\tr\left[f\left(\dv_\nu h-\hbar\wab\dv_\alpha h\Omega_{\nu\beta}
    (1-\frac{\hbar}{2}\wsl\Omega_{\sigma\lambda})+\frac{\hbar^2}{2}\wab\wsl\dv^2_{\alpha\sigma}h\dv_\nu
    g_{\beta\lambda}\right)\right]\\
    &+\hbar\wmn\wab\dv_\alpha\tr[f (b^s_{\nu\beta}+b^a_{\nu\beta})]
    +\frac{\hbar^2}{12}\wmn\wab\wsl\dv^2_{\alpha\sigma}\tr[f
    (b^s_{\nu\beta\lambda}+b^m_{\nu\beta;\lambda}+b^m_{\nu\lambda;\beta})]
  \end{split}
\end{equation}
where
\begin{equation}
  b_{\nu\beta}^s\equiv\frac{\hbar}{2}\wsl(\dv_\sigma h \dv_\lambda
  g_{\nu\beta}+\dv^2_{\sigma\nu}h g_{\beta\lambda}+\dv^2_{\sigma\beta}h g_{\nu\lambda})
\end{equation}
\begin{equation}
  b^a_{\nu\beta}\equiv -m_{\nu\beta}\left(1-\frac{\hbar}{2}
  \wsl\Omega_{\sigma\lambda}\right)-\frac{\hbar}{2}\wsl
  \left( \dv_\beta(\dv_\sigma h g_{\nu\lambda})
  -\dv_{\nu}(\dv_\sigma hg_{\beta\lambda})\right)
\end{equation}
\begin{equation}
  b_{\nu\beta\lambda}^s\equiv -\frac{1}{2}\dv^3_{\nu\beta\lambda}h
\end{equation}
\begin{equation}
  b^m_{\nu\lambda;\beta}\equiv
 \dv_\nu c_{\beta\lambda} -\dv_\lambda c_{\beta\nu}
  +2c_{\beta\lambda;\nu}-2c_{\beta\nu;\lambda}
\end{equation}
Then, by calculating $J_i\equiv -e \int_p \mathcal{J}_{x_i}$ we get \cref{eq:diagonalcurrent}.

\section{Chiral anomaly effect}\label{app:chiral anomaly}

The equilibrium real-space current has been shown in \eqref{eq:eq_magnetization} to be divergence-free in the absence of external fields. Here we want to demonstrate that it will break down in the presence of external fields at a Weyl node, known as the chiral anomaly effect. In this section, we focus on the $O(\hbar)$ contributions. 

According to \secref{sec:magnetic field}, the additional contribution to the equilibrium real-space current in the presence of a uniform magnetic field is due to $\omega_B^{p_ip_j} \neq 0$, where we relabeled the kinetic momentum $\pi_i\rightarrow p_i$. Such a constribution is given by

\begin{align}
    J'_{i,\mathrm{eq}} = -e\hbar \omega_B^{p_jp_k} \int_p \tr \left[\left(-\p_{p_j}h \Omega_{p_i p_k} + \frac{1}{2} \Omega_{p_j p_k} \p_{p_i}h \right)n_\F \right],
\end{align}
whose divergence is given by
\begin{align}
    \p_{x_i} J'_{i,\mathrm{eq}} & = -e\hbar \omega^{p_jp_k}_B \int_p \tr\left[ \left(-\p_{p_j}h\p_{x_i} \Omega_{p_i p_k} + \p_{x_i} h \p_{p_j}\Omega_{p_i p_k} + \frac{1}{2} \p_{x_i} \Omega_{p_j p_k} \p_{p_i}h - \frac{1}{2} \p_{p_i} \Omega_{p_j p_k} \p_{x_i}h \right) n_\F \right],
\end{align}
where we used integration by part. In $d=2$, we can write $\omega_B^{p_j p_k}=-e B \epsilon^{jk}$ and $\Omega_{p_i p_j} = \Omega \epsilon_{ij}$, and we find
\begin{align}
    \p_{x_i} J'_{i,\mathrm{eq}} &= 0.
\end{align}
In $d=3$, we can write $\omega_B^{p_j p_k}=-e B_l \epsilon^{jkl}$ and $\Omega_{p_i p_j} = \Omega_{p}^k \epsilon_{ijk}$, and we find
\begin{align}
    \p_{x_i} J'_{i,\mathrm{eq}} &=  e^2\hbar B_l \int_p(\p_{p_k} h \p_{x_l} \Omega_{p}^k - \p_{x_l} h \p_{p_k}\Omega_{p}^k)n_\F.
\end{align}
For a regular (twice differentiable) Hamiltonian, $\Omega_{p}^i$ will also be regular in real space, thus the only non-trivial term would be the second term; if $\Omega_p^i$ is regular in momentum space, however, $\nabla_p \cdot \Omega_p = \nabla_p \cdot (\nabla_p \times A_p) = 0$. At a Weyl node,  $\Omega_{p}^i = \pm p_i/2p^3$, thus, under an external electric field $\p_{x_i} h = -e E_i$, we find the breakdown of charge conservation
\begin{align}
    \p_{x_i} J'_{i,\mathrm{eq}} = \pm \frac{e^3\hbar}{4\pi^2} (E \cdot B), 
\end{align}
which is because at the gapless Weyl node, there is a Berry monopole $\nabla_p\cdot \Omega_p = 2\pi \delta^{(3)}(p)$.

\section{Deriving $f_r$ for the relaxation time approximation (\cref{eq:f_r})}
\label{app:f_r}
Here, we derive $f_r$ given by \cref{eq:f_r}, the distribution function which
$f$ relaxes to under the relaxation time approximation when the Hamiltonian
takes the form $\sfh=\sfh_0(p)-eV(x)I_N$. The relaxation time approximation of
the collision integral for the original Liouville-von Neumann equation given by
\cref{eq:eom f star} is $-\tau^{-1}(\sff-\sff_r)$ where
$\sff_r=\sff_\textrm{eq}= U^{(0)}n_F(\varepsilon) U^{(0)\dagger}$. To calculate
$f_r$, we need to first find an explicit expression for $\tilde
f_r=U^\dagger\star \sff_r\star U$. To do this, we use $T=\hbar U^\dagger\star
\dv_\hbar U$ which we defined when studying flow equations in
\cref{app:diagonalization}. Recall, $T= Y+\frac{\hbar}{4}\wab \dv_\alpha
\Lambda_\beta$ where $Y$ is the anti-Hermitian part and the second term is the
Hermitian part of $T$. If we expand $T$ in powers of $\hbar$ we find
\begin{align}
  T=&(U^{(0)\dagger}+U^{(1)\dagger}+\cdots)\star (U^{(1)}+2 U^{(2)}+\cdots)\\
  =&U^{(0)\dagger}U^{(1)}+2 U^{(0)\dagger}U^{(2)}+U^{(1)^\dagger}U^{(1)}
  +\frac{i\hbar}{2}\wab\dv_\alpha U^{(0)\dagger}\dv_\beta
  U^{(1)}+\mathcal{O}(\hbar^3)
\end{align}
For the Hamiltonian we are considering, $T^{(1)}=Y^{(1)}$ since $\Lambda_{x_i}=0$ and $\Lambda_{p_i}^{(0)}=\Lambda_{p_i}^{(0)}(p)$. Therefore,
$U^{(0)\dagger}U^{(1)}=Y^{(1)}$. In addition,
\begin{equation}
  U^{(0)\dagger}U^{(2)}=\frac{1}{2}\left(T^{(2)}+(Y^{(1)})^2+\frac{\hbar}{2}\Lambda_{p_i}^{(0)}\dv_{x_i}Y^{(1)}\right)
\end{equation}
so the Hermitian part of $U^{(0)\dagger}U^{(2)}$ is 
\begin{equation}
  \frac{1}{2}(U^{(0)\dagger}U^{(2)}+U^{(2)\dagger}U^{(0)})=-\frac{\hbar}{2}\comm{\dv_{x_i}Y^{(1)}}{\Lambda_{p_i}^{(0)}}
  +(Y^{(1)})^2,
\end{equation}
where we used the fact that
$A^{(1)}_{p_i}=\comm{\Lambda_{p_i}^{(0)}}{Y^{(1)}}^d$.

Let's now expand $\tilde f_r$ to second order in $\hbar$.
At zeroth order,  we trivially find $\tilde
f_r^{(0)}=n_F(\varepsilon)$. To first order it is given by $\tilde
f_r^{(1)}=\diag(T^{(1)}+T^{(1)\dagger}) n_F(\varepsilon)=0$. The second-order
contribution is
\begin{align}
  \tilde f_r^{(2)}=& U^{(2)\dagger}\sff_r U^{(0)}+U^{(0)\dagger}\sff_r U^{(2)}
  +U^{(1)\dagger}\sff_r U^{(1)}\\
  &+\frac{i\hbar}{2}(\dv_{x_i}U^{(1)\dagger}\dv_{p_i}\sff_r U^{(0)}
  -U^{(0)\dagger}\dv_{p_i}\sff_r\dv_{x_i}U^{(1)}
  +\dv_{x_i}U^{(1)\dagger}\sff_r \dv_{p_i}U^{(0)}
  -\dv_{p_i}U^{(0)\dagger}\sff_r \dv_{x_i}U^{(1)})\notag\\
  =&(U^{(2)\dagger}U^{(0)}+U^{(0)\dagger}U^{(2)})^dn_F(\varepsilon)
  -(Y^{(1)}n_F(\varepsilon)Y^{(1)})^d
  +\frac{\hbar}{2}\comm{\dv_{x_i}Y^{(1)}}{\Lambda_{p_i}^{(0)}}^dn_F(\varepsilon)\\
  =&\left(Y^{(1)}\comm{Y^{(1)}}{n_F(\varepsilon)}\right)^d
\end{align}
Here, we used $(\cdot)^d$ as short hand for $\diag(\cdot)$.
We took the diagonal of the right-hand side since $\tilde f_r$ is assumed to be
diagonal.
Since $(Y^{(1)})_{nm}=-\frac{e}{\varepsilon_n-\varepsilon_m} 
(\Lambda^{(0)}_{p_i})_{nm}\dv_{x_i}V$ for $n\neq m$, if we take just the $n$-th
element of $\tilde f_r^{(2)}$, we see we can rewrite it as
\begin{equation}
  \tilde f_{r,n}^{(2)}=-e^2\sum_{m\neq
  n}(\Lambda_{p_i}^{(0)})_{nm}(\Lambda_{p_j}^{(0)})_{mn}
  \frac{n_{F,n}-n_{F,m}}{(\varepsilon_n-\varepsilon_m)^2}\dv_{x_i}V\dv_{x_j}V.
\end{equation}
Now, $f_r$ is $f_r=\tilde f_r+\hbar \tilde
f_r\dv_{x_i}A_{p_i}+\mathcal{O}(\hbar^3)$ from \cref{eq:invtf}. Therefore,
\begin{equation}
  f_{r,n}=n_{F,n}(1-e\hbar^2 T_{n,ij}\dv^2_{x_i x_j}V)-e^2\sum_{m\neq
  n}\langle\dv_{p_i} u_n|u_m\rangle\langle u_m|\dv_{p_j} u_n\rangle
  \frac{n_{F,n}-n_{F,m}}{(\varepsilon_n-\varepsilon_m)^2}\dv_{x_i}V\dv_{x_j}V
\end{equation}
where we used $(\Lambda^{(0)}_{p_i})_{nm}=-i\langle u_n|\dv_{p_i} u_m\rangle$.

\section{Density-density correlation function}\label{app:density-density}

The density-density correlation function for non-interacting multi-band fermions is given by 
\begin{align}\label{eq:C2 full}
  \widehat{C}^{(2)}(q,\omega)=&i \hbar \sum_{\substack{n,m=1\\n\neq m}}^N\int_k
  (U^{(0)\dagger}(k-q) U^{(0)}(k))_{nm} (U^{(0)\dagger}(k) U^{(0)}(k-q))_{mn}
  \frac{n_{F,n}(k-q)-n_{F,m}(k)}{\hbar\omega-\varepsilon_{m}(k)+\varepsilon_{n}(k-q)}\\
  \approx&-i\sum_{n=1}^N\int_k\frac{\hbar^2 n_{F,n}'\vec{v}\cdot\vec{q}\,(1-{\sf g}_{n,ij} q_i q_j)}
  {\hbar\omega-\varepsilon_{n}(k)+\varepsilon_{n}(k-q)}
  +i\hbar\sum_{\substack{n,m=1\\n\neq m}}^N\int_k (n_{F,n}-n_{F,m})
  \frac{({\sf \Lambda}_i)_{nm}({\sf \Lambda}_j)_{mn}}
  {\hbar\omega-\varepsilon_{m}+\varepsilon_{n}}q_i q_j
  +\mathcal{O}(q^3)\notag
\end{align}
where $n_{F,n}=n_F(\varepsilon_n)$. The last line was obtained by expanding to second-order in $q$.
In the low-frequency limit $\omega\ll\omega_\textrm{gap}$, the frequency in the denominator of the last term
goes to zero and this term can be expressed using ${\sf t}_{ij}$. A straightforward calculation shows that this expression reduces to \eqref{eq:C2}. However, the full expression \eqref{eq:C2 full} is capable of describing an insulator, for which \eqref{eq:Sq intra} vanishes. In particular, the interband static structure factor due to the last term in \eqref{eq:C2 full} is given by
\begin{align}
    S_q^{\mathrm{inter}} = q_i q_j \int_k {\sf g}_{ij}^{\mathrm{occ}}(k),
\end{align}
where ${\sf g}_{ij}^{\mathrm{occ}}\equiv \sum_{m>n,n\in \mathrm{occ}} ({\sf \Lambda}_i)_{nm}({\sf \Lambda}_j)_{mn}$ is the quantum metric of the occupied bands \cite{souzaPolarizationLocalizationInsulators2000}.

\section{ Comparing to Blount's mixed representation} \label{app:blount}

In \cite{blount_1962_formalisms,Blount_extension,Blount_bloch}, Blount introduced the mixed representation of band theory. In this appendix, we compare our formalism to it.

Consider a separable Hamiltonian given in \eqref{eq:seperable H}. Using the quantities calculated in Sec.\ref{sec:application}, we find the band-diagonal but gauge-non-invariant corrections, \eqref{eq:tilde h 1} and \eqref{eq:tilde h 2}, as
\begin{align}
    \tilde h^{(1)} = -\hbar \omega^{\alpha\beta} \p_\alpha\tilde h^{(0)} A^{(0)}_\beta = e\hbar \p_i V A^{(0)}_{p^i},
\end{align}
and
\begin{align}
    \tilde h^{(2)}&=-\frac{\hbar}{2}\wab(\dv_\alpha\tilde h^{(1)} A_{\beta}^{(0)}
    +\dv_\alpha\tilde h^{(0)} A_{\beta}^{(1)})+\frac{1}{2}\tilde w^{(2)}\nonumber\\
    & = -\frac{\hbar}{2}(\dv_{x^i}\tilde h^{(1)} A_{p^i}^{(0)}
    +\dv_{x^i}\tilde h^{(0)} A_{p^i}^{(1)}) + \frac{\hbar}{2} m^{(1)}_{x^i p^i}\nonumber\\
    & = -\frac{e\hbar^2}{2}\p_{ij}^2 V A_{p^i}^{(0)} A_{p^j}^{(0)} - \frac{e^2\hbar^2}{2} t_{ij} \p_i V \p_j V - \frac{e\hbar^2}{2} \p_{ij}^2 V g_{ij}.
\end{align}
Gathering above, we have the gauge-non-invariant diagonal Hamiltonian to the second order as
\begin{align}\label{eq:tilde h blount}
    \tilde h &= \varepsilon(p) - e V(x) + e\hbar \p_i V A^{(0)}_{p^i}-\frac{e\hbar^2}{2}\p_{ij}^2 V A_{p^i}^{(0)} A_{p^j}^{(0)} - \frac{e^2\hbar^2}{2} t_{ij} \p_i V \p_j V - \frac{e\hbar^2}{2} \p_{ij}^2 V g_{ij} +\cO(\hbar^3)\nonumber\\
    & = \varepsilon(p) - e V\left(x^i - \hbar A_{p^i}^{(0)}  \right)  - \frac{e^2\hbar^2}{2} t_{ij} \p_i V \p_j V - \frac{e\hbar^2}{2} \p_{ij}^2 V g_{ij}+\cO(\hbar^3).
\end{align}
We can then define a gauge-invariant diagonal Hamiltonian through 
\begin{align}
    \underline{h}(x,p) \equiv \tilde h(\underline{x},p)
\end{align}
where $\underline{x}^i\equiv x^i+\hbar A_{p^i}^{(0)}$, following the notation in \cite{mangeolle2024quantum}. By identifying $x^i$ as the quantum operator $\ii \frac{\p}{\p p^i}$, $\underline{h}$ is consistent with the band-diagonal Hamiltonian studied in \cite{blount_1962_formalisms}. In particular, let us consider a Weyl fermion which has a similar quantum geometry as the Dirac fermion in \cite{Blount_extension} but is non-degenerate. For a right-handed Weyl fermion described by the Hamiltonian $H_0(\vec{p}) = \vec{p}\cdot \vec{\sigma}$ where $\vec{\sigma}=  (\sigma^x,\sigma^y,\sigma^z)$ are the Pauli matrices, the tensors $g_{ij},t_{ij}$ of the negative energy band are given by
\begin{align}
    g_{ij} = \frac{1}{4}\p_{p_i}\hat p \cdot \p_{p_j} \hat p = \frac{1}{4 p^2}\left( \delta_{ij} - \hat{p}_i  \hat{p}_j\right),
\end{align}
where $\hat p = \vec{p}/p,p=\abs{\vec{p}}$, and $t_{ij} = \frac{1}{p} g_{p_i p_j}$. Thus, the corrections that involves the gradients of the potential $V$ in $\underline{h}$ are given by
\begin{align}
    \Delta \underline{h} = -\frac{e^2\hbar^2}{8 p^3} \left( \delta_{ij} - \hat{p}_i  \hat{p}_j\right) \p_i V \p_j V - \frac{e\hbar^2}{8 p^2} \left( \delta_{ij} - \hat{p}_i  \hat{p}_j\right)\p_{ij}^2 V +\cO(\hbar^3). 
\end{align}
This matches Eq.(12) and Eq.(13) in \cite{Blount_extension} in the massless limit.
Notice that this $\underline{h}$ is not equivalent to our definition of a gauge-invariant diagonal Hamiltonian $h$ in \eqref{eq: separable h}. However, as we discussed in the main text, this is merely a choice of representation of the band theory without translational symmetry and will not cause any difference in physical quantities.

Under a uniform magnetic field, the Moyal algebra changes to the magnetic Moyal algebra as described in \cref{sec:magnetic field}. Assuming the system only depends on the kinetic momentum, i.e. ${\sf H} = {\sf H}(\pi_i)$, we can replace every $\omega^{\alpha\beta}$ with $\omega_B^{\pi_i\pi_j}\equiv -e\mathfrak{F}_{ij}$. Hence, the derivation of the band-diagonal gauge-invariant Hamiltonian follows from \eqref{eq:h1} and \eqref{eq:h2}. In order to compare with \cite{Blount_bloch}, it is instructive to derive the diagonal Hamiltonian using explicitly the gauge field $\Lambda_{p^i}$. To this end, we combine \eqref{eq:tilde h 1}, \eqref{eq:tilde h 2} and \eqref{eq:invth} and find that
\begin{align}\label{eq:h1 magnetic field}
    h^{(1)} =\frac{i\hbar}{4}\wab\diag[\acommu{\Lambda^{(0)}_\alpha}
  {\commu{\Lambda^{(0)}_\beta}{h^{(0)}}}] = -\frac{e\hbar}{2}\mathfrak{F}_{ij} m_{ij}^{\mathrm{orb}},
\end{align}
where $m_{ij}^{\mathrm{orb}}\equiv m_{p^i p^j}^{(0)}$ is the orbital magnetization. Next, the second order correction is given by
\begin{align}
    h^{(2)} &= \frac{\hbar}{2}\wab(\dv_\alpha\tilde h^{(1)} A_{\beta}^{(0)}
    +\dv_\alpha\tilde h^{(0)} A_{\beta}^{(1)})+\frac{1}{2}\tilde w^{(2)} \nonumber\\
    &\quad +\hbar^2\wab\wsl \dv_\alpha\tilde h^{(0)} A^{(0)}_\lambda
    \dv_\sigma A^{(0)}_\beta -\frac{\hbar^2}{4}\wab\wsl\dv^2_{\alpha\sigma}\tilde h^{(0)}
    \left(\diag(\acomm{\Lambda^{(0)}_\beta}{\Lambda^{(0)}_\lambda})-4A^{(0)}_\beta
    A^{(0)}_\lambda\right).
\end{align}
Let us temporarily ignore the terms proportional to $Y^{(1)}$ by assuming there is a huge band gap. Then, $h^{(2)}$ simplifies to
\begin{align}\label{eq:h2 magnetic field}
    h^{(2)}&\approx -(e\hbar)^2 \mathfrak{u}_i \p_{p_i}\left(\frac{1}{2}\mathfrak{F}\cdot m^{\mathrm{orb}} + 2 v_j \mathfrak{u}_j\right) - \frac{(e\hbar)^2}{2} v_i \diag\left(\{\mathfrak{U}_j,\p_{p_j} \mathfrak{U}_i\} - 8 \mathfrak{u}_j\p_{p_j}\mathfrak{u}_i\right) \nonumber\\
    &\quad\quad + \ii \frac{(e\hbar)^2}{4}\diag\left(\{\{\mathfrak{U}_j,\p_{p_j} \mathfrak{U}_i\},[\Lambda^{(0)}_{p_i},h_0]\}\right) - \ii \frac{(e\hbar)^2}{4}\diag  \left(\{\mathfrak{U}_j,[\Lambda^{(0)}_{p_j},\frac{1}{2}\mathfrak{F}\cdot m^{\mathrm{orb}} + 2 v_j \mathfrak{u}_j]\}\right) \nonumber\\
    &\quad\quad - \frac{(e\hbar)^2}{8}\mathfrak{F}_{ij} \diag\left(\p_{p_i}\mathfrak{U}_k\p_{p_j}[\Lambda^{(0)}_{p_k},h_0] + \p_{p_i}[\Lambda^{(0)}_{p_k},h_0] \p_{p_j}\mathfrak{U}_k  \right) - \frac{(e\hbar)^2}{4}\mathfrak{F}_{ij} \diag\left(\{\mathfrak{U}_k, \p_{p_i}\Lambda^{(0)}_{p_k} v_j \}\right) \nonumber\\
    &\quad\quad - (e\hbar)^2 \p_{p_i}v_j\left(\diag(\{\mathfrak{U}_i,\mathfrak{U}_j\}) - 4 \mathfrak{u}_i \mathfrak{u}_j\right)
\end{align}
where we defined $\mathfrak{U}_j \equiv \frac{1}{2}\mathfrak{F}_{ij}\Lambda^{(0)}_{p_i}$ and $\mathfrak{u}_j \equiv \diag(\mathfrak{U}_j)$, and $\mathfrak{F}\cdot m^{\mathrm{orb}} \equiv \mathfrak{F}_{ij} m^{\mathrm{orb}}_{ij}$. These new variables appeared in \cite{Blount_bloch}. First, the magnetization as the first order correction in \eqref{eq:h1 magnetic field} is consistent with \cite{Blount_bloch}. The second order corrections, however, does not apparently agree with \cite{Blount_bloch}. In fact, we find both the manipulations enigmatic and notations opaque in \cite{Blount_bloch}.  For example, if we look at Eq.(I.30) in \cite{Blount_bloch}, the term proportional to the velocity $\mathbf{v}_A$ seems to trivially vanish since the commutators between diagonal matrices are zero. Our result corresponding to Blount's $\mathbf{v}_A$ would be the second term in the second line of \eqref{eq:h2 magnetic field} that is, however, nonzero. Moreover, some of the quantities in their Eq.(I.30), like $H$ and $\mathfrak{B}_\mu$, are only defined before the unitary rotation, making the comparison rather difficult. Given above, we will not attempt to make a full comparison to Blount's \cite{Blount_bloch} in this work.

\end{document}